%% file: Boente_Martinez_2015.tex
\documentclass [11pt]{article}
\usepackage[update,prepend]{epstopdf} 

\usepackage{amsfonts}
\usepackage{amsmath}
\usepackage{amsthm}
\usepackage{color}
\usepackage{graphicx}
\usepackage{rotating}
\usepackage{verbatim}
\usepackage{float}
\usepackage{accents}
\usepackage{subfigure}
 \usepackage{url}

\theoremstyle{definition}

\DeclareMathOperator*{\median}{{\mathop{\mbox{\sc med}}}}

\include{definiciones}

\begin{document}


\title{Marginal integration  $M-$estimators for additive models}
 \author{Graciela Boente\\
    \small Facultad de Ciencias Exactas y Naturales, Universidad de Buenos Aires and CONICET, Argentina\\
    Alejandra Mart\'{\i}nez\\
    \small Facultad de Ciencias Exactas y Naturales, Universidad de Buenos Aires and CONICET, Argentina
    }
  \date{}
  \maketitle

\begin{abstract}
Additive regression models have a long history in multivariate nonparametric regression. They provide a  model in which each regression function depends only on a single explanatory variable allowing to obtain estimators  at the optimal univariate rate. Beyond backfitting, marginal integration is a common procedure to estimate each component.  In this paper, we propose a robust estimator of the additive components  which combines 
local polynomials on the component to be estimated and marginal integration. The proposed estimators are consistent and  asymptotically normally distributed.  A simulation study allows to show the advantage of the proposal over the classical one when outliers are present in the responses, leading to   estimators with good robustness and efficiency properties. 

\end{abstract}
 
\noi\textbf{Key Words:}  Additive models; Local $M-$estimation; Kernel weights; Marginal integration; Robustness

\noi \textbf{AMS Subject Classification:} {MSC 62G35; 62G20, 62G05}

\newpage

\normalsize

\section{Introduction}{\label{intro}}
Several authors have dealt with the dimensionality reduction problem in non--parametric regression models. In particular, additive models allow the modelling of a response $Y$ as a sum of smooth functions of individual covariates $\bX=(X_1,\dots, X_d)\trasp$. The advantage of additive models over   general non--parametric regression models is that they allow to circumvent  the so--called \textsl{curse of dimensionality}, which is caused by the fact that the expected
number of observations in local neighbourhoods decreases exponentially as a function of the dimension $d$ of the covariates. More precisely, Stone (1985) defined  the \textsl{curse of dimensionality} as \textsl{``being that the amount of data required to avoid an unacceptably
large variance increases rapidly with increasing dimensionality"}.  This results in the poor convergence rate of the estimators which, as it is well known, depends exponentially on the dimension and on the degree of smoothness of the regression function. 
To be more precise, let $(\bX\trasp,Y)$ be a random vector where $Y\in \real$ is the dependent variable and $\bX\in \real^d$ is the vector of explanatory variables. Consider the  non--parametric regression model $Y=g(\bX)+\sigma(\bX)\eps$ where the error $\eps$ is independent of $\bX$ and centered at zero and $g:\real^d\to\real$ is the function to be estimated. Stone (1980, 1982) showed that the optimal rate for estimating $g$ is $n^{-\ell/ (2\ell+d)}$ where $\ell$ is the degree of  of smoothness of $g$.

To face this problem, Stone (1985) and Hastie and Tibshirani (1990) considered additive models which generalize   linear models, solve the problem of the \textit{curse of dimensionality} and provide easily interpretable models. Additive models assume that $g(\bx)=\mu+\sum_{j=1}^d g_j(x_j)$ where $\mu$ is the location parameter and the additive   components $g_j:\real\to\real$ satisfy some additional condition to be identifiable such as $\esp g_j(X_j)=0$. One of the advantages of additive models is that they allow   for independent interpretation of the effect of each
variable on the regression function $g$, as in linear regression models. Besides, as shown by Stone (1985),  for such regression models the optimal rate for estimating $g$ is the one-dimensional rate of convergence $n^{-\ell/ (2\ell+1)}$ leading to dimensionality reduction through additive modelling.

Several estimation procedures to fit additive models have been proposed in the literature. The iterative method called \textsl{backfitting} proposed by Buja, Hastie and Tibshirani (1989) and Hastie and Tibshirani (1990) is one of the most popular procedures. Even if the procedure converges quickly, its iterative nature makes difficult to analyse its statistical properties. Besides, the backfitting algorithm does not allow to estimate   derivatives  since it does not give a closed form for the estimator.  On the other hand, the \textsl{marginal integration} procedure proposed by Tj{\o}stheim and Auestad (1994) and Linton and Nielsen (1995) and generalized by Chen \textsl{et al.} (1996)  allows for the derivation of a closed form
for the estimator and has been shown to work very well in simulation studies, see Sperlich \textsl{et al.} (1999).  In particular, Severance--Lossin and Sperlich (1999) combine the integration procedure with a local polynomial approach to estimate  simultaneously the additive components  and its derivatives. When first moments exist, the idea beyond  marginal integration  is to estimate  the marginal effects defined as the expectation of $Y$ with respect to the random error $\eps$ and all the covariates except the $X_\alpha$ which is fixed. The marginal effect says how $Y$ varies in average when $X_\alpha$ varies. If the true multivariate function $g$ is additive, the marginal effects match with the additive components $g_\alpha$, except for the constant $\mu$, allowing for precise estimations under an additive model. The estimators are obtained estimating, in a first step,  the multivariate function $g$  and then, using the marginal integration procedure to obtain the marginal effects.

As in other non--parametric settings, the estimators obtained through marginal integration can be seriously affected by a relatively small proportion of atypical observations if the smoother chosen to estimate the multivariate function $g$ is not resistant to outliers in the response variable. As is well known, in a non--parametric framework outlying observations can be even more dangerous than in a parametric model, since extreme points affect the scale and the shape of any estimate of the regression function based on local averaging, leading to possible wrong conclusions. This has motivated the interest in combining the ideas of robustness with those of smoothed regression, to develop procedures which will be resistant to deviations from the central model in non--parametric regression models. In this paper, we go further and we focus on robust estimators for additive models leading to  reliable non--parametric  regression estimators when atypical responses arise and which attain a univariate rate of convergence. Indeed, we seek  for consistent estimators of the regression function $g$ without requiring moment conditions on the errors $\eps_i$ so as to include the well--known  $\alpha$-contaminated neighbourhood for the errors distribution. More precisely, in a robust framework, one looks for procedures that remain valid when $\eps_i\sim F_0\in {\cal F}_{\alpha}= \left\{G: \;  G(y) = (1-\alpha)G_0(y) + \alpha H(y) \right\}$, with $H$ any symmetric distribution and $G_0$ a central model with possible first or second moments.  No moment conditions are required to the errors so that outliers correspond to deviations on the errors distribution. 

In this framework, some resistant procedures for additive models based on  $M-$smoothers have been considered previously in the literature. Bianco and Boente (1998) considered robust estimators for additive models using kernel regression. Their approach, which is a robust version of that considered in Baek and Wehrly (1993), has the drawback of assuming that $Y -g_j(X_j)$ is independent from $X_j$, which is difficult to justify or verify in practice. Robust estimators based on backfitting and penalized splines $M-$estimators have been proposed  for generalized additive models by Alimadad and Salibian--Barrera (2012) and Wong \textsl{et al.} (2014). In the particular case of the non--parametric regression model $Y=g(\bX)+\sigma(\bX)\eps$, with $g(\bx)=\mu+\sum_{j=1}^d g_j(x_j)$, the procedures considered in Alimadad and Salibian--Barrera (2012) and Wong \textsl{et al.} (2014) assume that the scale function is known. For generalized additive models with nuisance parameters, Croux \textsl{et al.} (2011) provides a robust fit using penalized splines, while recently Boente \textsl{et al.} (2015) combines the backfitting algorithm with robust univariate scale equivariant smoothers to provide robust estimators under an additive model with unknown scale. However, up to our knowledge,  except for the estimators considered in Bianco and Boente (1998), the asymptotic distribution of the  estimators mentioned above has not been obtained.   

On the other hand, Li \textsl{et al.} (2012) introduced robust estimators of the additive components $g_j$  using
local linear regression and marginal integration and derived their asymptotic behaviour. Besides assuming that the scale is known, the main disadvantage of the  procedure defined in Li \textsl{et al.} (2012) is that the estimators solve the curse of dimensionality only when $d\le 4$, since the local multivariate polynomial considered is of order one. This effect has also been described for the classical   estimators, based on a local least squares approach, by Hengartner and Sperlich (2005) and Kong \textsl{et al.} (2010) who noted that to solve the curse of dimensionality the order  of the local polynomial approximation should increase with the dimension of the covariates, leading to a  higher numerical  complexity. To avoid this problem,  Severance--Lossin and Sperlich (1999) modified the  initial estimators used in the integration procedure,  using higher order kernels and local polynomials that depend only on the covariate $X_j$ related to  the $j-$th additive component to be estimated. 

In this paper, we introduce robust estimators of the additive components using local polynomials on the component to be estimated and marginal integration. In this sense, our approach can be viewed as a robust version of the estimators defined in Severance--Lossin and Sperlich (1999). Besides, our proposal allows to provide also robust estimators of the derivatives of the marginal components. Taking into account that in some studies, specially in many biological situations, missing responses may arise, we will provide a unified approach for complete data sets and for data sets in which  responses are missing at random.  The rest of the paper is organized as follows. Section \ref{est} introduce the family of estimators to be considered. Consistency results and the asymptotic distribution are derived in Sections \ref{consist} and \ref{ASMq}, respectively. Finally, the results of a numerical experiment conducted to evaluate the performance of the proposed procedure with respect to its classical counterpart defined in Severance--Lossin and Sperlich (1999) are reported in Section \ref{monte}. Proofs relegated to the Appendix.

\section{The estimators}{\label{est}}
 We will consider robust inference with an incomplete data set $\left(\bX_i\trasp,Y_i,\delta_i\right)\trasp$, $1\le i \le n$, where $\delta_i=1$ if $Y_i$ is observed and $\delta_i=0$ if $Y_i$ is missing. Let $(\bX\trasp,Y,\delta)\trasp$ be a random vector with the same distribution as $\left(\bX_i\trasp,Y_i,\delta_i\right)\trasp$ and assume that  $(\bX\trasp,Y)\trasp$ satisfies the additive model 
$
Y=\mu+\sum_{j=1}^d g_j(X_j)+\sigma(\bX)\eps\;,
$
 where the error $\eps$ is independent of $\bX$ with {symmetric} distribution $F_0(\cdot)$, that is, we assume that the error's scale equals 1 to identify the scale function. Hence, when second moments exist, we have that  $\esp(Y|\bX)=g(\bX)=\mu+\sum_{j=1}^d g_j(X_j)$ and $\sigma^2(\bX)=\esp((Y-g(\bX))^2|\bX)$ is   the conditional variance
function. Some additional conditions to be discussed below on the marginal components need to be require in order to guarantee identifiability. 

Our aim is to estimate the non--parametric regression components $g_j$ and its derivatives in a robust way with the data set at hand. An ignorable missing mechanism will be imposed by assuming that $\delta$ and $Y$ are conditionally independent given $\bX$, i.e., 
\begin{equation}
\prob\left(\delta=1\vert Y,\bX \right)=\prob\left(\delta=1\vert \bX \right)=p\left(\bX \right)\,.
\label{MAR}
\end{equation}
To define the conditions needed for identifiability, we begin by fixing some notation.  We   will  partition $\bX_i$ and $\bx $   into a scalar and a $(d-1)-$dimensional sub--vectors. To avoid burden notation, we denote  $\bX_i= (X_{i, \alpha}, \bX_{i, \undalpha}\trasp)\trasp $ and $\bx = (x_{ \alpha}, \bx_{ \undalpha}\trasp)\trasp$, respectively where  $x_{ \alpha}$ and $ \bx_{ \undalpha}$ are the directions of interest and  not of interest, respectively. As in Linton and Nielsen (1995) and Nielsen and Linton (1998), let  $Q$ be a given  probability  measure with  density $q(\bx)$. Denote as $q_\alpha(x)\,dx=dQ_\alpha(x)$ and $q_\undalpha  d\bx_\undalpha =d Q_\undalpha (\bx_\undalpha )$ where $Q_\alpha$ stands for the $\alpha$-th marginal of the measure $Q$ and $Q_\undalpha   $   corresponds to the marginal of $\bx_{ \undalpha}$.  From now on, the additive components will be identified using the   condition
\begin{equation}
\int\!g_\alpha(x)q_\alpha(x)\,dx=0\quad \mbox{ for all }\quad \alpha=1,\dots,d\,.
\label{identif}
\end{equation}
In particular, when $q=f_{\bX}$ the density of $\bX$, equation (\ref{identif}) corresponds to $\esp g_\alpha(X_\alpha)=0$ for $\alpha=1,\dots,d$. However, to define the estimators we assume that the marginal $q_\undalpha $ is known and so the choice $q=f_{\bX}$ is not a valid one. It is worth noting that the location parameter $\mu$ equals $\int g(\bx) dQ(\bx)$, so it  can be estimated with a root$-n$ rate of convergence using a preliminary regression estimator. For that reason, throughout this paper,  we assume that $\mu=0 $, i.e., $\int g(\bx) dQ(\bx)=0$. Hence, the model to be considered throughout this paper is 
  \begin{equation}\label{modeloaditivobis}
Y=\sum_{j=1}^d g_j(X_j)+\sigma(\bX)\,\eps\;,
\end{equation}
 where the error $\eps$ is independent of $\bX$ and has a {symmetric} distribution $F_0$.

The estimators to be defined are based on   initial local polynomial $M-$estimators of  order $q$ for the regression function $g$, where the polynomial to be considered is expanded only   on the component of interest. More precisely, if  we are interested in estimating $g_\alpha(x)$ the $\alpha$-th additive component, the  estimator to be considered treat differently the covariate $X_\alpha$ which corresponds to the direction of interest and the other ones,  calculating  a robust local polynomial of order $q$ only on the $\alpha-$th direction. 
 As for the estimators introduced in  Severance--Lossin and Sperlich (1999), the use of   higher order kernels will allow to obtain resistant estimators of the additive component which achieve the optimal univariate rate of convergence. 

To reduce the effect of outliers on the regression estimates, we replace the square
loss function in Severance--Lossin and Sperlich (1999) by a function $\rho $ with bounded derivative. Usually, the loss function  depends on a tuning constant $c$ allowing to achieve a given efficiency, so that it can be written as $\rho(u)=\rho_c(u)=c^2\rho_1(u/c)$. Typical choices for the loss function are the Huber--loss function  defined as  $\rho_1(u)=\rho_{\hub} \left( u \right) \, = \,(u^2 / 2)\, \indica_{ |u| \le 1} +(|u| - 1 / 2)\,  \indica_{ |u| > 1} $ otherwise. The Tukey's loss defined as  $\rho_1(u)=\rho_{\tuk}(u) = \min\left(3 u ^2 - 3u ^4 +  u ^6, 1\right)$ provides an example of bounded loss function. The bounded  derivative of the loss function controls the effect of outlying values in the responses. As it is well known, to obtain robust scale invariant estimators, the residuals must be standardized using a robust scale estimator. For that reason, from now on, $\wese(\bx)$ stands for a preliminary robust consistent scale estimator which can be taken, for instance, as the local \textsc{mad} defined in Boente and Fraiman (1989). If the additive model has homoscedastic errors, i.e., if $\sigma(\bx)\equiv \sigma$ for all $\bx$, the estimator $\wese(\bx)\equiv \wese$ to be considered below can be defined as  the \textsc{mad} of the residuals obtained with a simple and robust nonparametric regression estimator, such as the local median. 

Assume that we are interested in estimating $g_\alpha$ which is assumed to be  a continuously differentiable function up to order $q$. Denote as $g_\alpha^{(\nu)}(x_\alpha)$   the derivative of order $\nu$ of the component $g_\alpha$ and let $\bbe^{(\alpha)}(\bx)=\bbe (\bx)=(g(\bx), g^\prime_\alpha(x_\alpha),\dots,g_\alpha^{(q)}(x_\alpha)/q\!)\trasp$. An estimator of $\bbe^{(\alpha)}(\bx)$ can be defined as the value $\wbbe^{(\alpha)}(\bx)=\wbbe(\bx)=(\wbeta_0(\bx),\wbeta_1(\bx),\dots,\wbeta_q(\bx))\trasp$  such that
\begin{equation}\label{probq}
\wbbe^{(\alpha)}(\bx)=\wbbe(\bx)=\argmin_{(\beta_0,\beta_1,\dots,\beta_q)}\sum_{i=1}^n\delta_i\, \itK_{\bH_d}(\bX_i-\bx) \, \rho\left(\frac{Y_i-\left[\beta_0+\sum_{j=1}^q \beta_j (X_{i\alpha}-x_{\alpha})^j\right]}{\wese(\bx)}\right)
\end{equation}
with $\itK_{\bH_d}(\bX_i-\bx)=(\mbox{det}(\bH_d))^{-1}\itK(\bH_d^{-1}(\bX_i-\bx))$,   $\itK(\bx)=\prod_{j=1}^d K_j(x_j)$ with $K_j:\real \to \real$  univariate kernels and $\bH_d=\mbox{diag}(h_1,\dots, h_q)$ is diagonal bandwidth matrix. When there is no confusion, we will avoid the superscript $ (\alpha)$ to avoid burden notation.

The preliminary estimator of the regression function $g(\bx)$ denoted $\wtg_{\eme_{q,\alpha}}(\bx)$ is defined as $\wtg_{\eme_{q,\alpha}}(\bx)=\wbeta_0(\bx)$, where the letter M indicates that we are using a local $M-$estimator and the subscripts \lq\lq $q,\alpha$\rq\rq\ indicate  the order of the local polynomial   used on the $\alpha$-th component of $\bx$.

Finally, the robust estimator of the $\alpha$-th component is obtained through the marginal integration procedure as 
\begin{equation}
{\wg}_{\alpha, {\eme_{q,\alpha}}}(x_{\alpha})=\int\!\wtg_{\eme_{q,\alpha}}(x_{\alpha},\bu_\undalpha )q_\undalpha (\bu_\undalpha )\,d\bu_\undalpha  =\int\!\be_{1}\trasp\,\wbbe(x_{\alpha},\bu_\undalpha )q_\undalpha (\bu_\undalpha )\,d\bu_\undalpha
\label{estrobq}
\end{equation}
where $\be_j\in\real^{q+1}$ is the vector with its $j$-th coordinate equals $1$ and the other ones equal $0$. Moreover, an estimator of the derivative of order $\nu$, $1\leq \nu\leq q$ of $g_\alpha$ is given by
$$\wg_{\alpha, \eme_{q,\alpha}}^{(\nu)}(x_\alpha)=\nu!\int\! \wbeta_\nu(x_{\alpha},\bu_\undalpha )q_\undalpha (\bu_\undalpha )\,d\bu_\undalpha  =\nu!\int\!\be_{\nu+1}\trasp\,\wbbe(x_{\alpha},\bu_\undalpha )q_\undalpha (\bu_\undalpha )\,d\bu_\undalpha \;.$$
Finally, the robust estimator of the multivariate regression function $g$ is defined as
$$\wg_{\eme_{q,\alpha}}(\bx)=\sum_{\alpha=1}^d \wg_{\alpha,\eme_{q,\alpha}}(x_\alpha)\,.$$

When $\mu\neq 0$, in the expressions of the   marginal component estimators   an estimator   $\wmu$ of $\mu$ should be substracted in order to obtain consistent estimators, that is, the estimator of $g_\alpha$ equals ${\wg}_{\alpha, {\eme_{q,\alpha}}}(x_{\alpha})=\int\!\wtg_{\eme_{q,\alpha}}(x_{\alpha},\bu_\undalpha )q_\undalpha (\bu_\undalpha )\,d\bu_\undalpha -\wmu$,
while the estimator of  the multivariate regression function $g$ is  $\wg_{\eme_{q,\alpha}}(\bx)=\wmu+\sum_{\alpha=1}^d \wg_{\alpha,\eme_{q,\alpha}}(x_\alpha)$. A possible choice for $\wmu$ is to compute a robust location estimator $\wa$ of the residuals $Y_i-\sum_{j=1}^d \int\!\wtg_{\eme_{q,\alpha}}(X_{i,\alpha},\bu_\undalpha )q_\undalpha (\bu_\undalpha )\,d\bu_\undalpha$ and to define $\wmu=\,-\,\wa/(d-1)$. The practitioner may also choose as location estimator   $\wmu=(1/d)\sum_{j=1}^d \wmu_j$ where $\wmu_j=\int\!\wtg_{\eme_{q, j}}(\bu)\,dQ(\bu) $. However,  this estimator does not necessary have a root$-n$ order of convergence, a fact which has already been  mentioned  by Sperlich \textsl{et al.} (1999) for the classical estimators.

It is worth noting that when $\rho$ is continuously differentiable with derivative $\rho^{\prime}=\psi$, $\wbbe^{(\alpha)}(\bx)$ satisfies the following system of equations
\begin{equation}\label{22alfa}
\bPsi_{n,\alpha}(\wbbe^{(\alpha)}(\bx), \bx,\wese(\bx))=\bcero_{d+1}\,,
\end{equation}
where $\bPsi_{n,\alpha}(\bbe, \bx,\sigma)=(\Psi_{n,\alpha,0}(\bbe, \bx,\sigma),\dots ,\Psi_{n,\alpha,q}(\bbe, \bx,\sigma))\trasp$ and $\Psi_{n,\alpha,\ell}(\bbe, \bx,\sigma)$ is defined  for $\ell=0,1,\cdots, d$ as
$$
\Psi_{n,\alpha,\ell}(\bbe, \bx,\sigma)=\sum_{i=1}^n \delta_i\, \itK_{\bH_d}(\bX_i-\bx) \, \psi\left(\frac{Y_i-\beta_0-\sum_{j=1}^q \beta_j(X_{i\alpha}-x_\alpha)^j}{\sigma}\right) \,(X_{i\alpha}-x_\alpha)^\ell   \;.  
$$

\section{Consistency}{\label{consist}}
In this section, we will show that the  estimators defined in Section \ref{est}  are   strongly consistent.  
Recall that the preliminary local  $M-$estimator based on local polynomials of order  $q$, $\wtg_{\eme_{q,\alpha}}(\bx)$, is adapted to the additive component $\alpha $ we want to estimate. Hence,  we   fix $\alpha=1,\dots,d$ and to derive strong consistency results for the estimator of the additive component $g_\alpha$, we  state  the conditions adapted to the choice of   $\alpha$. The kernels to be used are also adapted to this framework. However, in order to allow more flexibility, we will not restrict the bandwidth choice to  $h_{\alpha,n}=h_n$ and $h_{j,n}=\wth_n$ for $j\ne \alpha$ allowing different bandwidths for each component. 

In what follows, $\itC$ stands for any compact set  and for any function $m:\real^d \to \real$ we denote as $i(m)= \inf_{\bx\in \itC}m(\bx)$. Let $1\leq \alpha\leq d$ be fixed and denote as $s_{i,j}^{(\alpha)}=\int\,u_\alpha^{i+j} \itK(\bu)\,d\bu=\int\,u^{i+j} K_{\alpha}(u)du$,  $0\leq i,j\leq q$ with   $\bu=(u_1,\dots,u_d)\trasp$. The following set of assumptions will be needed.

\begin{enumerate}
\item[\textbf{A0}] The product measure $Q$ has compact support ${\itS}_Q$ contained in the support ${\itS}_f$ of $f_\bX$.  
\item[\textbf{A1}] $(\bX_i\trasp,Y_i,\delta_i)\trasp$, $1\le i\le n$ are i.i.d. vectors satisfying (\ref{MAR}). Moreover,  $(\bX_i\trasp,Y_i)\trasp$ fulfil the additive model (\ref{modeloaditivobis}) where the functions $g_{\alpha}$ verify (\ref{identif}).

\item[\textbf{A2}] The density function, $f_{\bX}(\bx)$, of  $\bX$ and the missingness probability   $p\left(\bx\right)$ are bounded over the compact $\itC\subset {\itS}_f$  and such that $i(p) >0$. $i(f_\bX) >0$. Moreover,    $p$   and $f_\bX$ are continuous in a neighbourhood of $\itC$. 

\item[\textbf{A3}] $\sigma(\bx)$ and $g(\bx)$ are continuous functions of $\bx$ in a neighbourhood of $\itC$ and $i(\sigma)>0$. 

\item[\textbf{A4}]  
\begin{itemize}
\item[a)] For all $j=1,\dots,d$, the marginal component  $g_j$ is continuously differentiable in a neighbourhood of the support, $\itS_j$, of the density of $X_j$ with derivative $g_j^{\prime}=g_j^{(1)}$ bounded. 
\item[b)] $g_\alpha$ is $(q+1)-$times continuously differentiable. 
\end{itemize}
 
\item[\textbf{A5}] 
\begin{itemize}
\item[a)] The kernel function $\itK:\real^d\to\real$ is such that $\itK(\bx)=\prod_{j=1}^d K_j(x_j)$, where $K_j:\real \to \real$ have bounded support, say $[-1,1]$ and  $\int K_j(u)du=1$. Besides, $K_j:[-1,1]\to \real$ are even, bounded functions and  Lipschitz continuous of order one. 
\item[b)] The matrix $\bS^{(\alpha)}=(S_{jk}^{(\alpha)})_{1\le i, j\le q+1}$  is positive definite, where $S_{ij}^{(\alpha)}=s_{i-1,j-1}^{(\alpha)}$   for $1\leq j,k,\leq q+1$.
\end{itemize}

\item[\textbf{A6}]  The bandwidth sequences   are such that  $h_{j,n}\to 0$ and ${n\prod_{j=1}^dh_{j,n}}/{\log n}\to \infty$.
 
 \item[\textbf{A7}]  The function $\rho$ is an  even and three times continuously differentiable function with bounded derivatives $\psi=\rho^{\prime}$,   $\psi^{\prime}$ and $\psi^{\prime\prime}$. Furthermore,    $\esp(\psi^{\prime}(\eps))> 0$ and $\zeta(u)=u\psi^{\prime}(u)$ and $\zeta_2(u)=u\psi^{\prime\prime}(u)$ are bounded.
 
 \item[\textbf{A8}] The scale estimator $\wese(\cdot)$ satisfies that $\sup_{\bx\in\itC}|\wese(\bx)-\sigma(\bx)|\convpp 0$.

\end{enumerate}

\noi \textbf{Remark \ref{consist}.1.} Assumptions \textbf{A3} to \textbf{A6} are standard conditions to derive consistency results in nonparametric regression models. Assumption \textbf{A1} establishes that the model is an additive one where the components are identifiable. On the other hand,   \textbf{A0} is a standard condition when using marginal integration procedures. It is worth noting  that \textbf{A2} implies that some response variables are observed for all $\bx\in\itC$, which is a common assumption in the literature of missing data.   Note that \textbf{A5} implies that $s_{0,0}^{(\alpha)}=1$ and $s_{i,j}^{(\alpha)}=0$ if $i+ j$ is odd. Assumptions \textbf{A1} and \textbf{A7} imply that $\esp \psi(\eps/\sigma) =0$ for any $\sigma>0$. Assumption \textbf{A7} is a standard condition on the score function when local polynomials and scale estimators are considered. Finally, \textbf{A8} requires uniform consistency of the preliminary scale estimator which is needed to derive uniform consistency of the initial regression function. Note that \textbf{A4} entails that    the derivative of $g_\alpha$ of order $q+1$, $g_\alpha^{(q+1)}$, is bounded in $\itS_\alpha$.

 \noi \textbf{Remark \ref{consist}.2.} It is easy to see that \textbf{A3} and \textbf{A8} imply that the robust scale estimator has upper and lower uniform  bounds almost surely. More precisely,  if $A=\inf_{\bx\in \itC}\sigma(\bx)/2$ and $B=(3/2)\sup_{\bx\in \itC}\sigma(\bx)$ we have that 
 \begin{equation}
 \prob\left(\exists \;n_0 \mbox{ such that for all } n\ge n_0 \mbox{ and for all }\, \bx \in \itC\quad A<\wese(\bx)<B\right)=1\,.
 \label{A7prob}
 \end{equation} 
 On the other hand, if we denote as $\wa_\sigma(\bx)=\sigma(\bx)/\wese(\bx)$,   \textbf{A3} and \textbf{A8} imply
\begin{equation}\label{conproba}
\sup_{\bx\in\itC}|\widehat{a}_\sigma(\bx)-1|\convpp 0\,.
\end{equation}

\vskip0.1in 
From now on, we denote as $\bH^{(\alpha)}$ the diagonal matrix given by $
\bH^{(\alpha)}=\mbox{diag}(1,h_\alpha,h_\alpha^2,\dots,h_\alpha^q)$.

\vskip0.1in

\noi \textbf{Proposition \ref{consist}.1.}\textsl{ Let $\itC\subset\itS_f$ be a compact set such that \textbf{A2} is satisfied. Assume that \textbf{A1}     to \textbf{A8} hold. Then, there exists a solution $\wbbe(\bx)$ of (\ref{22alfa}) such that $\sup_{\bx\in\itC}\|\bH^{(\alpha)}\{\wbbe(\bx)-\bbe(\bx)\}\|\convpp 0$ where $\bbe(\bx)=(g(\bx),g_\alpha^{(1)}(x_\alpha),\dots,g_\alpha^{(q)}(x_\alpha))\trasp$ and $g_\alpha^{(1)}=g_\alpha^\prime$.
}

\vskip0.2in
Theorem \ref{consist}.1 shows the consistency of the marginal integration estimator of the regression function and its derivatives when using local polynomials of order $q$ in the direction $\alpha$. We omit the proof of Theorem \ref{consist}.1 since it follows straightforwardly from Proposition \ref{consist}.1  using   similar arguments to those considered in the proof of Theorem 3.2.3 in Boente and Mart\'{\i}nez (2015).  

\vskip0.1in
\noi \textbf{Theorem \ref{consist}.1.} \textsl{Assume that \textbf{A0} to \textbf{A8} hold with $\itC=\itS_Q\subset\itS_f$ and some fixed $\alpha$. Denote as $\itC_\alpha$  the support of $q_\alpha$. Then, we have that
\begin{enumerate}
\item[a)] $\sup_{x \in \itC_{\alpha}} |\wg_{\alpha,\eme_{q,\alpha}}(x) -g_{\alpha}(x)|  \convpp 0$,
\item[b)] $\sup_{\bx \in \itC} h_\alpha^\nu\;|\wg_{\alpha,\eme_{q,\alpha}}^{(\nu)}(x)  -g_{\alpha}^{(\nu)}(x)|  \convpp 0$.
\end{enumerate}
Furthermore, if for any $\alpha=1,\dots, d$,  \textbf{A0} to \textbf{A8} hold for the kernels used to define the $\alpha-$th additive component estimators and $\itC=\itS_Q$, then  $\sup_{\bx \in \itC} |\wg_{\eme_q}(\bx) -g(\bx)|  \convpp 0$, where $\wg_{\eme_q}(\bx)=\sum_{j=1}^d\wg_{j,\eme_{q,\alpha}}(x_j)$.}

\section{Asymptotic distribution}\label{ASMq}

 In this section, we derive   the asymptotic distribution of the $\alpha-$th additive component estimator. As in Severance--Lossin and Sperlich (1999), we will assume that to compute the preliminary estimator $\wtg_{\eme_{q,\alpha}}(\bx)$  the diagonal bandwidth  matrix  $\bH_d$ is such that its $\alpha-$th diagonal element  equals   $h_\alpha$ and the remaining ones are $\wth$, i.e., we assume that $h_j=\wth$, for $j\ne \alpha$. Moreover,   we will consider two different univariate even and bounded  kernels, $K$ and $L$. The kernel $K$ is positive and used over the $\alpha-$th coordinate of $\bx$, i.e.,  $K_\alpha=K$. On the other hand, the kernel $L$ is used on the remaining components of $\bx$,  that is,  $K_j=L$, for $j\neq \alpha$. Furthermore,   to obtain a univariate rate of convergence for the $\alpha-$th additive component estimator  $L$ will be chosen as  a    kernel of order $\ell\ge 2$, that is, $\int\!L(u)\,du=1$, $\int\!u^sL(u)\,du=0$, for $s=1,\dots,\ell-1$ and $\int\!u^{\ell}L(u)\,du\neq 0$. Clearly, the choice of kernel and bandwidth   as well as the computation of the preliminary estimator need  to be done for each additive component to be estimated, making the method computationally expensive. Thus,  to gain in convergence rate some numerical complexity seems to be necessary.
 
Throughout this section, we will assume an homoscedastic model, that is,  $\sigma(\bx)\equiv \sigma$ so that the additive model can be written as 
$Y=\sum_{j=1}^d g_j(X_j)+\sigma \,\eps$ where the error  $\eps$ is independent of $\bX$ and has a {symmetric} distribution $F_0$ with scale 1, so as to identify $\sigma$. We will also assume that a robust root$-n$ convergent    scale estimator $\wese$ of $\sigma$ is available.  

Due to the kernels choice and the homoscedasticity assumption,   assumptions \textbf{A1}, \textbf{A4} and  \textbf{A5} will be replaced by the following ones.

\begin{itemize}
\item[\textbf{N1}]   $(\bX_i\trasp,Y_i,\delta_i)\trasp$, $1\le i\le n$ are i.i.d. vectors satisfying (\ref{MAR}). Moreover,  $(\bX_i\trasp,Y_i)\trasp$ are such that $Y_i=\sum_{j=1}^d g_j(X_{i,j})+\sigma \,\eps$ where the errors  $\eps_i$ are independent of $\bX_i$ with {symmetric} distribution $F_0$ and  the functions $g_{\alpha}$ verify (\ref{identif}).
\item[\textbf{N2}] For all $j=1,\dots,d$ and $j\neq\alpha$, the marginal component $g_j$ is $\ell$ times continuously differentiable in a neighbourhood of the support $\itS_j$ of the density $X_j$ and   $g_j^{(\ell)}$ is bounded. Besides, $g_\alpha$  is continuously differentiable until order $q+1$ and the derivative $q+1$, $g_\alpha^{(q+1)}$, is bounded in $\itS_\alpha$.
\item[\textbf{N3}] 
\begin{itemize}
\item[a)] The kernel function $\itK:\real^d\to\real$ is such that $K_j=L$ for $j\ne \alpha$. Moreover, $K$ and $L$ are bounded, even, compactly supported and Lipschitz continuous with  $\int\!K(u)\,du=\int\!L(u)\,du=1$. Without loss of generality, we assume that the support of $K$ and $L$ is $[-1,1]$.
\item[b)] The kernel $K_\alpha$ is such that the  matrix $\bS^{(\alpha)}=\left(\int\!u^{i+j}K_\alpha(u)\,du\right)_{0\leq i,j\leq q}$ defined in \textbf{A5}b) is non--singular. 

\item[c)]The kernel $L$ is  a kernel  of order $\ell\ge 2$, that is, $\int\!L(u)du=1$, $\int\!u^j L(u)\,du=0$ if $1\leq j\leq \ell-1$ and $\int\!u^\ell L(u)\,du\neq 0$.
\end{itemize}

\item[\textbf{N4}] The bandwidth sequences $h_j=h_{j,n}>0$ are such that  $h_{j,n}=\wth_n\to 0$ for $j\ne \alpha$,   $h_\alpha=\beta n^{-\frac{1}{2q+3}}$. Moreover, $\wth=\wth_n$ is such that $\wth=o\left(n^{-\frac{q+1}{\ell(2q+3)}}\right)$  and	$n^{\frac{q+1}{2q+3}}\wth^{d-1}/\log{n}\to\infty$. 
\item[\textbf{N5}] The function $q_\undalpha(\bu)$ is continuous and the functions $f_{\bX}(\bu)$ and $ p(\bu) $ are  continuously differentiable up to order $\ell$. Furthermore,    $\sup_{\bx\in\itS_Q} f_\bX(\bx)<\infty$,  $\inf_{\bx\in\itC} f_{\bX}(\bx)>0$    and $\inf_{\bx\in\itC} p(\bx)>0$, where  $\itC\subset\itS_f$ stands for some compact neighbourhood of $\itS_Q$. 
\end{itemize}

Assumptions \textbf{N2} to \textbf{N4} correspond to assumptions A3, A1 and A2 in Severance--Lossin and Sperlich (1999), respectively. Note that the order $\ell$ of  the kernel $L$ is an even number, since   $L$ is an even function. Also, notice that \textbf{N2} implies that $\sup_{\bx\in \itS_Q}|g(\bx)|<\infty$.  
The proof of the asymptotic distribution of the preliminary estimators $\wtg_{\eme_{q,\alpha}}(\bx)$ can be found in Mart\'{\i}nez (2014).
 
Denote as  $\lambda(a)=\esp\psi(\eps_1+a)$ and $\lambda_1(a)=\esp\psi^{\prime}(\eps_1+a)$. Given a symmetric matrix  $\bA\in \real^{m\times m}$,   $\nu_1(\bA)\le \dots\le \nu_m(\bA)$ stand for the eigenvalues of $\bA$.

\vskip0.2in
\noi \textbf{Theorem \ref{ASMq}.1.} \textsl{Assume that \textbf{A0}, \textbf{A2}, \textbf{A7} and \textbf{N1} to   \textbf{N5} hold and  that the function $\lambda(a)$ has bounded Lipschitz continuous derivatives up to order $\ell-1$, in a neighbourhood of $0$. Let  $\wese$ be a consistent estimator of $\sigma$ such that $\sqrt{n}(\wese-\sigma)=O_{\prob}(1)$.  Let $\bx$ be an interior point of $\itS_f$ and $\wbbe(\bx)$ be a solution of  (\ref{22alfa}) with     $\wese(\bx)=\wese$, for all $\bx$,  such that
$\sup_{\bx\in \itS_Q}\|\bH^{(\alpha)}[\wbbe(\bx)-\bbe(\bx)]\|\convprob 0$, where  $\bbe(\bx)=(g(\bx),g_\alpha^{(1)}(x_\alpha), \dots, g_\alpha^{(q)}(x_\alpha)/q!)\trasp$ and $g_\alpha^{(1)} =g_\alpha^{\prime}$.  Then, we have that
$$\sqrt{nh_\alpha}[\wg_{\alpha,\eme_{q,\alpha}}(x_\alpha)-g_\alpha(x_\alpha)]\convdist N\left(b_{q,\alpha}(x_\alpha),\sigma_{q,\alpha}^2(x_\alpha)\right)$$ where 
\begin{eqnarray*}
b_{q,\alpha}(x_\alpha)&=& \beta^{\frac{2q+3}{2}}\frac{1}{(q+1)!}\,g_\alpha^{(q+1)}(x_\alpha)\;\be_1\trasp\left(\bS^{(\alpha)}\right)^{-1}\bese_q^{(\alpha)}\,,
\\ 
\sigma_{q,\alpha}^2(x_\alpha)&=& \sigma^2 \,\frac{\esp \psi^2(\eps) }{\left[\esp\psi^{\prime}(\eps)\right]^2}\,\left(\int\!\frac{q_\undalpha^2(\bx_\undalpha)}{f_\bX(x_\alpha,\bx_\undalpha)p(x_\alpha,\bx_\undalpha)}\,d\bx_\undalpha\right)\, \be_1\trasp(\bS^{(\alpha)})^{-1}\bV_\alpha(\bS^{(\alpha)})^{-1}\be_1\;,
\end{eqnarray*} 
with $\bese_{q}^{(\alpha)}=(s_{q,1}^{(\alpha)},\dots, s_{q,q+1}^{(\alpha)})\trasp$ where $s_{q,j}^{(\alpha)}=\int\!K_\alpha(t)t^{q+j}\,dt$ for $j=1,\dots,q+1$ and $\bV_\alpha=\left(v_{sm}^{(\alpha)}\right)_{1\le s,m\le q+1}$  with $v_{sm}^{(\alpha)}=\int u^{s+m-2} K_\alpha^2 (u) du$.
}

\vskip0.2in
\noi\textbf{Remark \ref{ASMq}.1.} It is worth noting that as in other nonparametric settings, the asymptotic bias does not depend on the score function. Moreover,  the score function appears in the asymptotic variance through the quantity
$$V(\psi)= \frac{\esp \psi^2(\eps) }{\left[\esp\psi^{\prime}(\eps)\right]^2}$$
which is similar to that given in the location setting. Hence, to calibrate the estimators to attain a given efficiency it is enough to choose the same tuning constant as in a location model.
  
Assume that the smoothing parameter $\wth$ in the directions not of interest is such that $\wth=\gamma n^{-\tau}$. Then,    $\wth=o\left(n^{-\frac{q+1}{\ell(2q+3)}}\right)$    if and only if $\tau>(q+1)/(\ell(2q+3))$. On the other hand,   $n^{\frac{q+1}{2q+3}}\wth^{d-1}/\log{n}\to\infty$ when $\tau<(q+1)/((2q+3)(d-1))$. Hence,    bandwidth rate of $\wth$ must satisfy
\begin{equation}\label{condiciones}
\frac{q+1}{\ell(2q+3)}<\tau<\frac{q+1}{(2q+3)(d-1)}\,,
\end{equation} 
which implies that the practitioner must choose a kernel $L$ with order  at least the dimension of the covariates, i.e., $ \ell\ge d$.

\vskip0.2in
\noi \textbf{Theorem \ref{ASMq}.2.} \textsl{Assume that \textbf{A0}, \textbf{A2}, \textbf{A7} and \textbf{N1} to   \textbf{N5} hold and  that the function $\lambda(a)$ has bounded Lipschitz continuous derivatives up to order $\ell-1$, in a neighbourhood of $0$. Let  $\wese$ be a consistent estimator of $\sigma$ such that $\sqrt{n}(\wese-\sigma)=O_{\prob}(1)$. Let $\bx$ be an interior point of $\itS_f$ and $\wbbe(\bx)$ be a solution of  (\ref{22alfa}) such that
$\sup_{\bx\in \itS_Q}\|\bH^{(\alpha)}[\wbbe(\bx)-\bbe(\bx)]\|\convprob 0$ where  $\bbe(\bx)=(g(\bx),g_\alpha^{(1)}(x_\alpha), \dots, g_\alpha^{(q)}(x_\alpha)/q!)\trasp$ and $g_\alpha^{(1)} =g_\alpha^{\prime}$.  Then,
we have that for $\nu=1,\dots,q$
 $$\sqrt{nh_\alpha}h_{\alpha}^\nu[\wg^{(\nu)}_{\alpha,\eme_{q,\alpha}}(x_\alpha)-g^{(\nu)}_\alpha(x_\alpha)]\convdist N\left(b^{(\nu)}_{q,\alpha}(x_\alpha),\sigma^{2}_{\nu,q,\alpha}(x_\alpha)\right)$$ where
\begin{eqnarray*}
b^{(\nu)}_{q,\alpha}(x_\alpha)&=&\nu! \,\beta^{\frac{2q+3}{2}}\,\frac{1}{(q+1)!}\,g_\alpha^{(q+1)}(x_\alpha)\;\be_{\nu+1}\trasp\left(\bS^{(\alpha)}\right)^{-1}\bese_q^{(\alpha)}\,,
\\ 
\sigma^{ 2}_{\nu,q,\alpha}(x_\alpha)&=&(\nu!)^2\,\sigma^2 \,\frac{\esp \psi^2(\eps) }{\left[\esp\psi^{\prime}(\eps)\right]^2}\,\left(\int\!\frac{q_\undalpha^2(x_\undalpha)}{f_\bX(\bx)p(\bx)}\,dx_\undalpha\right)\be_{\nu+1}\trasp(\bS^{(\alpha)})^{-1}\, \bV_\alpha\, (\bS^{(\alpha)})^{-1}\be_{\nu+1}\,,
\end{eqnarray*} 
with $\bese_{q}^{(\alpha)}$ and $\bV_\alpha$ given in Theorem \ref{ASMq}.1.}

It is worth noting that $\bese_q^{(\alpha)}$ is such that its $j-$th component, $1\le j\le q+1$, equals 0  when $ q+j $ is odd since $  K_\alpha$ is an even function. Hence if $q+\nu+1$ is odd, or equivalently, when $q-\nu$ is even, the bias will be 0. Hence, the bias term in the estimation of $g^{(\nu)}_\alpha$ appears only when $q-\nu$ is odd.

\section{Monte Carlo Study}{\label{monte}}

This section contains the results of 
a simulation study conducted with the aim of comparing the performance of   estimator defined in Section \ref{est} with that of its  classical counterpart introduced in Severance--Lossin and Sperlich (1999), which corresponds to the choice $\rho(u)=u^2$. 
We have performed $N = 500$ replications taking samples of size $n = 500$ when the dimension of the covariates is $d=2$ and $d=4$.
We considered samples without outliers and also samples
contaminated in   different ways. For $d=2$, we also included in our experiment cases where the response
variable may be missing. All computations were carried out using an  \texttt{R} implementation of our algorithm, which can be provided upon request.

To generate missing responses, we first generated observations $(\bX_{i}\trasp, Y_{i})\trasp$ satisfying the additive model 
$Y=g_0(\bX)+u= \mu_0+\sum_{j=1}^d g_{0,j}(X_j)+u\,,$
where $u=\sigma_0\,\eps$. 
Then, we generate $\{\delta_i\}_{i=1}^n$ independent 
Bernoulli random variables such that 
$\prob\left(\delta_i=1\vert Y_i,\bX_i \right)=\prob\left(\delta_i=1\vert \bX_i \right)=p\left(\bX_i \right)$.  
When $d=2$ we used two different missing probabilities: $p(\bx)\equiv 1$, which corresponds to  the case where all the responses are observed, and $p(\bx)=p_{2}(\bx)= \mbox{0.4}+\mbox{0.5}(\cos(x_{1}+0.2))^2$,
which yields around 31.5\% of missing responses. For $d = 4$, we only report the results for $p(\bx)\equiv 1$.

In all cases, we considered polynomials of order $q=1$. The smoothers were computed using the Epanechnikov kernel $K_1(u)=K_2(u)=0.75 \, (1-u^{2}) \indica_{[-1,1]}(u)$ when $d=2$, while for $d=4$ we choose  $K_\alpha $ as the  Epanechnikov kernel and $L$ the fourth order kernel $L(u)= (15/32)(1-u^2)(3-7u^2)\indica_{[-1,1]}(u)$ when estimating $g_\alpha$.  We compared the classical marginal integration estimator  denoted $\wg_{\cla}$ with the robust marginal integration estimator, denoted $\wg_{\rob}$, using the Huber's loss function with tuning constant $c=1.345$. To identify  the marginal component estimators,   we added a subscript  indicating the additive component label.

The performance of each estimator $\wg_j$ of $g_j$, $1\leq j\leq d$, was measured through the following approximated 
integrated squared error (\textsc{ise}):
$${\mbox{\textsc{ise}}}(\wg_j) =  \frac 1{\sum_{i=1}^n \delta_i} \sum_{i=1}^n \left(  g_{j}\left( X_{ij}\right)  - \wg_j\left(  X_{ij}\right)  \right) ^2 \, \delta_i \,,
$$ 
where $X_{ij}$ is the $j$th component of $\bX_{i}$ and $\delta_i = 0$ if the 
$i$-th response was missing and $\delta_i = 1$ otherwise. A similar measure was used to compare the estimators of the regression function $g = \mu + \sum_{j=1}^d g_{j}$.

\subsection{Monte Carlo study with $d=2$ additive components} \label{monted2}

In this case, the covariates were generated from a uniform distribution on the unit square,
$\bX_i = (X_{i,1}, X_{i,2})\trasp \sim U([0,1]^2)$,
the error scale was $\sigma_0=\mbox{0.5}$ and the overall location $\mu = 0$.  We choose as measure in the integration procedure $Q=U([0,1]^2)$ and the integral in (\ref{estrobq}) was approximated as the mean over  {500} points generated according to $Q$.
 
The additive components were chosen to be
\begin{eqnarray*}
g_{1}(x_1) \,  =   \, 24\left( x_1 - 0.5 \right)^2-2 \, ,
\qquad \qquad g_{2}(x_{2}) \,   =   \, 2\pi\sin(\pi x_{2})-4 \, . 
\end{eqnarray*}
We have fixed both bandwidths $h_1$ and $h_2$ in $0.1$. These are values close to the optimal ones  with respect to the integrated mean square error for the bandwidth $h_\alpha=\beta n^{-1/5}$ given in \textbf{N3} (see Severance--Lossin and Sperlich, 1999).

For the errors, we considered the following settings:
\begin{itemize}
\item $C_0$: $u_i \sim N(0, \sigma_0^2)$.
\item $C_1$:  $u_i\sim (1-0.15) \, N(0,\sigma_0^2)+0.15 \, N(15, 0.01)$. 
\item $C_2$: $u_i\sim N(10, 0.01)$ for all  $i$'s such that $\bX_i\in {\cal D}_{0.09}$, 
where $\itD_\eta$ is as above. 
\item $C_3$:  $u_i\sim (1-0.30) \, N(0,\sigma_0^2)+0.30 \, N(15, 0.01)$ for all $i$'s such that $\bX_i\in {\cal D}_{0.3}$. 
\end{itemize}

Case $C_0$ corresponds to samples without outliers and they will 
illustrate the loss of efficiency incurred by using a robust estimator when
it may not be needed. The contamination setting $C_1$ corresponds to
a \textit{gross-error model} where all observations have the same chance
of being contaminated. On the other hand, case $C_2$ is   pathological in the 
sense that all observations with covariates in the square $[0.2, 0.29] \times
[0.2, 0.29]$
are severely affected. Note that we choose an area where the interval length is smaller than the bandwidth, otherwise, the initial estimator will be severely affected. Finally, case $C_3$ is
a gross-error model 
with a higher probability of observing an outlier, but 
these are restricted to the square 
$[0.2, 05] \times [0.2, 0.5]$. 

To summarize the values of ${\mbox{\textsc{ise}}}(\wg_j)$ and ${\mbox{\textsc{ise}}}(\wg)$  over replications, we report  an approximation of the mean integrated squared error, denoted \textsc{mise}, which is obtained by averaging de \textsc{ise} over all replications, and    a more robust measure, denoted \textsc{medise},  that corresponds to the median over replications of the \textsc{ise}. The obtained results are given in 
Table \ref{tab:tabla_dim2}, for the different errors distributions as well as for data sets with and without missing responses. 
 
 \begin{table}[ht!]
    \centering
\footnotesize
 \begin{tabular}{|c|c|c|c|c|c|c||c|c|c|c|c|c|}
 \hline
 & \multicolumn{6}{|c||}{$p(\bx)\equiv 1$} & \multicolumn{6}{|c|}{$p_2(\bx)= \mbox{0.4}+\mbox{0.5}\cos^2(x_{1}+\mbox{0.2})$}\\
 \hline
  & $\wg_{\cla}$ & $\wg_{1,\cla}$ & $\wg_{2,\cla}$ & $\wg_{\rob}$ & $\wg_{1,\rob}$ & $\wg_{2,\rob}$ & $\wg_{\cla}$ & $\wg_{1,\cla}$ & $\wg_{2,\cla}$ & $\wg_{\rob}$ & $\wg_{1,\rob}$ & $\wg_{2,\rob}$    \\\hline
    & \multicolumn{12}{|c|}{\textsc{mise}}\\\hline
 $C_0$ & 0.0188 & 0.0216 & 0.0174 & 0.0172 & 0.0200 & 0.0183 & 0.2506 & 0.1278 & 0.1451 & 0.2540 & 0.1270 & 0.1500 \\
 $C_1$ & 6.4543 & 0.7739 & 0.5208 & 0.9348 & 0.4706 & 0.2517 & 16.3902& 6.2902 & 4.8353 & 8.5197 & 4.5263 & 3.3571 \\
 $C_2$ & 0.1005 & 0.0590 & 0.0532 & 0.0557 & 0.0374 & 0.0363 & 0.3353 & 0.1661 & 0.1823 & 0.2987 & 0.1470 & 0.1708 \\
 $C_3$ & 0.8652 & 0.3662 & 0.3472 & 0.1557 & 0.0811 & 0.0792 & 1.1434 & 0.4921 & 0.4928 & 0.4734 & 0.2252 & 0.2443 \\
 \hline
     & \multicolumn{12}{|c|}{\textsc{medise}}\\\hline
 $C_0$ & 0.0103 & 0.0113 & 0.0111 & 0.0108 & 0.0118 & 0.0115 & 0.0286 & 0.0220 & 0.0220 & 0.0307 & 0.0232 & 0.0234 \\
 $C_1$ & 6.0024 & 0.4251 & 0.4179 & 0.5153 & 0.1474 & 0.1335 & 6.9604 & 0.8206 & 0.7700 & 1.8799 & 0.5898 & 0.5317 \\
 $C_2$ & 0.0850 & 0.0477 & 0.0481 & 0.0300 & 0.0234 & 0.0264 & 0.1174 & 0.0641 & 0.0640 & 0.0708 & 0.0415 & 0.0454 \\
 $C_3$ & 0.8030 & 0.3456 & 0.3249 & 0.0483 & 0.0338 & 0.0363 & 0.8886 & 0.3728 & 0.3476 & 0.1520 & 0.0798 & 0.0702 \\
 \hline
 \end{tabular}
\caption{\label{tab:tabla_dim2} \textsc{mise} and \textsc{medise} of the estimators of 
the regression functions $g$, $g_1$ and $g_2$ under different contaminations, for the complete data and for sets with missing responses.}
\end{table}

As expected, when the data do not contain outliers or missing responses, the robust estimators shows larger \textsc{medise} values than the classical estimators based on the square loss function. In a few cases, the \textsc{mise} values of the robust estimators are slightly smaller than those of the classical ones. However, all these differences are well within the Monte Carlo margin of error. For contaminated errors, the behaviour of the classical and robust estimators are quite different. The contamination  setting $C_1$ is the worst for the estimators defined in Severance--Lossin and Sperlich (1999), since a 15\% of the observations are contaminated with a large residual. Effectively, under $C_1$, the \textsc{mise} of the classical estimator of the regression function $g$ is more than 6 times larger than those of its robust counterpart, while the \textsc{medise} is  $10$ times larger. This difference is smaller when estimating the additive components, but is still important. On the other hand, $C_2$ seems to affect less the  classical estimator. Indeed,  under $C_2$ the \textsc{mise} and \textsc{medise} of $\wg_{\cla}$ are  twice those of $\wg_{\rob}$, while for each additive component, the \textsc{mise} and \textsc{medise} of the classical estimators are a 50\% larger than those of the robust ones. Finally, contamination $C_3$ seems to be more harmful than $C_2$. Effectively,  the reported \textsc{medise} values for the classical regression estimator $\wg_{\cla}$ are  more than $15$ times larger than those of the robust estimator $\wg_{\rob}$, while when estimating each additive component the classical estimators \textsc{medise} is $10$ times larger than those obtained with its robust counterpart. It is worth noting that the ratio between the classical and robust estimators  \textsc{mise} is smaller than when using  the \textsc{medise}, although large values are still obtained. This fact may be explained by the presence of a few samples where the estimators, specially the robust estimator, perform differently from the majority of the samples.
  
When missing responses arise, as one would expect, all estimators have larger \textsc{mise} and   \textsc{medise} values than when $p\equiv 1$ due to the loss of about 31.5\% of responses.  Beyond this fact, similar conclusions can be drawn regarding the advantage of the robust procedure over the classical estimators.

\subsection{Monte Carlo study with $d=4$ additive components} \label{monted4}

For this model we generated covariates $\bX_i=(X_{i1},X_{i2},X_{i3},X_{i4})\sim U([-3,3]^4)$,
independent errors  $\eps_i \sim N(0, 1)$ and $\sigma_0=\mbox{0.15}$. Similarly to what we have set for $d=2$, we chose as measure in the integration procedure $Q=U([-3,3]^4)$ and, as in Section \ref{monted2}, the integral in (\ref{estrobq}) was also approximated as the mean over $500$ points generated according to $Q$.

The additive components chosen are related to those in the numerical study in Severance--Lossin and Sperlich (1999) and correspond to
$$\begin{array}{ll}
g_{0,1}(x_1) = \frac{1}{12}x_1^3,\qquad &
g_{0,2}(x_2) = \sin(-x_2),\\
\\
g_{0,3}(x_3) = \frac{1}{2}x_3^2 - 1.5, \qquad&
g_{0,4}(x_4)=\frac{1}{4}e^{x_4} - \frac{1}{24}(e^3-e^{-3}).
\end{array}$$

In this numerical experiment, the bandwidths were selected using a $K-$fold cross--validation procedure as follows.  As usual, we first randomly partition the data set into $K$ 
disjoint subsets of approximately equal sizes  $\itG_k$, $1 \le k \le K$, so that $\bigcup_{k=1}^K\itG_k
= \{ 1, \ldots, n \}$. For each fixed $(h,\widetilde{h})$, let  $\bh=(h,\widetilde{h})$. Note that when estimating the $\alpha-$th additive component,  the bandwidth used for the $\alpha-$th component is $h$, while on the nuisance directions we use $\widetilde{h}$. Moreover, the kernels are also modified depending on the component to be estimated. More precisely, when estimating $g_\alpha$, for $\ell\ne \alpha$, $K_\ell=L$ the fourth order kernel described above, while $K_\alpha$ is the Epanechnikov kernel. 

Denote as $\wg_{\cla,\bh}^{(-k)}(\bx)$  and  $\wg_{\rob,\bh}^{(-k)}(\bx)$  the classical and robust  marginal integration estimators  computed with the bandwidths $h$ and $\widetilde{h}$, without using the observations with indices in $\itG_k$. The classical $K$--fold cross--validation criterion   given by 
$$L_{\ls}(h,\widetilde{h})=\frac 1n\sum_{k=1}^K \sum_{i\in \itG_k}(Y_i-\wg_{\cla,\bh}^{(-k)}(\bX_i))^2\,,$$
is minimized over a set  $\itH\times \widetilde{\itH}$ of possible bandwidths $(h,\widetilde{h})$. 

On the other hand, as is well known, a robust cross--validation criterion needs to be considered when using robust estimators. The robust  $K-$fold cross-validation method used in this numerical study is related to the procedure defined in Boente \textsl{et al.} (2010) and minimizes over   $\itH\times \widetilde{\itH}$ the robust criterion
$$L_{\rob}(h,\widetilde{h})=\sum_{k=1}^{K}\left\{\left(\median_{i\in \itG_k}\{Y_i-\wg_{\rob,\bh}^{(-k)}(\bX_i)\}\right)^2+\left(\mad_{i\in \itG_k}\{Y_i-\wg_{\rob,\bh}^{(-k)}(\bX_i)\}\right)^2\right\}\,.$$

The number of folds $K$ was set equal to $K=5$. Due to the computational complexity involved, we only considered the contamination schemes  $C_0$ and $C_1$  defined  in Section \ref{monted2}.

To obtain bandwidths satisfying (\ref{condiciones}) with $q=1$, the set $\itH\times \widetilde{\itH}$ of possible values for  $(h,\widetilde{h})$ was chosen satisfying $h=C\,\,n^{-1/5}$ and $\widetilde{h}=C\,n^{-\tau}$ with $\tau=0.12$. The constant $C$ took initially five possible values leading to $\widetilde{\itH}=\{1, 1.5, 2, 2.5, 3\}$. When the minimum, was attained at $\widetilde{h}=3$, the grid was enlarged to include values of  $\widetilde{h}\in\{3.5, 4, 4.5, 5, 5.5\}$. Note that when $\widetilde{h}=1$, $C\simeq 2.11$ and $h=C\,n^{-1/5}$, so we expect in average $3$ observations in each $4-$dimensional neighbourhood. For that reason, to obtain a reliable estimate of the residual scale $\sigma_0$,   independently of the choice of $(h, \widetilde{h}), $a preliminary regression estimator was computed using as bandwidth $\bh_\sigma=(0.93,0.93,0.93,0.93)$. With these bandwidths, we expect an average of 5 points in each 4-dimensional neighbourhood. It is also worth noting that the optimal bandwidth $h$ to estimate $g_\alpha$ in this model lead to very small values and were not taken as possible values of the grid.

In this numerical study, the \textsc{ise} of  few samples was very different to most of the data sets, probably due to the fact that the bandwidth search was not exhaustive. Hence, to provide summary values for ${\mbox{\textsc{ise}}}(\wg_j)$ and ${\mbox{\textsc{ise}}}(\wg)$  over replications, we report  the  median over replications  as well as the trimmed mean over replications of the \textsc{ise}  with  $1\%$ and $5\%$  trimming. Note that the \textsc{medise} corresponds to a $50\%$ trimming.  The  results obtained under $C_0$ and $C_1$ are given in 
Table \ref{tab:tabla_dim4}.

\begin{table}[ht!]
 \centering
 \small
 \begin{tabular}{|c|c|c|c|c|c|c||c|c|c|c|c|}
 \hline
 $\nu$ & & $\wg_{\cla}$ & $\wg_{1,\cla}$ & $\wg_{2,\cla}$ & $\wg_{3,\cla}$ & $\wg_{4,\cla}$ & $\wg_{\rob}$ & $\wg_{1,\rob}$ & $\wg_{2,\rob}$ & $\wg_{3,\rob}$ & $\wg_{4,\rob}$  \\\hline
1\% & $C_0$ & 1.0969 & 0.2076 & 0.3019 & 0.1358 & 0.1824 & 1.4550 & 0.0962 & 0.1360 & 0.2527 & 0.0638 \\
 & $C_1$ & 10.2064 & 1.0854 & 1.1212 & 0.6874 & 0.5816 & 0.3268 & 0.0945 & 0.1080 & 0.1098 & 0.0653 \\\hline
 5\% & $C_0$ & 0.3589 & 0.0788 & 0.1029 & 0.0916 & 0.0547 & 0.3612 & 0.0462 & 0.0498 & 0.0464 & 0.0370 \\
 & $C_1$ & 5.2254 & 0.2199 & 0.2324 & 0.2594 & 0.1994 & 0.3210 & 0.0929 & 0.1060 & 0.1087 & 0.0639 \\\hline
 50\% & $C_0$ & 0.1536 & 0.0577 & 0.0674 & 0.0808 & 0.0371 & 0.1526 & 0.0391 & 0.0415 & 0.0303 & 0.0277\\
 & $C_1$ & 5.2118 & 0.1875 & 0.2033 & 0.2202 & 0.1738 & 0.3109 & 0.0926 & 0.1040 & 0.1072 & 0.0621\\
\hline 
\end{tabular}
\caption{\label{tab:tabla_dim4} Trimmed mean  of the   \textsc{ise} for the estimators of 
the regression functions $g$ and $g_j$, $1\le j\le 4$ under different contaminations. The  trimming values $\nu$ considered equal   $1\%$, $5\%$ and $50\%$.}
\end{table}

The numerical experiment for $d=4$ yields similar conclusions regarding the advantage of the robust procedure over the classical one than in dimension $d=2$. As  expected,   the robust marginal integration estimator is less efficient than the classical estimator for clean data.  Under $C_1$, the  \textsc{ise} trimmed means  of  $\wg_{\cla}$ are more than 15 times  larger than those obtained with $\wg_{\rob}$. Besides, when considering a $1\%$ trimming, the classical estimators of $g_1$, $g_2$, $g_3$ and $g_4$ gives trimmed mean values more than  11, 10, 6 and 8 times larger  than those corresponding to the robust estimator. When considering the $5\%$ trimmed mean and the \textsc{medise}, the difference is not so noticeable as with a 1\% trimming but it is still large, since in all cases the summary measure of classical estimator is at least the double of that corresponding to  the robust estimator.

 \small
 \noi \textbf{Acknowledgements.} This research was partially supported by Grants  \textsc{pip} 112-201101-00339 from \textsc{conicet}, \textsc{pict} 0397 from \textsc{anpcyt} and   20120130100279BA from the Universidad de Buenos Aires  at Buenos Aires, Argentina. 
\normalsize

\renewcommand{\thesection}{\Alph{section}}
\setcounter{section}{0}
{\setcounter{equation}{0}
\renewcommand{\theequation}{A.\arabic{equation}}

\section{Appendix}{\label{append}}
We begin by fixing some notation which will be useful in the sequel.  Denote as
\begin{equation}
 R_{\alpha}(X_{i,\alpha},x_\alpha)=g_\alpha(X_{i,\alpha})-g_\alpha(x_\alpha)-\sum_{j=1}^q g_\alpha^{(j)}(x_\alpha)\frac{(X_{i,\alpha}-x_\alpha)^j}{j!}
 \label{erreX}
 \end{equation}
and  $R(\bX_i,\bx)=\sum_{j\neq\alpha}\{g_j(X_{i,j})-g_j(x_j)\}+R_{\alpha }(X_{i,\alpha},x_\alpha)$.  Furthermore, define 
$$\breve{\bx}_{i,\alpha}=\left(1,\frac{X_{i,\alpha}-x_\alpha}{h_\alpha},\frac{(X_{i,\alpha}-x_\alpha)^2}{h_\alpha^2},\dots,\frac{(X_{i,\alpha}-x_\alpha)^q}{h_\alpha^q}\right)\trasp=(\breve{x}_{i,1,\alpha}, \dots, \breve{x}_{i,q+1,\alpha})\trasp\,.$$ 
Then, $Y_i-\breve{\bx}_{i,\alpha}\bH^{(\alpha)}\bbe(\bx)=\epsilon_i+R(\bX_i, \bx)/\sigma$. Let $U_i=\sigma(\bX_i)\eps_i$ so that $Y_i=g(\bX_i)+U_i$. Denote  $V_i=\sigma(\bX_i)\eps_i/\sigma(\bx)=U_i/\sigma(\bx)$ and for $\br=(\beta_0, h_\alpha \beta_1, h_\alpha^2 \beta_2,\dots,  h_\alpha^q\beta_q)\trasp$ define
  \begin{eqnarray}
  \ell_n(\br)&=&\frac{1}{n }\sum_{i=1}^n \delta_i \rho\left(\frac{Y_i-\beta_0-\sum_{j=1}^q \beta_j(X_{i\alpha}-x_\alpha)^j}{\wese(\bx)}\right)\itK_{\bH_d}(\bX_i-\bx)   \nonumber\\
&=&\frac{1}{n }\sum_{i=1}^n \delta_i \rho\left(\frac{Y_i-\br\trasp \breve{\bx}_{i,\alpha}}{\wese(\bx)}\right)\itK_{\bH_d}(\bX_i-\bx) \,,\label{lnralpha}\\
\bJ_{n,\alpha}(\bx,a)&=& \frac{1}{n } \sum_{i=1}^n \itK_{\bH_d}(\bX_i-\bx)\xi_i  \psi\left(\frac{\sigma(\bX_i)\eps_i}{\sigma(\bx)} \,a\right) \breve{\bx}_{i,\alpha}
\nonumber\\ 
&=&\frac{1}{n } \sum_{i=1}^n \itK_{\bH_d}(\bX_i-\bx)\xi_i  \psi\left(V_i \,a\right) \breve{\bx}_{i,\alpha}\,.\label{jnalpha}
\end{eqnarray}
Given a compact set $\itC\subset \real^d$, we denote as $N_\rho(\itC)$ the minimum number of balls of radius $\rho$ needed to cover $\itC$. Then, we have that  $\itC\in\bigcup_{k=1}^{N_\rho(\itC)}\itB_d(\bx_k,\rho)$ where $\itB_d(\bx_k,\rho)=\{\by\in \real^d: \|\by-\bx_k\|\le \rho\}$ stands for the ball of center $\bx_k$ and radius $\rho$. It is well known that $N_\rho(\itC)\le A_1/\rho^d$ where the constant $A_1$ does not depend on $\rho$. We also denote as $\itB_{d,\rho}= \itB_d(\bcero,\rho)$ the ball centered at $\bcero$ and as $\itV_{d,\rho}=\{\by\in \real^d: \|\by \|= \rho\}$ the sphere of center $\bcero$ and radius $\rho$.

\subsection{Proof of  Proposition \ref{consist}.1}{\label{apend1}}
We begin by proving some Lemmas that will be helpful in the sequel.

The following Lemma corresponds to the well known exponential inequality for bounded variables and can be seen, for instance, in Pollard (1984) or   Ferraty and Vieu (2006). Lemma \ref{apend1}.1 is needed to derive Lemma \ref{apend1}.2 which is a previous step to prove Lemma \ref{apend1}.3.

\noi \textbf{Lemma \ref{apend1}.1.} \textsl{Let $\{Z_i\}_{i\ge 1}$ be   independent random variables such that $\esp Z_i=0$, $|Z_1|\leq M$ and   $\sigma^2=\esp Z_i^2<\infty $. Then, for all $\epsilon>0$ we have that
$$\prob\left(\left|\sum_{i=1}^n Z_i\right|>\epsilon\, n\right)\leq 2\exp\left\{-\frac{\epsilon^2 n}{2\sigma^2\left(1+\epsilon\frac{M}{\sigma^2}\right)}\right\}\,.$$
}

\vskip0.2in

\noi \textbf{Lemma \ref{apend1}.2.} \textsl{Let $\itC\subset \real^d$ be a compact set with non-empty interior  and  $\itI_\delta=[1-\delta,1+\delta] $  where  $\delta\le 1/2$. Let $W_i=W_i(a)=f(Y_i,\bX_i,\delta_i,a)$ be a sequence of random variables such that $|W_i|\leq M$, for all $i$ and $|W_i(a_1)-W_i(a_2)|\le M_1 |a_1-a_2|$. Define   $S_n(\bx,a)=(1/n)\sum_{i=1}^n \left(G_i(\bx,a)-\esp G_i(\bx,a)\right)$ where  $G_i(\bx,a)=\itK_{\bH_d}(\bX_i-\bx)W_i(a) \breve{x}_{i,j, \alpha}^m \breve{x}_{i,\ell, \alpha}^{\wtm}$ where $ m,\wtm=0, 1$ and $1\leq j,\ell\leq q+1$ are fixed. Assume also that      \textbf{A2},  \textbf{A6},  \textbf{A5} hold and denote $A_h=1/\min_{1\leq j\leq d}h_j$. 
\begin{enumerate}
\item[a)]  Let $\theta_n$ and $\rho=\rho_n$ be non-negative numerical sequences converging to zero, such that $\theta_n^{-1}\,A_h\,\rho\leq M_2$, for all $n\ge 1$ and $ {\rho_n}\,\left\{\prod_{j=1}^n h_j\right\}^{-1}\to 0$. Then, there exist $b_1>0$ and $b_2>0$ and a constant $C_0>0$ such that for all $C> C_0$ and for all $n\geq n_0$, 
\begin{eqnarray*}
\prob\left(\theta_n^{-1}\sup_{\bx\in \itB(\bx_k,\rho)\cap\itC} \sup_{a\in \itI_s\cap\itI_\delta}\left|\wtS_{k,s}(\bx,a)\right|>C\right)\leq 4\exp\left\{ -\frac{C^2\theta_n^2 n\prod_{j=1}^d h_j}{b_1A_h^2\rho^2+b_2C\theta_n A_h\rho}\right\}
\end{eqnarray*}
where $\wtS_{k,s}(\bx,a)= S_n(\bx,a)-S_n(\bx_k,a_s)$, $\itC\subset\bigcup_{k=1}^{N_\rho(\itC)} \itB_d(\bx_k,\rho)$ and  $\itI_\delta=[1-\delta,1+\delta]\subset \bigcup_{k=1}^{N_\rho(\itI_\delta)} \itI_s$  with $\itI_s=[a_s-\rho,a_s+\rho]$.
\item[b)]  Let $\theta_n$ and $\rho=\rho_n$ be non-negative numerical sequences converging to zero, such that $\theta_n^{-1}\,A_h\,\rho\leq M_2$ and $ {\rho_n}\,\left\{\prod_{j=1}^n h_j\right\}^{-1}\to 0$. Then, there exist $b_j>0$, $1\le j\le 4$ and a constant $C_0>0$ such that for any $C> C_0$ and for all $n\geq n_0$,
\begin{eqnarray*}
\prob\left(\sup_{\stackrel{\bx\in\itC}{a\in \itI_\delta}}\left|S_n(\bx,a)\right|>C\theta_n\right)\leq \frac{4 A_1}{\rho^{d+1}}\,\left\{\exp\left\{-\frac{C^2\theta_n^2 n \prod_{j=1}^d h_j}{{4 b_3+ 2 b_4  C\theta_n}}\right\}+\exp\left\{ -\frac{C^2\theta_n^2 n\prod_{j=1}^d h_j}{4 b_1 A_h^2\rho^2+2 b_2\, C\theta_n  A_h\rho}\right\}\right\}
\end{eqnarray*}
\item[c)] Let $\theta_n=\sqrt{{\log n}/({n\prod_{j=1}^n h_j})}$. Then, there exists $C$ such that
$$\sum_{n\ge 1}\prob\left(\theta_n^{-1}\sup_{\bx\in\itC} \sup_{a\in \itI_\delta}\left|S_n(\bx,a)\right|>C\right)< \infty\,,$$ 
that is, $\sup_{\bx\in\itC} \sup_{a\in \itI_\delta}\left|S_n(\bx,a)\right|=O_{\aco}\left(\theta_n\right)$.
\end{enumerate}
}

\vskip0.1in
\noi \textbf{Proof.} 
a)  For a fixed $1\le k \le N_\rho(\itC)$, let 
$$\breve{\bx}_{k,i,\alpha}=\left(1,\frac{X_{i,\alpha}-x_{k,\alpha}}{h_\alpha},\frac{(X_{i,\alpha}-x_{k,\alpha})^2}{h_\alpha^2},\dots,\frac{(X_{i,\alpha}-x_{k,\alpha})^q}{h_\alpha^q}\right)\trasp=(\breve{x}_{k,i,1 },\dots, \breve{x}_{k,i,q+1 })\trasp\,,$$
where we avoid the subscript $\alpha$ to simplify the notation.
For $1\leq j,\ell\leq d+1$ define $\itK^{(j\ell)}(\bu)= \itK(\bu) u_{\alpha}^{m(j-1)} u_{\alpha}^{\wtm(\ell-1)}$. Then, we have that $\itK_{\bH_d}(\bX_i-\bx)\breve{x}_{i,j, \alpha}^{m}\breve{x}_{i,\ell, \alpha}^{\wtm}=\itK_{\bH_d}^{(j\ell)}(\bX_i-\bx)$ and $\itK_{\bH_d}(\bX_i-\bx_k)\breve{x}_{k,i,j}^{m}\breve{x}_{k,i,\ell}^{\wtm}=\itK_{\bH_d}^{(j\ell)}(\bX_i-\bx_k)$. Hence, using that the kernels $K_j$ have compact support in $[-1,1]$ we obtain that
\begin{eqnarray*}
|\wtS_{k,s}(\bx,a)|&\leq& \frac{1}{n}\left|\sum_{i=1}^n (G_i(\bx,a)-G_i(\bx_k,a_s))\right|+\frac{1}{n}\sum_{i=1}^n \left|\esp G_i(\bx_k,a_s)-\esp G_i(\bx,a) \right|\\
&\le & \wttS_{k}(\bx,a) +\wttS_{k,s}(\bx_k,a)
\end{eqnarray*}
where
\begin{eqnarray*}
\wttS_{k}(\bx,a)&=& \frac{1}{n}\left|\sum_{i=1}^n (G_i(\bx,a)-G_i(\bx_k,a))\right|+\frac{1}{n}\sum_{i=1}^n \left|\esp G_i(\bx_k,a)-\esp G_i(\bx,a) \right|\\
&\le&\frac{M}{n}\sum_{i=1}^n \left\{\left|\itK_{\bH_d}^{(j\ell)}(\bX_i-\bx) -\itK_{\bH_d}^{(j\ell)}(\bX_i-\bx_k)\right|\right.\\
&&+ \left.\left|\esp \itK_{\bH_d}^{(j\ell)}(\bX_i-\bx) -\esp \itK_{\bH_d}^{(j\ell)}(\bX_i-\bx_k) \right|\right\}\indica_{ B(\bx,\bh)\cup  B(\bx_k,\bh)}(\bX_i)\\
\wttS_{k,s}(\bx_k,a)& = &  \frac{1}{n}\left|\sum_{i=1}^n (G_i(\bx_k,a)-G_i(\bx_k,a_s))\right|+\frac{1}{n}\sum_{i=1}^n \left|\esp G_i(\bx_k,a_s)-\esp G_i(\bx_k,a) \right| 
\end{eqnarray*}
with $\bh=(h_1,\dots,h_d)\trasp$ and $ B(\bx,\bh)=\{\by\in\real^{d}: |y_j-x_j|\le h_j\mbox{ for }1\leq j\leq d\}$. 
Using that the kernels $K_j$ are Lipschitz of order one and that if $\itK(\bX_i-\bx_k)\ne 0$, then $\breve{x}_{k,i,j}^{m}\breve{x}_{k,i,\ell}^{\wtm}\le 1$, we  get easily that for any $\bx\in \itB_d(\bx_k,\rho)$
$$\left|\itK_{\bH_d}^{(j\ell)}(\bX_i-\bx) -\itK_{\bH_d}^{(j\ell)}(\bX_i-\bx_k) \right|\leq c_1\frac{\|\bH_d^{-1}(\bx-\bx_k)\|}{ \prod_{j=1}^d h_j }\le  c_1 \frac{A_h \rho}{\prod_{j=1}^d h_j}\,,$$
which leads to
$$ \theta_n^{-1}\sup_{\bx\in \itB_d(\bx_k,\rho)\cap\itC}\sup_{a\in\itI_s\cap\itI_\delta}|\wttS_{k}(\bx,a)|\leq \theta_n^{-1}c_2\frac{A_h \rho}{\prod_{j=1}^d h_j}\frac{1}{n}\sum_{i=1}^n \indica_{B_d(\bx_k,\bh+\rho)}(\bX_i)=\wtA_{k,n}\,,
$$
where $c_2=2Mc_1$.

On the other hand, since $ {\rho}/{\prod_{j=1}^n h_j}\to 0$, we get that there exists $n_1\in \natu$ such that for all $n\geq n_1$, $\prod_{j=1}^d(h_j+\rho)\le\prod_{j=1}^d h_j + c_3\rho\leq 2\prod_{j=1}^d h_j$. Observe that, since $n$ is   large enough, we may assume that $h_j<1$ for all $j$, so $A_h\rho\leq  {\rho}/{\prod_{j=1}^d h_j}\to 0$. 

Let $Z_i=\indica_{B_d(\bx_k,\bh+\rho)}(\bX_i)\,{A_h \rho}/{\prod_{j=1}^d h_j}$. Then, using that $\esp\, \indica_{B_d(\bx_k,\bh+\rho)}(\bX_i)\le c_4 \prod_{j=1}^d (h_j+\rho)$ where $c_4=\|f_\bX\|_\infty$, we obtain that for $n\geq n_1$,
$$|Z_i|\leq \frac{A_h\rho}{\prod_{j=1}^d h_j}\hspace{1cm}\esp Z_i\leq c_4\prod_{j=1}^d (h_j+\rho)\frac{A_h\rho}{\prod_{j=1}^d h_j}\leq 2c_4 A_h\rho\to 0\,.$$
Therefore, $|Z_i-\esp Z_i|\leq 2\,{A_h\rho}/{\prod_{j=1}^d h_j}$ and
$$\var(Z_i)\leq \esp Z_i^2\leq c_4\prod_{j=1}^d (h_j+\rho)\left(\frac{A_h\rho}{\prod_{j=1}^d h_j}\right)^2\leq 2c_4\frac{A_h^2\rho^2}{\prod_{j=1}^d h_j}\,.$$
Then, we have that
\begin{equation}\label{ig2}
\frac{A_h\rho}{\prod_{j=1}^d h_j}\left(\frac{1}{n}\sum_{i=1}^n \indica_{B_d(\bx_k,\bh+\rho)}(\bX_i)\right)=\frac{1}{n}\sum_{i=1}^n Z_i\leq \frac{1}{n}\left|\sum_{i=1}^n (Z_i-\esp Z_i)\right|+2c_4 A_h\rho\,.
\end{equation}
On the other hand, the fact that  $\theta_n^{-1}A_h\rho\leq M_2$, for any $C>C_0=4c_2\,c_4\,M_2$ lead us to
$$\prob\left(\theta_n^{-1}\sup_{\bx\in \itB(\bx_k,\rho)} \left|\wtS_k(\bx)\right|>C\right)\leq \prob\left(\frac{1}{n}\left|\sum_{i=1}^n (Z_i-\esp Z_i)\right|>\frac{C\theta_n}{2c_2}\right)$$
Finally, Lemma \ref{apend1}.1 for all $n\geq n_1$ implies that, for all $n\geq n_1$, if $C>C_{0}$ we have that there exist $b_1$ and $b_2$
\begin{equation}
\prob\left(\theta_n^{-1}\sup_{\stackrel{\bx\in \itB_d(\bx_k,\rho)\cap\itC}{a\in\itI_s\cap\itI_\delta}}|\wttS_{k}(\bx,a)|>\frac{C}2\right) \leq 
 \prob\left(\theta_n^{-1} \wtA_{k,n} >\frac{C}2\right) \leq 2\exp\left\{ -\frac{C^2\theta_n^2 n\prod_{j=1}^d h_j}{b_1A_h^2\rho^2+b_2C\theta_n A_h\rho}\right\}\,.
 \label{parte1}
\end{equation}
 On the other hand, using that $|W_i(a_1)-W_i(a_2)|\le M_1 |a_1-a_2|$ and the fact that $|a-a_s|<\rho$, we get that
$$|G_i(\bx_k,a)-G_i(\bx_k,a_s)|\le M_1 \rho\, \itK_{\bH_d}^{(j\ell)}(\bX_i-\bx_k)\le  \frac{M_1\,\rho}{\prod_{j=1}^d h_j}\indica_{ B_d(\bx_k,\bh)}(\bX_i)\le   \frac{M_1\,\rho}{\prod_{j=1}^d h_j}\indica_{ B_d(\bx_k,\bh+\rho)}(\bX_i)$$
Then, if $n\ge n_2$ we have that
\begin{eqnarray*}
\wttS_{k,s}(\bx_k,a)& \le &   2\, M_1 \rho \frac{1}{n} \sum_{i=1}^n \indica_{ B_d(\bx_k,\bh+\rho)}(\bX_i) \le \wtA_{k,n}
\end{eqnarray*}
since $A_h\to \infty $. Therefore, if $n\ge \max\{n_1,n_2\}$ 
$$
\prob\left(\theta_n^{-1}\sup_{\stackrel{\bx\in \itB(\bx_k,\rho)\cap\itC}{a\in\itI_s\cap\itI_\delta}}|\wttS_{k,\ell}(\bx,a)|>\frac{C}2\right) \leq 
 \prob\left(\theta_n^{-1} \wtA_{k,n} >\frac{C}2\right) \leq 2\exp\left\{ -\frac{C^2\theta_n^2 n\prod_{j=1}^d h_j}{b_1A_h^2\rho^2+b_2C\theta_n A_h\rho}\right\}\,,$$
which together with (\ref{parte1}) concludes the proof of a).

\vskip0.2in
\noi b) Recall that  $\itC\subset\bigcup_{k=1}^{N_\rho(\itC)} \itB_d(\bx_k,\rho)$ with $N_\rho(\itC)\le A_1/\rho^d$ and $\itI_\delta=[1-\delta,1+\delta]\subset \bigcup_{k=1}^{N_\rho(\itI_\delta)} \itI_s$ with $N_\rho(\itI_\delta)\le 2/\rho$. Then, given $\bx\in\itC$ and $a\in \itI_\delta$ there exist $k,s$ such that $\bx\in \itB(\bx_k,\rho)$, $a\in \itI_s$. Besides, for any $\bx\in \itB(\bx_k,\rho)$, we have that
$ \left|S_n(\bx,a)\right| \leq  \left|S_n(\bx_k,a_s)\right|+ \left|\wtS_{k,s}(\bx,a)\right|$,  so
\begin{eqnarray*}
\sup_{\bx\in \itC} \sup_{a\in \itI_\delta} \left|S_n(\bx,a)\right| \leq \max_{\stackrel{1\le k\le N_\rho(\itC)}{1\le s\le N_\rho(\itI_\delta)}}\left|S_n(\bx_k,a_s)\right|+\max_{\stackrel{1\le k\le N_\rho(\itC)}{1\le s\le N_\rho(\itI_\delta)}} \sup_{\stackrel{\bx\in \itB(\bx_k,\rho)\cap\itC}{a\in\itI_s\cap\itI_\delta}}\left|\wtS_{k,s}(\bx,a)\right|\,,
\end{eqnarray*}
which entails that $\prob\left(\theta_n^{-1}\sup_{\bx\in \itC} \sup_{a\in \itI_\delta} \left|S_n(\bx,a)\right|>C\right)\leq \beta_n+\gamma_n$ where 
\begin{eqnarray*}
\beta_n&=& \prob\left(\max_{\stackrel{1\le k\le N_\rho(\itC)}{1\le s\le N_\rho(\itI_\delta)}}\left|S_n(\bx_k,a_s)\right|>\frac{C\theta_n}{2}\right) \le   N_\rho(\itC)\, N_\rho(\itI_\delta)\sup_{\stackrel{\bx \in \itC}{a\in \itI_\delta } }\prob\left(\left|S_n(\bx,a)\right|>\frac{C\theta_n}{2}\right)\\
 \gamma_n&=&\prob\left(\max_{\stackrel{1\le k\le N_\rho(\itC)}{1\le s\le N_\rho(\itI_\delta)}} \sup_{\stackrel{\bx\in \itB(\bx_k,\rho)\cap\itC}{a\in\itI_s\cap\itI_\delta}}\left|\wtS_{k,s}(\bx,a)\right|>\frac{C\theta_n}{2}\right) 
\end{eqnarray*}
Using Lemma \ref{apend1}.1, straightforward calculations (see Mart\'{\i}nez (2014) for details)  allow to show that there exists $b_3,b_4>0$ such that 
\begin{equation}\label{betan}
\beta_n \leq  2 N_\rho(\itC)N_\rho(\itI_\delta)\exp\left\{-\frac{C^2\theta_n^2 n \prod_{j=1}^d h_j}{4 b_3+ 2 b_4  C\theta_n} \right\}\le  2\frac{A_1}{\rho^{d+1}}\exp\left\{-\frac{C^2\theta_n^2 n \prod_{j=1}^d h_j}{4 b_3+ 2 b_4  C\theta_n} \right\}\,.
\end{equation}
 Using a), it follows that for all $C>2\,C_0> 0$ and all $n\geq n_0$
\begin{equation}\label{gamman}
\gamma_n  \leq  N_\rho(\itC)N_\rho(\itI_\delta) \prob\left(\sup_{\stackrel{\bx\in \itB(\bx_k,\rho)\cap\itC}{a\in\itI_s\cap\itI_\delta}} \left|\wtS_{k,s}(\bx,a)\right|>\frac{C\theta_n}{2}\right)\leq 4\, \frac{A_1}{\rho^{d+1}} \exp\left\{ -\frac{C^2\theta_n^2 n\prod_{j=1}^d h_j}{4 b_1 A_h^2\rho^2+2 b_2\, C\theta_n  A_h\rho }\right\}\,.
\end{equation}
The bound given in b) follows now from (\ref{betan}) and (\ref{gamman}).

c) Observe that,   \textbf{A6} implies that $\theta_n\to 0$. Define $\rho={\log n}/{n}$. We will show that the conditions in b) are fulfilled. It is clear that $\rho \to 0$ and besides by \textbf{A6}, ${\rho_n}\,\left\{\prod_{j=1}^n h_j\right\}^{-1}=\theta_n^2\to 0$. On the other hand,
\begin{eqnarray*}
 \left(\theta_n^{-1} A_h\rho\right)^2=\left(\frac{1}{\min_{1\leq j\leq d}{\{h_j\}}}\frac{\log n}{n}\sqrt{\frac{n\prod_{j=1}^d h_j}{\log n}}\right)^2\leq \frac{\log n}{n\prod_{j=1}^d h_j}\to 0\,,
 \end{eqnarray*}
so $\theta_n^{-1} A_h\rho\le 1 $ if $n$ is  large enough. Noticing that $\theta_n^2 n \prod_{j=1}^d h_j=\log(n)$ and that there exists $A_1$ such that $N_\rho(\itC)\leq A_1\rho^{-d}$, we have that b) implies that
\begin{eqnarray*}
\prob\left(\theta_n^{-1}\hskip-0.1in\sup_{\stackrel{\bx\in \itB(\bx_k,\rho)\cap\itC}{a\in\itI_s\cap\itI_\delta}} \left|S_n(\bx,a)\right|>C\right)&\hskip-0.1in \leq &\hskip-0.1in \frac{4A_1}{\rho^{d}}\left[\exp\left\{-\frac{C^2\theta_n^2 n \prod_{j=1}^d h_j}{{4 b_3+ 2 b_4  C\theta_n}}\right\}+ \exp\left\{ -\frac{C^2\theta_n^2 n\prod_{j=1}^d h_j}{4 b_1 A_h^2\rho^2+2 b_2\, C\theta_n  A_h\rho}\right\}\right]\\
&\hskip-0.1in \leq & \hskip-0.1in\frac{4A_1}{\rho^{d}}\left[\exp\left\{-\frac{C^2\log(n)}{{4 b_3+ 2 b_4  C\theta_n}}\right\}+ \exp\left\{ -\frac{C^2\log(n)}{4 b_1 A_h^2\rho^2+2 b_2\, C\theta_n  A_h\rho}\right\}\right]
\end{eqnarray*}
Finally, using that $\theta_n\to 0$ and $A_h\rho\to 0$, we obtain that there exists $n_1$ such that for all $n\geq n_1$, $2b_4{C\theta_n}\leq 4b_3$ and $4 b_1 A_h^2\rho^2+2 b_2\, C\theta_n  A_h\rho\le 8 b_3$, then
\begin{eqnarray*}
\prob\left(\theta_n^{-1}\sup_{  \bx\in \itB(\bx_k,\rho)\cap\itC} \sup_{a\in\itI_s\cap\itI_\delta} \left|S_n(\bx,a)\right|>C\right) 
& \leq & 8\,A_1\rho^{-d} \exp\left\{-\frac{C^2\log(n)}{8 b_3}\right\}\\
&\le & 8\,A_1\left(\frac 1{\log(n)}\right)^d n^{d-\frac{C^2}{8 b_3}}\le  84\,A_1\; n^{d-\frac{C^2}{8 b_3}}
\end{eqnarray*}
Therefore, for any $C>C_1=\max\{C_0,\sqrt{8b_3 d +3}\}$, we get that $\sum_{n\ge 1} \prob\left(\theta_n^{-1}\sup_{\bx\in\itC} \left|S_n(\bx))\right|>C\right)<\infty$  concluding the proof.  \square

\vskip0.2in
\noi \textbf{Remark \ref{apend1}.1.} Taking $m=\wtm=0$ and $W_i\equiv 1$, Lemma \ref{apend1}.2c) entails the uniform convergence for the kernel density estimator, that is, we obtain that
$$\sup_{\bx\in\itC}\frac{1}{n}\left|\sum_{i=1}^n (\itK_{\bH_d}(\bX_i-\bx)-\esp \itK_{\bH_d}(\bX_i-\bx))\right|=O_{\aco}(\theta_n)\,.$$
Besides, using that $\sup_{\bx\in\itC}|f_\bX(\bx)|<\infty$, $\itK$ has compact support and $f_\bX$ is uniformly continuous in $\itC$, we get that
 $$\sup_{\bx\in\itC}\left|\frac{1}{n}\sum_{i=1}^n \esp\itK_{\bH_d}(\bX_i-\bx)-f_\bX(\bx)\right|=\sup_{\bx\in\itC}\left|\int \itK(\bu)\left(f_\bX(\bH_d\bu+\bx)-f_\bX(\bx)\right)\right|\to 0$$ 
which together with the above result implies that
\begin{equation}\label{obs3lem3}
\sup_{\bx\in\itC}\left|\frac{1}{n}\sum_{i=1}^n \itK_{\bH_d}(\bX_i-\bx)-f_\bX(\bx)\right|=o_{\as}(1)\,.
\end{equation}

\vskip0.2in
\noi \textbf{Lemma  \ref{apend1}.3.} \textsl{Assume that \textbf{A1}, \textbf{A2} and \textbf{A4}  to  \textbf{A7}  hold, then
$$\sup_{\bx\in\itC}|J_{n,j}(\bx,\widehat{a}_\sigma(\bx))|=O_{\as}(\theta_n)$$
with $ \theta_n=\sqrt {\log{n}/({n\prod_{j=1}^d h_j})}$, where $j=1,\dots,d+1$ and $J_{n,j}$ is the $j$th component of vector $\bJ_n$ and $\wa_\sigma(\bx)=\sigma(\bx)/\wese(\bx)$.
}

\vskip0.1in
\noi \textbf{Proof}. By (\ref{conproba}), taking $\delta=1/2$, we get that there exists $\itN$ such that $\prob(\itN)=0$ and for any $\omega\notin\itN$, there exists $n_1\in \natu$ such that for all $n\geq n_1$ 
$ \sup_{\bx\in\itC}|\widehat{a}_\sigma(\bx)-1|\leq \delta$.
Therefore, for any $\omega\notin\itN$ and $n\geq n_1$, we have that $\sup_{\bx\in\itC}\left| J_{n,j}(\bx,\widehat{a}_\sigma(\bx))\right| \le \sup_{\bx\in\itC}\sup_{a\in[1-\delta,1+\delta]}\left| J_{n,j}(\bx,a)\right|$, so to conclude the proof it is enough to see that $\sup_{\bx\in\itC}\sup_{a\in[1-\delta,1+\delta]}\left| J_{n,j}(\bx,a)\right|=O_{\as}(\theta_n)$.
Recall that
$$
\bJ_n(\bx,a)= \frac{1}{n } \sum_{i=1}^n \delta_i \itK_{\bH_d}(\bX_i-\bx)  \psi\left(\frac{\sigma(\bX_i)\eps_i}{\sigma(\bx)} \,a\right) \breve{\bx}_{i, \alpha}=\frac{1}{n } \sum_{i=1}^n \delta_i \itK_{\bH_d}(\bX_i-\bx)   \psi\left(V_i \,a\right) \breve{\bx}_{i, \alpha}\;.
$$
then, $J_{n,j}(\bx,a)=(1/{n }) \sum_{i=1}^n \delta_i \itK_{\bH_d}(\bX_i-\bx)   \psi\left(V_i \,a\right) \breve{x}_{i,j, \alpha}$ and the proof follows now from  Lemma \ref{apend1}.2c) taking $m=1$, $\wtm=0$ and $W_i(a)=\delta_i  \psi\left(V_i \,a\right)$ and noting that
$|W_i(a)|\le \|\psi\|_\infty /i(t)$ and $|W_i(a_1)-W_i(a_2)|\le  (\|\zeta\|_\infty /i(t))\, |a_1-a_2|$.  \square

\vskip0.2in
\noi \textsc{Proof of Proposition \ref{consist}.1.}
Let $\br=(\beta_0,h_\alpha \beta_1,h_\alpha^2 \beta_2,\cdots,h_\alpha^q \beta_q)\trasp $, $\ell_n(\br)$ be defined in (\ref{lnralpha}) and
$\br_0(\bx)=\left(g(\bx),h_\alpha g_\alpha^{(1)}(x_\alpha),\dots,h_\alpha^q g_\alpha^{(q)}(x_\alpha)\right)\trasp $. For the sake of simplicity denote $\itV_\tau=\itV_{q+1,\tau}=\{\br: \|\br\|=\tau\}$. To prove Proposition \ref{consist}.1, we will first show that it is enough to see that there exists $\itN$ such that $\prob(\itN)=0$ and such that for all $\omega \notin \itN$,  given $\nu>0$  there exists  $0<\tau_\nu<1$  small enough such that for any $0<\tau<\tau_\nu$ and $n\ge n_0$, 
\begin{equation}
 \inf_{\br\in \itV_\tau}\inf_{\bx\in\itC}\left\{\ell_n(\br+\br_0(\bx))-\ell_n(\br_0(\bx))\right\}> 0\,. 
\label{infellnueq}
\end{equation}
Indeed, in the set $ \{\inf_{\br\in \itV_\tau}\inf_{\bx\in\itC}\left[\ell_n(\br+\br_0(\bx))-\ell_n(\br_0(\bx))\right]>0\}$ for all $\bx\in \itC$ we have that $\inf_{\br\in \itV_\tau}\ell_n(\br+\br_0(\bx))>\ell_n(\br_0(\bx))$, which implies that the function $L_n(\br)=\ell_n(\br+\br_0(\bx))-\ell_n(\br_0(\bx))$ has a local minimum  $\wtbr(\bx)$ in   $\accentset{\circ}{\itB}_{q+1,\tau}$, where $\accentset{\circ}{\itR}$ stands for the interior of the set $\itR$. Then, for all $\bx\in \itC$,  $\wtbr(\bx)+\br_0(\bx)$ is a local minimum of $\ell_n(\br)$ and $\wtbr(\bx)+\br_0(\bx)$ belongs to  $\accentset{\circ}{\itB}_{q+1}(\br_0(\bx),\tau)=\{\br: \|\br-\br_0(\bx)\|< \tau\}$, as a result of which  $\wbbe(\bx)=\wtbr(\bx)+\br_0(\bx)$ is a solution of (\ref{22alfa}). That is, with probability 1, for all $\bx\in \itC$, there exists a solution $\wbbe(\bx)$ of (\ref{22alfa}) in the interior of ${\itB}_{q+1}(\br_0(\bx),\tau)$. Hence, for any $\omega \notin \itN$, given $\nu>0$, and $\tau>0$   small enough $
 \sup_{\bx\in\itC}\|\bH^{(\alpha)}[\wbbe(\bx)-\bbe(\bx)]\|\leq \tau$,  $n\ge n_0$, 
which implies that $\sup_{\bx\in\itC}\|\bH^{(\alpha)}[\wbbe(\bx)-\bbe(\bx)]\|\convpp 0$ as desired.

In order to prove (\ref{infellnueq}), observe that $Y_i-(\br+\br_0(\bx))\trasp \breve{\bx}_{i,\alpha}=U_i+g(\bX_i)-\br\trasp \breve{\bx}_{i,\alpha}-\br_0\trasp \breve{\bx}_{i,\alpha}=U_i+R(\bX_i,\bx) \breve{\bx}_{i,\alpha}-\br\trasp \breve{\bx}_{i,\alpha}$. Denote as $\wZ_i(\bx)=(U_i+R(\bX_i,\bx))/\wese(\bx)=\wV_i(\bx)+R(\bX_i,\bx){\wese(\bx)}^{-1}$ with $\wV_i(\bx)= U_i/\wese(\bx)=\sigma(\bX_i)\eps_i/\wese(\bx)$ and $\wDelta_i(\bx)=\br\trasp \breve{\bx}_{i,\alpha}/{\wese(\bx)}$. Then, using that $\rho(b)-\rho(a)=\int_a^b \psi(u)du$ we obtain  that for all $\br\in \itV_\tau$ 
\begin{equation}
\ell_n(\br+\br_0(\bx))-\ell_n(\br_0(\bx))=\frac{1}{n }\sum_{i=1}^n \itK_{\bH_d}(\bX_i-\bx)\xi_i\int_{\wZ_i}^{\wZ_i-\wDelta_i}\hskip-0.1in\psi\left(t\right)\,dt= K_{n1}(\bx)+K_{n2}(\bx)+K_{n3}(\bx)\,,
\label{62nueq}
\end{equation}
with
\begin{eqnarray*}
K_{n1}(\bx)&=&\frac{1}{n }\sum_{i=1}^n\delta_i \itK_{\bH_d}(\bX_i-\bx)\int_{\wZ_i(\bx)}^{\wZ_i(\bx)-\wDelta_i(\bx)} \psi(\wV_i)dt=\,-\frac{1}{\wese(\bx)}\br\trasp\frac{1}{n } \sum_{i=1}^n \delta_i \itK_{\bH_d}(\bX_i-\bx)   \psi(\wV_i(\bx)) \breve{\bx}_{i,\alpha}\;,\\
 K_{n2}(\bx)&=&\frac{1}{n }\sum_{i=1}^n\delta_i  \itK_{\bH_d}(\bX_i-\bx) \int_{\wZ_i(\bx)}^{\wZ_i(\bx)-\wDelta_i(\bx)} \psi^{\prime}(\wV_i(\bx))(t-\wV_i(\bx))dt\\
 &=&\frac{1}{2\wese^{\,2}(\bx)}\frac{1}{n }\sum_{i=1}^n \delta_i \itK_{\bH_d}(\bX_i-\bx)   \psi^{\prime}(\wV_i(\bx))\left[\left(\br\trasp \breve{\bx}_{i,\alpha}\right)^2-2 R(\bX_i,\bx)\br\trasp \breve{\bx}_{i,\alpha}\right]\;,\\
 K_{n3}(\bx)&=&\frac{1}{n }\sum_{i=1}^n \delta_i \itK_{\bH_d}(\bX_i-\bx) \int_{\wZ_i(\bx)}^{\wZ_i(\bx)-\wDelta_i(\bx)}\left[\psi\left(t\right)-\psi(\wV_i(\bx))-\psi^{\prime}(\wV_i(\bx))(t-\wV_i(\bx))\right] \,dt\;.
\end{eqnarray*}
The proof of (\ref{infellnueq}) will be done in several steps. Let us assume that the following approximations hold
\begin{eqnarray}
\label{eqq5q}
\sup_{\br\in \itV_\tau}\sup_{\bx\in\itC}\|K_{n1}(\bx)\|&=&\tau\sqrt{\frac{\log n}{n\prod_{j=1}^d h_j}}\wUps_{1,n}\;,\\
K_{n2}(\bx)&=&\frac{1}{2\sigma^2(\bx)}\esp(\psi^{\prime}(\eps))f_\bX(\bx)  p(\bx)\;\br\trasp  \bS^{(\alpha)}\br(1+\wzeta_n)+\tau\,\left( h_\alpha^{q+1}+\sum_{j\neq\alpha}{h_j}\right)\,\wUps_{2,n}\,,\qquad \qquad\label{eqkn2q}\\
 \sup_{\br\in \itV_\tau}\sup_{\bx\in\itC}\|\itK_{n3}(\bx)\|&=& \tau  \left(\tau+h_\alpha^{q+1}+\sum_{j\neq\alpha} h_j\right)^2\wUps_{3,n} \;,
 \label{eqq14q}
\end{eqnarray}
with $\wUps_{1,n}=O_{\aco}(1)$, $\wUps_{3,n}=O_{\aco}(1)$, $\wUps_{1,n}$ and $\wUps_{3,n}$   not depending on  $\tau$ and where 
$\wzeta_n=o_{\aco}(1)$ and $\wUps_{2,n}=O_{\aco}\left(1\right)$ do not depend on  $\bx\in\itC$ nor $\br$ and, consequently, neither on $\tau$. 

We begin by showing that (\ref{eqq5q}) to (\ref{eqq14q}) imply  (\ref{infellnueq}). 

Let us denote as $\nu_1>0$ the minimum eigenvalue of $\bS^{(\alpha)}$ which is a symmetric and positive definite   matrix. Using that $\esp\psi^\prime(\eps)>0$, $i(f_\bX)>0$, $i(p)>0$ and that the scale function $\sigma$ is bounded over $\bx\in\itC$, if $M=\nu_1\esp\psi^\prime(\eps)i(f_\bX)i(p)/(2\sup_{\bx\in\itC}\sigma^2(\bx))$, we obtain that $\bx\in\itC$ and $\br\in\itV_\tau$
$$
Q(\br,\bx)=\frac{1}{2\sigma(\bx)}\esp(\psi^{\prime}(\eps))f_\bX(\bx)\, p(\bx)\, \br \trasp  \bS^{(\alpha)} \br \ge \frac{\nu_1}{2\sigma(\bx)}\esp(\psi^{\prime}(\eps))f_\bX(\bx)\,p(\bx)\,\tau^2\geq M\tau^2>0\;.
$$  
As $\wUps_{2,n}=O_{\aco}\left(1\right)$ and $\wzeta_n=o_{\aco}(1)$, given $\nu>0$ there exists $\wtA_1$ such that 
$$\sum_{n\ge 1}\prob\left(|\wUps_{2,n}|> \wtA_1\right)<\infty\qquad \qquad\qquad \sum_{n\ge 1}\prob\left(|\wzeta_n|> \frac 12\right)<\infty \,.$$
Let $n_\tau$ be such that  $h_\alpha^{q+1}+\sum_{j\neq\alpha} h_j\leq \tau \min\{M/(4\wtA_1),1\}$, for   $n\ge n_\tau$. Hence, there exists a set $\itN_1$ satisfying that $\prob(\itN_1)=0$ and for all $\omega \notin \itN_1$, there exists $n_1\in \natu$ such that for all $n\ge n_1$, we have that $|\wUps_{2,n}|< \wtA_1 $ and $ |\wzeta_n|< \frac 12 $.

Since $K_{n2}(\bx)\ge Q(\br,\bx)(1-|\wzeta_n|)- \tau\, \left(h_\alpha^{q+1}+\sum_{j\neq\alpha}{h_j}\right)\,|\wUps_{2,n}|$, if $\omega \notin \itN_1$ and $n\ge n_{1,\tau}=\max(n_\tau,n_1)$ we have that
\begin{equation}\label{65nue}
  \inf_{\br\in \itV_{\tau}}\inf_{\bx\in\itC}K_{n2}(\bx)>   \frac{M}{4}\tau^2 \;.
\end{equation}
From (\ref{eqq5q}) and the fact that $ n\prod_{j=1}^d h_j/\log n\to \infty$,we get easily that there exist a positive constant $\wtA_2$ and a set $\itN_2$ such that $\prob(\itN_2)=0$ and for all $\omega \notin \itN_2$ there exists $n_{2,\tau}$ such that  
$$ |\wUps_{1,n}|\le \wtA_2 \qquad \mbox{ and } \qquad \sqrt{\frac {\log n}{n\prod_{j=1}^d h_j}}\le \tau\min\left\{\frac{M}{8\wtA_2},1\right\}\;,$$
for $n\ge n_{2,\tau}$. Thus, using (\ref{65nue}), we obtain that if $\omega \notin \itN_1\cup\itN_2$ and $n\ge \max(n_{1,\tau},n_{2,\tau})$
\begin{equation}\label{66nue}
 \inf_{\br\in \itV_{\tau}}\inf_{\bx\in\itC}(K_{n1}(\bx)+K_{n2}(\bx))>   \frac{M}{8}\tau^2 \;.
\end{equation}
On the other hand,  $K_{n3}$ satisfies (\ref{eqq14q}) with $\wUps_{3,n}=O_{\aco}(1)$, then there exist a positive $\wtA_3$ and a set $\itN_3$ such that $\prob(\itN_3)=0$ and for all $\omega \notin \itN_3$ there exists $n_{3}$ such that   $ |\wUps_{3,n}|\le \wtA_3$, for $n\ge n_{3}$. Besides, there exists $n_{3,\tau}\in \natu$ such that $h_\alpha^{q+1}+\sum_{j\neq\alpha} h_j\leq \tau$, for  $n\ge n_{3,\tau}$. Therefore, we obtain that for all $\omega \notin \itN_3$ and for any $n\geq n_{4,\tau}=\max\{n_{3,\tau},n_3\}$,
$ \sup_{\br\in \itV_\tau}\sup_{\bx\in\itC}|K_{n3}(\bx)|\le \wtA_3\tau^3$. 
Taking $\tau_{\nu}<\min\{1,M/(16\,\wtA_3) \}$, we get that for $\tau<\tau_\nu$, $\omega \notin \itN_3$ and $n\geq n_{4,\tau}$
\begin{equation}\label{eqq13}
 \sup_{\br\in \itV_\tau}\sup_{\bx\in\itC}|K_{n3}(\bx)|\leq \frac{M}{16}\tau^2  \,.
\end{equation}
Therefore, we have shown that for all $0<\tau<\tau_{\nu}<1$,  $\omega \notin \itN_3$ and $n\ge  \max(n_{1,\tau},n_{2,\tau},n_{4,\tau})$,  the assertions (\ref{eqq13}) and (\ref{66nue}) hold which together with (\ref{62nueq}) lead us to (\ref{infellnueq}).
 
It remains to show (\ref{eqq5q}), (\ref{eqkn2q}) and (\ref{eqq14q}). 

\noindent $\bullet$ \underline{Let us  begin by proving (\ref{eqq5q})}. Note that
$$K_{n1}(\bx)=\,-\frac{1}{\wese(\bx)}\br\trasp\bJ_{n,\alpha}(\bx,\wa_\sigma(\bx))\;,$$
where  $\bJ_{n,\alpha}(\bx,a)$ is defined in (\ref{jnalpha}). Note that (\ref{A7prob}) implies that
$$\prob\left(\exists n_0 \mbox{ tal que } \forall n\ge n_0 \sup_{\br\in \itV_\tau}\sup_{\bx\in\itC}\|K_{n1}(\bx)\|<\frac{\tau}{A}\sup_{\bx\in\itC}\|\bJ_{n,\alpha}(\bx,\wa_\sigma(\bx))\|\right)= 1\,.$$
Lemma \ref{apend1}.3 entails that $\sup_{\bx\in\itC}\|\bJ_{n,\alpha}(\bx,\wa_\sigma(\bx))\|=O_{\as}(\theta_n)$, which concludes the proof of (\ref{eqq5q}).

\noindent $\bullet$ \underline{Let us show that (\ref{eqq14q}) holds}. Recall that $\wZ_i(\bx)=\wV_i(\bx)+R(\bX_i,\bx){\wese(\bx)}^{-1}$ and $\wDelta_i(\bx)=\br\trasp \breve{\bx}_{i,\alpha}/{\wese(\bx)}$. By the integral mean value theorem,
\begin{eqnarray*}
K_{n3}(\bx)&=&\frac{1}{n }\sum_{i=1}^n \delta_i\itK_{\bH_d}(\bX_i-\bx)\int_{\wV_i(\bx)+R(\bX_i,\bx){\wese(\bx)}^{-1}}^{\wV_i(\bx)+R(\bX_i,\bx){\wese(\bx)}^{-1}-\wDelta_i(\bx)}\left[\psi\left(t\right)-\psi(\wV_i(\bx))-\psi^{\prime}(\wV_i(\bx))(t-\wV_i(\bx))\right] \,dt\\
&=& -\br\trasp \frac{1}{n \wese(\bx)}\sum_{i=1}^n \delta_i \itK_{\bH_d}(\bX_i-\bx) \left[\psi\left(\wV_i(\bx)+\wtheta_i(\bx)\right)-\psi(\wV_i(\bx))-\psi^{\prime}(\wV_i(\bx))\wtheta_i(\bx)\right]\breve{\bx}_{i,\alpha}\,,
\end{eqnarray*}
where $\wtheta_i(\bx)$ is an intermediate point between  $R(\bX_i,\bx)/{\wese(\bx)}$ and $\{R(\bX_i,\bx)-\br\trasp \breve{\bx}_{i,\alpha}\}/{\wese(\bx)}$. By simplicity, if $i\in \{i: \itK_{\bH_d}(\bX_i-\bx)= 0\}$, we define $\wtheta_i(\bx)=0$ since it does not change the sum. Note that $|X_{i,j}-x_j|\leq h_j$ for $i=1,\cdots,n$ and $j=1,\cdots,d$, when $\itK_{\bH_d}(\bX_i-\bx)\ne 0$, thus \textbf{A4} implies that
\begin{equation}\label{eqq10q}
\sup_{\bx\in\itC}\max_{{i: \itK_{\bH_d}(\bX_i-\bx)\ne 0}}|R(\bX_i,\bx)|\leq A_g  \left(\sum_{j\neq\alpha}h_j + h_\alpha^{q+1}\right)\,,
\end{equation} 
where $A_g$ is a constant only depending on $\|g_j^{(1)}\|_\infty$ for $j\neq\alpha$ and $\|g_\alpha^{(q+1)}\|_\infty$. Then, using (\ref{A7prob}), we obtain that
  $$\prob\left(\exists n_0 \mbox{ such that } \forall n\ge n_0 \quad\sup_{\bx\in\itC}\max_{1\le i\le n}|\wtheta_i(\bx)|\le A_1\left(\tau+h_\alpha^{q+1}+\sum_{j\neq\alpha} {h_j}\right)\right)=1\,.$$
Let $\itK^{\star}(\bu)=|\itK(\bu)|/\int\!|\itK(\bu)|d\bu$, then,   Remark \ref{apend1}.1 implies that the density estimator based on $\itK^{\star}$,  $\wefe(\bx)=(1/n)\sum_{j=1}^n \itK_{\bH_d}^{\star}\left(\bx-\bX_j\right)$ converges uniformly and almost surely to $f_\bX$ (see (\ref{obs3lem3})). Hence, using  \textbf{A2}, we obtain that $\sup_{\bx\in \itC}\wefe(\bx)=O_{\as}(1)$ which together with the fact that $\psi^{\prime\prime}$ is bounded and that each component of $\breve{\bx}_{i,\alpha}$ is smaller or equal to 1 when $\itK_{\bH_d}(\bX_i-\bx)\ne 0$,  leads to
$$
\sup_{\br\in \itV_\tau}\sup_{\bx\in\itC}\|\itK_{n3}(\bx)\|\leq \tau \frac{\|\psi^{\prime\prime}\|_\infty}{A\,i(t)} \sup_{\bx\in\itC}\max_{1\leq i\leq n}|\wtheta_i(\bx)|^2\sup_{\bx\in\itC} \frac{1}{n}\sum_{i=1}^n |\itK_{\bH_d}(\bX_i-\bx)|\;,
$$
whenever   $A<\wese(\bx)$ for all $\bx\in\itC$. Hence, from (\ref{A7prob}) we obtain that
$$
 \prob\left(\exists n_0 \mbox{ such that } \forall n\ge n_0\quad \sup_{\br\in \itV_\tau}\sup_{\bx\in\itC}\|\itK_{n3}(\bx)\|\leq \tau \frac{\|\psi^{\prime\prime}\|_\infty}{A\,i(t)} A_1\left(\tau+h_\alpha^{q+1}+\sum_{j\neq\alpha} h_j\right)^2\wUps_{3,n}^{\star}\right)=1
$$
where $\wUps_{3,n}^{\star}=\sup_{\bx\in \itC}(1/{n})\sum_{i=1}^n |\itK_{\bH_d}(\bX_i-\bx)|=O_{\as}(1)$  does not depend on $\tau$, concluding the proof of (\ref{eqq14q}).

\noindent $\bullet$ \underline{Finally, to conclude the proof we will obtain (\ref{eqkn2q})}. We have that
\begin{eqnarray}
K_{n2}(\bx)&=&\frac{1}{2\wese^{\,2}(\bx)}\frac{1}{n }\sum_{i=1}^n \itK_{\bH_d}(\bX_i-\bx)\delta_i  \psi^{\prime}(\wV_i(\bx))\left[\left(\br\trasp \breve{\bx}_{i,\alpha}\right)^2-2 R(\bX_i,\bx)\br\trasp \breve{\bx}_{i,\alpha}\right]\nonumber\\
&=& \frac{1}{2\wese^{\,2}(\bx)}\left(\br\trasp\wbM_{n1}\br-2\br\trasp\wbM_{n2}\right) \, .\label{eqq12q}
\end{eqnarray} 
 For $\bx\in\itC$ and $a\in\itI_\delta=[1-\delta,1+\delta]$ (with $0<\delta<1$),  define $\bM(\bx,a)=\esp \psi^\prime(\eps a) f_\bX(\bx)p(\bx)\bS^{(\alpha)}$ and
$$
\bM_{n1}(\bx,a)=\frac{1}{n}\sum_{i=1}^n  \itK_{\bH_d}(\bX_i-\bx)\delta_i \psi^{\prime}( V_i(\bx)a) \breve{\bx}_{i,\alpha}\breve{\bx}_{i,\alpha}\trasp\; .
$$ 
Then, $\bM (\bx,1)= \esp \psi^\prime(\eps  ) f_\bX(\bx)p(\bx)\bS^{(\alpha)}$. We want to show that
\begin{eqnarray}\label{eqqbM1q}
\sup_{\bx\in\itC}\|\wbM_{n1 }(\bx)-\esp \psi^\prime(\eps  ) f_\bX(\bx)p(\bx)\bS^{(\alpha)}\|&=& o_{\as}(1)\\
\sup_{\bx\in\itC}\|\wbM_{n2}(\bx)\|&=&O_{\as}\left(h_\alpha^{q+1}+\sum_{j\neq\alpha} h_j\right)\,.
\label{eqqbM2q}
\end{eqnarray}
Indeed, if (\ref{eqqbM1q}) and (\ref{eqqbM2q}) hold, using (\ref{A7prob}) and that $\sup_{\bx\in\itC}|\wese(\bx)-\sigma(\bx)|\convpp 0$, $\sigma$ is bounded in $\itC$, $i(\sigma)>0$ and replacing (\ref{eqqbM1q})  and (\ref{eqqbM2q}) in (\ref{eqq12q}) we get that
$$K_{n2}(\bx)=\frac{1}{2\sigma^2(\bx)}\esp(\psi^{\prime}(\eps))f_\bX(\bx)  p(\bx)\;\br\trasp  \bS^{(\alpha)}\br(1+\wzeta_n)+\tau\, \left(h_\alpha^{q+1}+\sum_{j\neq\alpha}{h_j}\right)\,\wUps_{2,n}\,,$$ 
where $\wzeta_n=o_{\as}(1)$ and $\wUps_{2,n}=O_{\as}\left(1\right)$ do not depend on $\br$ and therefore, neither on $\tau$, nor $\bx\in\itC$ since the convergences are uniform over $\bx$, which would conclude the proof of (\ref{eqkn2q}).

In order to prove (\ref{eqqbM1q}) it is enough to show that for all $1\leq j,k\leq d+1$
\begin{equation}\label{eqq9q}
\sup_{\bx\in\itC}|\wM_{n1,{jk}}(\bx)-M_{jk}(\bx,1)|=o_{\as}(1)\,,
\end{equation}
where $\wM_{n1,{jk}}(\bx)$, $M_{n1,{jk}}(\bx,a)$ and $M_{jk}(\bx,a)$ are the components $(j,k)$ of matrices $\wbM_{n1}(\bx,a)$, $\bM_{n1}(\bx,a)$ and $\bM(\bx,a)$, respectively.

Note that $\wM_{n1,{jk}}(\bx)=M_{n1,{jk}}(\bx, \wa_\sigma(\bx))$ where $\wa_\sigma(\bx)=\sigma(\bx)/\wese(\bx)$. Hence, from the bounds 
\begin{eqnarray*}
\sup_{\bx\in\itC}|M_{n1,{jk}}(\bx,a)-M_{jk}(\bx,a)|&\hskip-0.1in \leq &\hskip-0.1in \sup_{\bx\in\itC}|M_{n1,{jk}}(\bx,a)-\esp M_{n1,{jk}}(\bx,a)|+\sup_{\bx\in\itC}|\esp M_{n1,{jk}}(\bx,a)-M_{jk}(\bx,a)|\;,\quad\\
\sup_{\bx\in\itC}|\wM_{n1,{jk}}(\bx)-M_{jk}(\bx,1)| 
 &\hskip-0.1in \leq &\hskip-0.1in  \sup_{\bx\in\itC}|M_{n1,{jk}}(\bx, \wa_\sigma(\bx))-M_{jk}(\bx,\wa_\sigma(\bx))|+\sup_{\bx\in\itC}|M_{jk}(\bx,\wa_\sigma(\bx))-M_{jk}(\bx,1)|\,,\qquad
\end{eqnarray*}
we obtain that in order to prove (\ref{eqq9q}), it is enough to see that
\begin{itemize}
\item[(i)] $\sup_{\bx\in\itC}\sup_{a\in\itI_\delta}|M_{n1,{jk}}(\bx,a)-\esp M_{n1,{jk}}(\bx,a)|=O_{\as}\left(\sqrt{{\log n}/({n\prod_{j=1}^d h_j}})\right)$.
\item[(ii)] $\sup_{\bx\in\itC}\sup_{a\in\itI_\delta}|\esp M_{n1,{jk}}(\bx,a)-M_{jk}(\bx,a)|=o(1)$
\item[(iii)] $\sup_{\bx\in\itC}|M_{n1,{jk}}(\bx, \wa_\sigma(\bx))-M_{jk}(\bx,\wa_\sigma(\bx))|=o_{\as}(1)$
\item[(iv)] $\sup_{\bx\in\itC}|M_{jk}(\bx,\wa_\sigma(\bx))-M_{jk}(\bx,1)|=o_{\as}(1)$.
\end{itemize}

(i) can be obtained immediately from Lemma \ref{apend1}.2 taking   $m=\wtm=1$ and considering the sequence of independent random variables $W_i(a)=\psi^{\prime}\left({\sigma(\bX_i)\eps_i a}/{\sigma(\bx)}\right)\delta_i$ and noting that $|W_i(a)|\leq  {\|\psi^{\prime}\|_\infty}/{i(t)}$ for all $a$ and that $|W_i(a_1)-W_i(a_2)|\le \|\zeta_2\|_{\infty} |a_1-a_2|$.

To show (ii), let $\itC_0$ be the compact neighbourhood of $\itC$ given in assumptions  \textbf{A2} and \textbf{A3}. Then, $f_\bX$, $p$ and $\sigma$ are uniformly continuous functions in $\itC_0$. Define
$$\gamma_{\bx,a}(\bt)=\esp\left(\psi^{\prime}\left(\frac{\sigma(\bX_1)\eps_1 }{\sigma(\bx)}a\right)|\bX_1=\bt\right)\,.$$
Using that $\eps_i$ are independent from covariates, we obtain that
\begin{eqnarray*}
|\gamma_{\bx,a}(\bt)-\esp \psi^{\prime}(\eps_1a)|&\le&\esp\left|\psi^{\prime}\left(\frac{\sigma(\bt)}{\sigma(\bx)}\eps_1a\right)-\psi^{\prime}(\eps_1a)\right|=\esp\left|{\zeta}_2(\eps_1\,a\,\theta) \right|\left|\frac{\sigma(\bt)}{\sigma(\bx)}-1\right|\frac{1}{\theta}\,,
\end{eqnarray*}
 where $\theta$ is an intermediate point between $\sigma(\bt)/{\sigma(\bx)}$ and $1$, so that
$$\frac 1\theta\le \max\left\{1, \frac{\sigma(\bx)}{\sigma(\bt)}\right\}\;.$$
Using that $ {\zeta}_2(u)=u\,\psi^{\prime\prime}(u)$ is bounded, we obtain the upper bound
$$
|\gamma_{\bx,a}(\bt)-\esp \psi^{\prime}(\eps_1a)| \le  \|{\zeta}_2\|_{\infty}\left|\frac {\sigma(\bt)}{\sigma(\bx)}-1\right|\max\left\{1, \frac{\sigma(\bx)}{\sigma(\bt)}\right\}=\|{\zeta}_2\|_{\infty}\left| {\sigma(\bt)} -\sigma(\bx)\right|\max\left\{\frac{1}{\sigma(\bx)}, \frac{1}{\sigma(\bt)}\right\}\,.
$$
Using that $\inf_{\bt\in \itC_0}\sigma(\bt)>0$ and $\sup_{\bt\in \itC_0}\sigma(\bt)<\infty$, we have that there exists $c_1$ such that for all $\bx\in \itC$ and $\bt\in  \itC_0$
\begin{equation}\label{eqq6}
|\gamma_{\bx,a}(\bt)-\esp \psi^{\prime}(\eps_1a)|\leq c_1\left| {\sigma(\bt)} -\sigma(\bx)\right|\,.
\end{equation}
Observe that
\begin{eqnarray*}
 \esp M_{n1,{jk}}(\bx,a)  &=& \esp \itK_{\bH_{d}}(\bX_1-\bx)p(\bX_1)\gamma_{\bx,a}(\bX_1)\breve{\bx}_{1,\alpha_j}\breve{\bx}_{1,\alpha_k} \\
&=& \int \!\itK_{\bH_d}^{(\alpha)}(\bu-\bx)p(\bu)\gamma_{\bx,a}(\bu)\left[\frac{u_{\alpha}-x_{\alpha}}{h_{\alpha}}\right]^{j+k-2} 
 f_\bX(\bu)\,d\bu\,.
\end{eqnarray*}
 Changing variables   $\by=\bH_d^{-1}(\bu-\bx)$, we obtain 
\begin{eqnarray*}
 \esp M_{n1_{jk}}(\bx,a)-M_{jk}(\bx,a)  &=& \int \!p(\bH_d\by+\bx)\gamma_{\bx,a}(\bH_d\by+\bx)f_\bX(\bH_d\by+\bx)y_{\alpha}^{j+k-2} \itK (\by)\,d\by \\
&& -\esp \psi^{\prime}(\eps_1 a) f_\bX(\bx)p(\bx)S^{(\alpha)}_{jk} \,,
\end{eqnarray*}
where   $S_{jk}^{(\alpha)}=\int y_{\alpha}^{j+k-2} \itK(\by)d\by$ for $1\leq j,k,\leq q+1$ is defined in \textbf{A5}. Then, if we denote as $r(\by, \bx)=p(\bH_d\by+\bx)\gamma_{\bx,a}(\bH_d\by+\bx)f_\bX(\bH_d\by+\bx)- \esp \psi^{\prime}(\eps_1 a)\;f_\bX(\bx)p(\bx)$ we obtain that
\begin{eqnarray*}
 \esp M_{n1,{jk}}(\bx,a)-M_{jk}(\bx,a)   
&=& \int r(\by, \bx)  y^{j+k-2}_{\alpha}\itK(\by)\,d\by\,.
\end{eqnarray*}
Using the uniform continuity of  $f_\bX$, $\sigma$ and $p$  in $\itC_0$, we obtain that given $\epsilon>0$ there exists $\eta>0$ such that for any $\bx\in\itC$, $\bu\in \itC_0$ such that  $\|\bu-\bx\|<\eta$ implies
$|f_\bX(\bu)p(\bu)-f_\bX(\bx)p(\bx)|<\epsilon$ and $|\sigma(\bu)-\sigma(\bx)|<\epsilon$. The fact that $K_j$ has compact support $[-1,1]$, entails that   $\|\by\|\le \sqrt{d}$ for any $\by$ such that  $\itK(\by)\ne 0$. Therefore, using that $\max_{1\le j\le d}h_{j,n}\to 0$, we obtain that there exists $n_0$ such that if $n\ge n_0$, $\|\bH_d\by\|<\eta$ for all $\by$ such that $\itK(\by)\ne 0$ and  $\bH_d\by+\bx\in \itC_0$. Hence, if $n\ge n_0$ for all $\bx\in\itC$ and for all $\by$ such that $\itK(\by)\ne 0$, 
we have that $|f_\bX(\bH_d\by+\bx)p (\bH_d\by+\bx)-f_\bX(\bx)p (\bx)|<\epsilon$  and $|\sigma(\bH_d\by+\bx)-\sigma(\bx)|\leq \epsilon$. Using that  $\gamma_{\bx,a}(\bx)=\esp \psi^{\prime}(\eps_1 a)$,  (\ref{eqq6}) and   $|\gamma_{\bx,a}(\bx)|\le \|\psi^{\prime}\|_{\infty}$, we obtain that for all $\bx\in \itC$, ${a\in \itI_\delta}$ and $\by$ such that $\itK(\by)\ne 0$
\begin{eqnarray*}
 |r(\by, \bx)|
&\le & c_1  \sup_{\bu\in \itC_0}p( \bu)f_\bX(\bu)\, \left|\sigma(\bH_d\by+\bx)-\sigma( \bx)\right|+ \|\psi^{\prime}\|_{\infty} \left|p(\bH_d\by+\bx)f_\bX(\bH_d\by+\bx)- f_\bX(\bx)p(\bx)\right| \\
&\le & \left(c_1  \sup_{\bu\in \itC_0}p( \bu)f_\bX(\bu)+  \|\psi^{\prime}\|_{\infty}\right)\epsilon=c_2\,\epsilon\,.
\end{eqnarray*}
Then, for $n\ge n_0$, we have that
\begin{eqnarray*}
\sup_{\bx\in\itC}\sup_{a\in\itI_\delta}|\esp M_{n1,{jk}}(\bx,a)-M_{jk}(\bx,a)| &\leq& c_2\epsilon\int\!  y_{\alpha}^{j+k-2} \itK (\by)\,d\by= c_3\epsilon\,,
\end{eqnarray*}
concluding the proof of (ii).

Note that from (i) and (ii) it follows that
\begin{equation}\label{eqq7}
\sup_{\bx\in\itC}\sup_{a\in\itI_\delta}|M_{n1,{jk}}(\bx,a)-M_{jk}(\bx,a)|=O_{\as}\left(\sqrt{\frac{\log n}{n\prod_{j=1}^d h_j}}\right)+o\left(1\right)=o_{\as}(1)\,,
\end{equation}
which in particular leads to 
\begin{equation}\label{eqq8}
\sup_{\bx\in\itC}|M_{n1,{jk}}(\bx,1)-M_{jk}(\bx,1)|=O_{\as}\left(\sqrt{\frac{\log n}{n\prod_{j=1}^d h_j}}\right)+o\left(1\right)=o_{\as}(1)\,.
\end{equation}

Let us prove (iii). By (\ref{conproba}) and (\ref{eqq7}), given $\eta>0$, there exists a set $\itN$ such that $\prob(\itN)=0$ and for any $\omega\notin\itN$, 
there exists $n_1$ satisfying that for all $n\geq n_1$, $|\wa_\sigma(\bx)-1|<1/2$ and $\sup_{\bx\in\itC}\sup_{a\in\itI_\delta}|M_{n1,{jk}}(\bx,a)-M_{jk}(\bx,a)|<\eta$, with $\delta=1/2$. Then, for all $\omega\notin\itN$ and $n\geq n_1$ we get that
$$\sup_{\bx\in\itC}|M_{n1,{jk}}(\bx, \wa_\sigma(\bx))-M_{jk}(\bx,\wa_\sigma(\bx))|\le \sup_{\bx\in\itC}\sup_{a\in\itI_\delta}|M_{n1,{jk}}(\bx,a)-M_{jk}(\bx,a)|<\eta \;,$$
which concludes the proof of (iii).

Let now prove (iv). Denote $c_4=\sup_{\bx\in\itC} f_\bX(\bx)p(\bx)\max(1,s_{1,1}^{(\alpha)})$. Using that $f_\bX$ and $p$ are bounded in $\itC$, we get that
$$\sup_{\bx\in\itC}|M_{jj}(\bx,\wa_\sigma(\bx))-M_{jj}(\bx,1)|\leq c_4\sup_{\bx\in\itC}\left|\esp\psi^{\prime}\left(\eps \wa_\sigma(\bx)\right)-\esp\psi^{\prime}(\eps)\right|\;.$$ 
Using similar arguments to those considered to bound $\gamma_{\bx,a}$ in (ii), we obtain  that
$$|\lambda_1(a)-\lambda_1(1)|=\left|\esp\left[\psi^{\prime}(\eps a)-\psi^{\prime}(\eps)\right]\right|=\left|\esp\psi^{\prime\prime}(\eps \theta)\eps (a-1)\right|\leq \| {\zeta}_2\|_\infty \frac{1}{\theta}(a-1)\,,$$
where   $\lambda_1(a)=\esp\psi^{\prime}(\eps a)$ and $\theta$ is an intermediate point between $a$ and $1$. Hence, 
$$|\lambda_1(a)-\lambda_1(1)| \leq \| {\zeta}_2\|_\infty  (a-1)\max\left(1,\frac 1a\right)\,,$$
which implies that
$$\sup_{\bx\in\itC}|M_{jj}(\bx,\wa_\sigma(\bx))-M_{jj}(\bx,1)|\leq c_4 \| {\zeta}_2\|_\infty \sup_{\bx\in\itC}\left|\wa_\sigma(\bx)-1\right| \; \sup_{\bx\in\itC} \left(1+ \frac{1}{\wa_\sigma(\bx)}\right)\;.$$ 
Now (iv) follows from the fact that $\sup_{\bx\in\itC}\left|\wa_\sigma(\bx)-1\right|\convpp 0$.  
That is, we have concluded the proof of (\ref{eqqbM1q}).

By (\ref{eqq10q}) and using that $\psi^{\prime}$ is bounded and that $i(p)>0$, we obtain that
$$\sup_{\bx\in\itC}\|\wbM_{n2}(\bx)\|\leq \frac{\|\psi^{\prime}\|_\infty}{i(t)} A_g\left(h_\alpha^{q+1}+\sum_{j\neq\alpha} h_j\right)\sup_{\bx\in\itC}\frac{1}{n}\sum_{i=1}^n \left|\itK_{\bH_d}(\bX_i-\bx)\right|\,.$$ 
Using that $\wefe(\bx)= (1/n )\sum_{j=1}^n  \itK_{\bH_d}^{\star}\left(\bx-\bX_j\right)$ converges to $f_\bX$ uniformly and  \textbf{A2}, we obtain that $\sup_{\bx\in \itC}\wefe(\bx)=O_{\as}(1)$, so we obtain (\ref{eqqbM2q})  and the proof is concluded. \square

\subsection{Proof of Theorem \ref{ASMq}.1.}{\label{apend2}}

We begin by proving the following Lemma which will be useful in the proof of Theorem \ref{ASMq}.1.

\noi \textbf{Lemma \ref{apend2}.1.} \textsl{Assume that \textbf{A0}, \textbf{A2}, \textbf{A7} and \textbf{N1} to   \textbf{N5} hold and  that the function $\lambda(a)$ has bounded Lipschitz continuous derivatives up to order $\ell-1$, in a neighbourhood of $0$.  Let $\bx$ be an interior point of $\itS_f$. Define
$$\wtbA_{1,n}(\bx)=\frac{1}{n}\sum_{i=1}^n \itK_{\bH_d}(\bX_i- \bx)p(\bX_i)\lambda\left(\frac{R(\bX_i, \bx)}{\sigma}\right)\breve{\bx}_{i,\alpha}\;,$$
where  $R(\bX_i,\bx)=\sum_{j\neq\alpha}\{g_j(X_{i,j})-g_j(x_j)\}+R_{\alpha }(X_{i,\alpha},x_\alpha)$ and  $ R_{\alpha}(X_{i,\alpha},x_\alpha)$ is given in  in (\ref{erreX}). Then, we have that
$$
\esp \wtbA_{1,n}(\bx)=\frac{A_0(\psi)}{\sigma}h_\alpha^{q+1}p(\bx)f_\bX(\bx)\frac{1}{(q+1)!}g_\alpha^{(q+1)}(x_\alpha)\bese_q^{(\alpha)}+\bnu_n(\bx)\,,
$$
where   $\sup_{\bx\in\itS_Q}\|\bnu_n(\bx)\|= h_\alpha^{q+1}o(1)$.
}
 
\noi \textsc{Proof.}   Using that $\lambda$ is $\ell-1$ times differentiable, a $(\ell -1)$th order Taylor's expansion together with the facts   that $\lambda(0)=0$,  $\lambda^\prime(0)=\esp\psi^\prime(\eps_1)=A_0(\psi)$ entail  that
\begin{eqnarray*}
\lambda\left(\frac{R(\bu,\bx)}{\sigma}\right)&=&\lambda(0)+\sum_{k=1}^{\ell-1}\frac{1}{k!}\lambda^{(k)}(0)\left(\frac{R(\bu,\bx)}{\sigma}\right)^{k}+\frac{\lambda^{(\ell-1)}(\theta(\bu))-\lambda^{(\ell-1)}(0)}{(\ell-1)!} \left(\frac{R(\bu,\bx)}{\sigma}\right)^{\ell-1}\\
&=&A_0(\psi)\frac{R(\bu,\bx)}{\sigma}+ \sum_{k=2}^{\ell-1}\frac{1}{k!}\lambda^{(s)}(0)\left(\frac{R(\bu,\bx)}{\sigma}\right)^{k}+\frac{1}{(\ell-1)!}\wtlam\left(\bu,\bx\right)\left(\frac{R(\bu,\bx)}{\sigma}\right)^{\ell-1}
\end{eqnarray*}
where $\wtlam(\bu,\bx)=\lambda^{(\ell-1)}(\theta(\bu))-\lambda^{(\ell-1)}(0)$ with $\theta(\bu)$ an intermediate point between $0$ and $R(\bu,\bx)/\sigma$. On the other hand, we also have that $|\wtlam(\bu,\bx)|\leq C|\theta(\bu)|\leq C|R(\bu,\bx)|/\sigma$ since $\lambda^{(\ell-1)}$ is Lipschitz.  Note that \textbf{N2} and the fact that $h_j=\wth$ if $j\ne \alpha$ and $R(\bx,\bx)=0$ imply that
\begin{eqnarray}
R(\bx+\bH_d\bu, \bx)&=& \sum_{j\ne \alpha} \sum_{s=1}^\ell \wth^s \frac{g_j^{(s)}(x_j)}{s!} u_j^s+\wth^\ell \sum_{j\ne \alpha}   \frac{g_j^{(\ell)}(\xi_j)g_j^{(\ell)}(x_j)}{\ell!} u_j^\ell \nonumber\\
&&+h_\alpha^{q+1}\; \frac{g_\alpha^{(q+1)}(x_\alpha)}{(q+1)!} u_\alpha^{q+1}+h_\alpha^{q+1} \;\frac{g_\alpha^{(q+1)}(\xi_\alpha)-g_\alpha^{(q+1)}(x_\alpha)}{(q+1)!} u_\alpha^{q+1}\,.
\label{desarrolloR}
\end{eqnarray}
Then,  we have that
\begin{eqnarray*}
 \esp\left(\wtbA_{1,n}(\bx)\right) &=&\esp\left[\itK_{\bH_d}(\bX_1-\bx)p(\bX_1)\lambda\left(\frac{R(\bX_1,\bx)}{\sigma}\right)\breve{\bx}_{1,\alpha}\right]\\
&=&\frac{A_0(\psi)}{\sigma} \bA_{11,n} +\sum_{k=2}^{\ell-1} \frac{\lambda^{(k)}(0)}{k!\sigma^{k}} \bA_{1k,n} + \frac{1}{(\ell-1)!\;\sigma^{\ell-1}} \bA_{1\ell,n} 
\end{eqnarray*}
where  for $k=1,\dots,\ell-1$
\begin{eqnarray*}
\bA_{1k,n}&=&\frac{1}{h_\alpha\wth^{d-1}}\esp\left[\prod_{s=1}^d K_s\left(\frac{X_{1,s}-x_s}{h_s}\right)p(\bX_1)R^{k}(\bX_1,\bx)\breve{\bx}_{1,\alpha}\right]\\
&=&\int\!\itK(\bu)v(\bx+\bH_d\bu)R^{k}(\bx+\bH_d\bu,\bx)\breve{\bu}_\alpha\,d\bu\\
\bA_{1\ell,n}&=&\frac{1}{h_\alpha\wth^{d-1}}\esp\left[\prod_{s=1}^d K_s\left(\frac{X_{1,s}-x_s}{h_s}\right)p(\bX_1)\wtlam (\bX_1,\bx) R^{\ell-1}(\bX_1,\bx)\breve{\bx}_{1,\alpha}\right]\\
&=&\int\!\itK(\bu)v(\bx+\bH_d\bu)\wtlam (\bx+\bH_d\bu,\bx) R^\ell(\bx+\bH_d\bu,\bx)\breve{\bu}_\alpha\,d\bu\,,
\end{eqnarray*}
where $v(\bu)=p(\bu)f_\bX(\bu)$ and $\breve{\bu}_\alpha=(1,u_\alpha,\dots,u_\alpha^q)\trasp\in\real^{q+1}$. Using an $\ell-$th order Taylor's expansion of $v(\bx+\bH_d\bu)$ around  $\bx$, we get that
\begin{equation}
v(\bx+\bH_d\bu)=v(\bx)+\sum_{0<|\beme|\le \ell} D^{\beme}v(\bx)\bh^\beme\bu^\beme+\sum_{|\beme|=\ell} \frac{1}{\beme!}\left[D^{\beme}v( \bxi^{(1)})-D^{\beme}v(\bx)\right]\bh^\beme\bu^\beme
\label{desarrollov}
\end{equation}
where we have used the notation of Bourbaki for the expansion, $\bh= (h_1,\dots,h_d)\trasp$, where $h_j=\wth$ for $j\ne \alpha$,  $\beme=(m_1,\dots, m_d)$ with $m_i\in \natu$, $|\beme|=\sum_{j=1}^d m_j$, $\bu^{\beme}=\prod_{j=1}^d u_j^{m_j}$, $\beme!=m_1!\dots m_d!$ and $D^{\beme}v=\partial^{|\beme|} v/\partial u_1^{m_1}\dots \partial u_d^{m_d}$.

Using (\ref{desarrolloR}) and (\ref{desarrollov}) in $\bA_{1k,n}$, $k=1,\dots, \ell$, we obtain that $\bA_{11,n}$ can be written as $\bA_{11,n}=\sum_{j=1}^{18}\bA_{11,j,n}$ where
\begin{eqnarray*}
\bA_{11,1,n}&\hskip-0.1in=&\hskip-0.1in v(\bx)\left\{\sum_{j\neq\alpha}g_j^\prime(x_j)\wth\int\!\prod_{t=1}^d K_t(u_t)u_j\breve{\bu}_\alpha\,d\bu\right\}\\
\bA_{11,2,n}&\hskip-0.1in=&\hskip-0.1in v(\bx)\frac{1}{(q+1)!}g_\alpha^{(q+1)}(x_\alpha)h_\alpha^{q+1}\,\int\!\prod_{t=1}^d K_t(u_t)u_\alpha^{q+1}\breve{\bu}_\alpha\,d\bu\\
\bA_{11,3,n}&\hskip-0.1in=&\hskip-0.1in\sum_{k=1}^{\ell-1}\sum_{|\beme|=k}\frac{1}{\beme!} D^{\beme}v(\bx)\sum_{j\neq\alpha}g_j^\prime(x_j)\,\wth\,\bh^\beme\int\!\prod_{t=1}^d K_t(u_t)u_j\bu^\beme\breve{\bu}_\alpha\,d\bu
\end{eqnarray*}
\begin{eqnarray*}
\bA_{11,4,n}&\hskip-0.1in=&\hskip-0.1in\sum_{k=1}^{\ell-1}\sum_{|\beme|=k}\frac{1}{\beme!} D^{\beme}v(\bx)\frac{1}{(q+1)!}g_\alpha^{(q+1)}(x_\alpha)h_\alpha^{q+1}\bh^\beme\int\!\prod_{t=1}^d K_t(u_t)u_\alpha^{q+1}\bu^\beme\breve{\bu}_\alpha\,d\bu\\
\bA_{11,5,n}&\hskip-0.1in=&\hskip-0.1in\sum_{|\beme|=\ell}\frac{1}{\beme!} D^{\beme}v(\bx)\sum_{j\neq\alpha}g_j^\prime(x_j)\,\wth\,\bh^\beme\,\int\!\prod_{t=1}^d  K_t(u_t)u_j\bu^\beme\breve{\bu}_\alpha\,d\bu\\
\bA_{11,6,n}&\hskip-0.1in=&\hskip-0.1in \sum_{|\beme|=\ell}\frac{1}{\beme!} D^{\beme}v(\bx)\frac{1}{(q+1)!}g_\alpha^{(q+1)}(x_\alpha)h_\alpha^{q+1}\bh^\beme
\int\!\prod_{t=1}^d  K_t(u_t)u_\alpha^{q+1}\bu^\beme\breve{\bu}_\alpha\,d\bu\\
\bA_{11,7,n}&\hskip-0.1in=&\hskip-0.1in\sum_{|\beme|=\ell}\sum_{j\neq\alpha}g_j^\prime(x_j) \,\wth\,\frac{1}{\beme!}\bh^\beme
\int\!\left[ D^{\beme}v(\bxi^{(1)})- D^{\beme}v(\bx)\right]\prod_{t=1}^d  K_t(u_t)u_j\bu^\beme\breve{\bu}_\alpha\,d\bu\\
\bA_{11,8,n}&\hskip-0.1in=&\hskip-0.1in \sum_{|\beme|=\ell}\frac{1}{\beme!}\frac{1}{(q+1)!}g_\alpha^{(q+1)}(x_\alpha)h_\alpha^{q+1}\bh^\beme\,\int\!\left[ D^{\beme}v(\bxi^{(1)})- D^{\beme}v(\bx)\right]\prod_{t=1}^d K_t(u_t)u_\alpha^{q+1}\bu^\beme\breve{\bu}_\alpha\,d\bu\\
\bA_{11,9,n}&\hskip-0.1in=&\hskip-0.1in v(\bx) \left\{\sum_{j\neq\alpha}\sum_{m=2}^{\ell-1}\frac{1}{m!}g_j^{(m)}(x_j) \wth^m\,\int\!\prod_{t=1}^d K_t(u_t)u_j^{m}\breve{\bu}_\alpha\,d\bu\right\}\\
\bA_{11,10,n}&\hskip-0.1in=&\hskip-0.1in v(\bx) \left\{\sum_{j\neq\alpha}\frac{1}{\ell!}g_j^{(\ell)}(\bxi_j) \wth^\ell\,\int\!\prod_{t=1}^d K_t(u_t)u_j^{\ell}\breve{\bu}_\alpha\,d\bu\right\}\\
\bA_{11,11,n}&\hskip-0.1in=&\hskip-0.1in v(\bx) \left\{\frac{1}{(q+1)!}h_\alpha^{q+1}\,\int\!\left[g_\alpha^{(\ell)}(\bxi_\alpha)-g_\alpha^{(\ell)}(x_\alpha)\right]\prod_{t=1}^d K_t(u_t)u_\alpha^{q+1}\breve{\bu}_\alpha\,d\bu\right\}\\
\bA_{11,12,n}&\hskip-0.1in=&\hskip-0.1in \sum_{k=1}^{\ell-1}\sum_{|\beme|=k}\frac{1}{\beme!} D^{\beme}v(\bx) \sum_{j\neq\alpha}\sum_{m=2}^{\ell}\frac{1}{m!}g_j^{(m)}(x_j) \wth^m\bh^\beme\int\!\prod_{t=1}^d K_t(u_t)u_j^{m}\bu^\beme\breve{\bu}_\alpha\,d\bu\\
\bA_{11,13,n}&\hskip-0.1in=&\hskip-0.1in\sum_{k=1}^{\ell-1}\sum_{|\beme|=k}\frac{1}{\beme!} D^{\beme}v(\bx) \sum_{j\neq\alpha} \wth^{\ell}\bh^\beme\int\![g_j^{(\ell)}(\bxi_j)-g_j^{(\ell)}(x_j)]\prod_{t=1}^d K_t(u_t)u_j^{\ell}\bu^\beme\breve{\bu}_\alpha\,d\bu\\
\bA_{11,14,n}&\hskip-0.1in=&\hskip-0.1in\sum_{|\beme|=\ell}\frac{1}{\beme!} \sum_{j\neq\alpha} \wth^{\ell}\bh^\beme\,\int\! D^{\beme}v( \bxi^{(1)})\left[g_j^{(\ell)}(\bxi_j)-g_j^{(\ell)}(x_j)\right]\prod_{t=1}^d K_t(u_t)u_j^{\ell}\bu^\beme\breve{\bu}_\alpha\,d\bu
\end{eqnarray*}
\begin{eqnarray*}
\bA_{11,15,n}&\hskip-0.1in=&\hskip-0.1in\sum_{k=1}^{\ell-1}\sum_{|\beme|=k}\frac{1}{\beme!} D^{\beme}v(\bx)\frac{1}{(q+1)!}h_\alpha^{q+1}\bh^\beme\int\!\left[g_\alpha^{(q+1)}(\bxi_\alpha)-g_\alpha^{(q+1)}(x_\alpha)\right]\prod_{t=1}^d K_t(u_t)u_\alpha^{q+1}\bu^\beme\breve{\bu}_\alpha\,d\bu\\
\bA_{11,16,n}&\hskip-0.1in=&\hskip-0.1in\sum_{|\beme|=\ell}\frac{1}{\beme!}\frac{1}{(q+1)!}h_\alpha^{q+1}\bh^\beme\,\int\! D^{\beme}v( \bxi^{(1)})\left[g_\alpha^{(q+1)}(\bxi_\alpha)-g_\alpha^{(q+1)}(x_\alpha)\right]\prod_{t=1}^d K_t(u_t)u_\alpha^{q+1}\bu^\beme\breve{\bu}_\alpha\,d\bu\\
\bA_{11,17,n}&\hskip-0.1in=&\hskip-0.1in\sum_{|\beme|=\ell}\frac{1}{\beme!} D^{\beme}v(\bx)\sum_{j\neq\alpha}\sum_{m=2}^\ell\frac{1}{m!}g_j^{(m)}(x_j)\wth^m\bh^\beme\int\!\prod_{t=1}^d K_t(u_t)u_j^m\bu^\beme\breve{\bu}_\alpha\,d\bu\\
\bA_{11,18,n}&\hskip-0.1in=&\hskip-0.1in\sum_{|\beme|=\ell}\frac{1}{\beme!}\sum_{j\neq\alpha}\sum_{m=2}^\ell \frac{1}{m!}g_j^{(m)}(x_j)\wth^m\bh^\beme\int\!\left[ D^{\beme}v( \bxi^{(1)})- D^{\beme}v(\bx)\right]\prod_{t=1}^d K_t(u_t)u_j^m\bu^\beme\breve{\bu}_\alpha\,d\bu\;,
\end{eqnarray*}
with $\bxi^{(1)}$ an intermediate point between $\bx$ and $\bx+\bH_d\bu$. 
The fact that $\int\!K_j(t)t\,dt=0$ entails that $\bA_{11,1,n}=\bcero$. On the other hand, using that $K_j=L$ is a kernel of order $\ell$, if $j\ne \alpha$, we get that $\int\!\itK(\bu)u_j\bu^\beme\breve{\bu}_\alpha\,d\bu=0$   for  $|\beme|\leq\ell-1$. Moreover, we also have that  $\int\!\itK(\bu)u_j\bu^\beme\breve{\bu}_\alpha\,d\bu=0$ if $|\beme|=\ell-1$ and $\beme\neq (\ell-1)\be_j$. On the other hand, using again   that $L$ is a kernel of order $\ell$, we obtain that $\int\!\itK(\bu)\bu^{\beme}u_\alpha^{q+1}\breve{\bu}_\alpha\,d\bu=0$ for  all $\beme$ with at least one component $m_j\ne 0$, for  $j\neq \alpha$. Thus, $\bA_{11,9,n}=\bcero$.

On the other hand, we have that
\begin{eqnarray*} 
\bA_{11,2,n}&=&h_\alpha^{q+1}v(\bx)\frac{1}{(q+1)!}g_\alpha^{(q+1)}(x_\alpha)\,\int\!K_\alpha(u_\alpha)u_\alpha^{q+1}\breve{\bu}_\alpha\,du_\alpha\\
&=&h_\alpha^{q+1}v(\bx)\frac{1}{(q+1)!}g_\alpha^{(q+1)}(x_\alpha)\bese_q^{(\alpha)}\,.
\end{eqnarray*}
Since $D^{\beme}v(\bu)$ with $|\beme|=k$ and $g_\alpha^{(q+1)}$ are continuous and bounded functions  and $g_j$ is $\ell$ times differentiable, with bounded derivatives for all $j\neq\alpha$, it follows that
\begin{eqnarray*}
\sup_{\bx\in\itS_Q}\|\bA_{11,3,n}\|=\wth^\ell O(1)\,,\qquad 
\sup_{\bx\in\itS_Q}\|\bA_{11,4,n}\|=h_\alpha^{q+2}O(1)\qquad\mbox{y}\quad
\sup_{\bx\in\itS_Q}\|\bA_{11,10,n}\|=\wth^\ell O(1)\,.
\end{eqnarray*} 
Similarly, using that the kernels are even we have that $\int\!\prod_{t=1}^d  K_t(u_t)u_j\bu^\beme\breve{\bu}_\alpha\,d\bu=0$ when $\beme$ has a component different from $\alpha$ and $j$ different from 0. Moreover, when $m_s=0$ for $s\ne j, \alpha$, using that $L$ is a kernel of order $\ell$ we get that the integral equals 0 except when $m_j\ne \ell -1$ and $m_\alpha= 1$. Arguing similarly with $\int\!\prod_{t=1}^d  K_t(u_t)u_\alpha^{q+1}\bu^\beme\breve{\bu}_\alpha\,d\bu$, we get that 
\begin{eqnarray*}
\bA_{11,5,n}&=&\sum_{j\neq\alpha}g_j^\prime(x_j)\,\frac{1}{(\ell-1)!}\left.\frac{\partial^\ell v(\bu)}{\partial u_j^{\ell-1}\partial u_\alpha}\right|_{\bu=\bx}\wth^{\ell}h_\alpha\int\!\itK(\bu)u_j^{\ell}u_\alpha\breve{\bu}_\alpha\,d\bu\\
\bA_{11,6,n}&=&\frac{1}{(q+1)!}g_\alpha^{(q+1)}(x_\alpha)h_\alpha^{q+1}\sum_{j\neq\alpha}\frac{1}{\ell!}\left.\frac{\partial^\ell v(\bu)}{\partial u_j^\ell}\right|_{\bu=\bx}h_j^\ell\int\!\itK(\bu)u_\alpha^{q+1}u_j^\ell\breve{\bu}_\alpha\,d\bu\\
&&+\frac{1}{(q+1)!}g_\alpha^{(q+1)}(x_\alpha)h_\alpha^{q+1}\frac{1}{\ell!}\left.\frac{\partial^\ell v(\bu)}{\partial u_\alpha^\ell}\right|_{\bu=\bx}h_\alpha^\ell\int\!\itK(\bu)u_\alpha^{q+1+\ell}\breve{\bu}_\alpha\,d\bu\,,
\end{eqnarray*}
which implies that
$$ \sup_{\bx\in\itS_Q}\|\bA_{11,5,n}\|=\wth^\ell h_\alpha \,O(1)\qquad \sup_{\bx\in\itS_Q}\|\bA_{11,6,n}\|=h_\alpha^{q+1}(\wth^\ell+h_\alpha^\ell)\,O(1)\,.$$

On the other hand,   $D^{\beme}v(\bu)$ for $|\beme|=k\le \ell$  is uniformly continuous, so using that $K_\alpha$ and $L$ have compact support in $[-1,1]$ and that $\bxi^{(1)}$ is an intermediate point between $\bx$ and $\bx+\bH_d\bu$, we have that
$$\sup_{\bx\in\itS_Q}\left| D^{\beme}v(\bxi^{(1)})-D^{\beme}v(\bx)\right|=o(1)$$
which leads to
\begin{eqnarray*}
\sup_{\bx\in\itS_Q}\|\bA_{11,7,n}\|=\wth^\ell o(1)\,,\qquad
\sup_{\bx\in\itS_Q}\|\bA_{11,8,n}\|=h_\alpha^{q+1}o(1)\qquad \mbox{and}\qquad
\sup_{\bx\in\itS_Q}\|\bA_{11,18,n}\|&=&\wth^\ell O(1)\,.
\end{eqnarray*} 
Similarly, using that $g_j^{(\ell)}$ is uniformly continuous and bounded, we get that $\sup_{\bx\in\itS_Q}|g_j^{(\ell)}(\bxi_j)-g_j^{(\ell)}(x_j)|=o(1)$ which implies that
\begin{eqnarray*}
\sup_{\bx\in\itS_Q}\|\bA_{11,11,n}\|=h_\alpha^{q+1}o(1)\quad && \quad
\sup_{\bx\in\itS_Q}\|\bA_{11,12,n}\|=\wth^\ell O(1)\\
\sup_{\bx\in\itS_Q}\|\bA_{11,13,n}\|=\wth^\ell(\wth+h_\alpha)o(1)\quad && \quad
\sup_{\bx\in\itS_Q}\|\bA_{11,14,n}\|=\wth^\ell o(1)\\
\sup_{\bx\in\itS_Q}\|\bA_{11,15,n}\|=h_\alpha^{q+2}o(1)\quad && \quad
\sup_{\bx\in\itS_Q}\|\bA_{11,16,n}\|=h_\alpha^{q+1} o(1)\\
\sup_{\bx\in\itS_Q}\|\bA_{11,17,n}\|=\wth^\ell O(1)\quad && \quad
\sup_{\bx\in\itS_Q}\|\bA_{11,18,n}\|=\wth^\ell o(1)\,.
\end{eqnarray*}
Using  (\ref{desarrolloR}) and the fact that, for $j\ne \alpha$, $K_j=L$ is a kernel of order $\ell$  and that $\wth^\ell=o(h_\alpha^{q+1})$, using analogous arguments, we obtain that for all $k=2,\dots,\ell-1$ 
$$
\sup_{\bx\in\itS_Q}\|\bA_{1k,n}\|= h_\alpha^{q+1}o(1)\;.
$$
Let $A_{1\ell,n,s}$ indicate the $s$th coordinate of $\bA_{1\ell,n}$. Using that $|\wtlam(\bu,\bx)|\leq C|R(\bu,\bx)|/\sigma$, $K_j$ has support in $[-1,1]$, $v$ is bounded and   (\ref{desarrolloR}), we get that, for $s=1,\dots q+1$,
\begin{eqnarray*}
\sup_{\bx\in\itS_Q}|A_{1\ell,n,s}|&\leq &\sup_{\bx\in\itS_Q}\int\!|\itK(\bu)||v(\bx+\bH_d\bu)||\wtlam (\bx+\bH_d\bu,\bx)|\;|R(\bx+\bH_d\bu,\bx)|^{\ell-1}|u_\alpha|^{s-1}\,d\bu\\
&\leq&\frac{C}{\sigma}\int\!|\itK(\bu)||v(\bx+\bH_d\bu)|\, |R(\bx+\bH_d\bu,\bx)|^{\ell}\,d\bu\leq c_2(\wth+h_\alpha^{q+1})^{\ell}=o\left(h_\alpha^{q+1}\right)\;,
\end{eqnarray*}
thus,
$$
\sup_{\bx\in\itS_Q}\|\bA_{1\ell,n}\|= h_\alpha^{q+1}o(1)\;.
$$
Hence, using that  $h_\alpha\to 0$ and $\wth\to 0$, we get that 
\begin{eqnarray*}
\esp \wtbA_{1,n}(\bx)&=&\frac{A_0(\psi)}{\sigma}\bA_{11,n}
+\sum_{k=2}^{\ell-1}\frac{\lambda^{(k)}(0)}{k!\sigma^{k}}\bA_{1k,n}+\frac{1}{(\ell-1)!\;\sigma^{\ell-1}}\bA_{1\ell,n}\nonumber\\
&=&\frac{A_0(\psi)}{\sigma}h_\alpha^{q+1}v(\bx)\frac{1}{(q+1)!}g_\alpha^{(q+1)}(x_\alpha)\bese_q^{(\alpha)}+\bnu_n(\bx)\,,
\end{eqnarray*}
where $\sup_{\bx\in\itS_Q}\|\bnu_n(\bx)\|=\wth^\ell O(1)+h_\alpha^{q+1}o(1)=h_\alpha^{q+1}o(1)$ and the proof is concluded. \square

\vskip0.2in
\noi \textsc{Proof of Theorem \ref{ASMq}.1.} The proof will be carried out in several steps. In a first step, we will show that it is enough to assume that, since the scale estimator has a root$-n$ rate of convergence, it is enough to prove the result in the situation in which scale is known to obtain the conclusion of Theorem \ref{ASMq}.1.
In a second step, we obtain an expansion for the estimator computed when scale is known into two terms. The first one will converge to the asymptotic bias and the second one  to a centered normal distribution from which the conclusion follows. To obtain these two last results some intermediate approximations will be needed.

\vskip0.1in
\noi \textbf{Step 1.}
For any $s>0$, define $\bPsi_{n,\alpha}^{\star}(\bbe,\bx,s)=\left(\Psi_{n,\alpha,0}^{\star}(\bbe,\bx,s), \dots, \Psi_{n,\alpha,q}^{\star}(\bbe,\bx,s)\right)$ where
\begin{eqnarray*}
\bPsi_{n,\alpha}^{\star}(\bb,\bx,s)&=&\frac{1}{n}\sum_{i=1}^n \psi\left(\frac{Y_i-b_0-\sum_{m=1}^q b_m(X_{i,\alpha}-x_\alpha)^m}{s}\right)\itK_{\bH_d}(\bX_i-\bx)\delta_i\breve{\bx}_{i,\alpha}\\
&=&\frac{1}{n}\sum_{i=1}^n \psi\left(\frac{Y_i-\breve{\bx}_{i,\alpha}\trasp \bH_d\bb}{s}\right)\itK_{\bH_d}(\bX_i-\bx)\delta_i\breve{\bx}_{i,\alpha}\,.
\end{eqnarray*}
 Using  that $\sqrt{n}(\wese-\sigma)=O_\prob(1)$, $\psi$ is Lipschitz and   $\zeta(u)=u\psi^\prime(u)$ is bounded, it is easy to see that for $j=0,\dots,q$, $\wD_{n,j}=\sup_{\bx\in\itS_Q}\sup_\bb|\wese\Psi_{n,\alpha,j}^{\star}(\bb,\bx,\wese)-\sigma\Psi_{n,\alpha,j}^{\star}(\bb,\bx,\sigma)|=O_\prob(1/\sqrt{n})$. On the other hand,  $\wbbe(\bx)$ is a solution of (\ref{22alfa}) with $\wese(\bx)=\wese$, that is, $\bPsi_{n,\alpha}^{\star}(\wbbe(\bx),\bx,\wese)=\bcero_{q+1}$, which implies that
 \begin{equation}
 \bPsi_{n,\alpha}^{\star}(\wbbe(\bx),\bx, \sigma)=O_\prob(1/\sqrt{n})
 \label{ecua1}
 \end{equation}

Denote as $\wtbbe(\bx)$ the solution of $\bPsi_{n,\alpha}^{\star}(\bb,\bx, \sigma)=\bcero_{q+1}$. Then,   Proposition \ref{consist}.1 entails that $\sup_{\bx\in\itS_Q}\left\|\bH^{(\alpha)}\left[\wtbbe(\bx)-\bbe(\bx)\right]\right\|=o_\prob(1)$, so using that  $\sup_{\bx\in\itS_Q}\left\|\bH^{(\alpha)}\left[\wbbe(\bx)-\bbe(\bx)\right]\right\|=o_\prob(1)$, we get that $\wD_n=\sup_{\bx\in\itS_Q}\left\|\bH^{(\alpha)}\left[\wbeta(\bx)-\wtbeta(\bx)\right]\right\|=o_\prob(1)$. We will further show that
\begin{equation}\label{klx4alpha} 
\wD_n=O_\prob(1/\sqrt{n})\,.
\end{equation}
To prove (\ref{klx4alpha}), denote as
$$\wbD_{1,n}(\bx, \bxi)=-\frac{1}{\sigma}\frac{1}{n}\sum_{i=1}^n \psi^\prime\left(\frac{Y_i-\breve{\bx}_{i,\alpha}\trasp\bH^{(\alpha)}\bxi}{\sigma}\right)\itK_{\bH_d}(\bX_i-\bx)\delta_i\breve{\bx}_{i,\alpha}\breve{\bx}_{i,\alpha}\trasp\,.$$ 
Then, a first order Taylor expansion and the fact  that $\bPsi_{n,\alpha}^{\star}(\wtbbe(\bx),\bx, \sigma)=\bcero$  lead us to 
\begin{equation}
\bPsi_{n,\alpha}^{\star}(\wbbe(\bx),\bx, \sigma)=\bPsi_{n,\alpha}^{\star}(\wtbbe(\bx),\bx, \sigma)+\wbD_{1,n}(\bx)\bH^{(\alpha)}(\wbbe(\bx)-\wtbbe(\bx))=\wbD_{1,n}(\bx, \bxi_n)\bH^{(\alpha)}(\wbbe(\bx)-\wtbbe(\bx))\,,
\label{ecua2}
\end{equation} 
where   $\bxi_n=\bxi_n(\bx)$ stands for an intermediate point between  $\wbbe(\bx)$ and $\wtbbe(\bx)$, so   $\sup_{\bx\in\itS_Q}\|\bH^{(\alpha)}[\bxi_n(\bx)-\beta(\bx)]\|=o_\prob(1)$. 
Denote as $A_0(\psi)=\esp(\psi^\prime(\eps))$ and
$$\bD_0(\bx)=\,-\,\frac 1{\sigma}A_0(\psi)p(\bx)f_\bX(\bx)\bS^{(\alpha)}\,.$$
Then, using that from \textbf{N3}b) $\bS^{(\alpha)}$ is non--singular, $\inf_{\bx\in\itC} f_{\bX}(\bx)>0$, $\inf_{\bx\in\itC} p(\bx)>0$ and $A_0(\psi)\ne 0$  we get that $\inf_{\bx\in\itC} \nu_1(\bD_0(\bx))>0$, with $\nu_1(\bA)$ the smallest eigenvalue of the matrix $\bA$. Hence, (\ref{ecua1})  and (\ref{ecua2})  implies that to show (\ref{klx4alpha}) it is enough to see that
\begin{equation}\label{klx5alpha}
\sup_{\bx\in\itS_Q}\|\wbD_{1,n}(\bx, \bxi_n)-\bD_{0}(\bx)\|=o_\prob(1)\,.
\end{equation}
We get that $\sigma\wbD_{1,n}(\bx, \bxi_n)= \wbD_{11,n}(\bx, \bxi_n)+\wbD_{12,n}(\bx)+\wbD_{13,n}(\bx)$ where
\begin{eqnarray*}
\wbD_{11,n}(\bx, \bxi_n)&=&\frac{1}{n}\sum_{i=1}^n \itK_{\bH_d}(\bX_i-\bx)\delta_i\left[\psi^\prime\left(\frac{Y_i-\breve{\bx}_{i,\alpha}\trasp\bH^{(\alpha)}\bbe(\bx)}{\sigma}\right)-\psi^\prime\left(\frac{Y_i-\breve{\bx}_{i,\alpha}\trasp\bH^{(\alpha)}\bxi_n}{\sigma}\right)\right]\breve{\bx}_{i,\alpha}\breve{\bx}_{i,\alpha}\trasp\\
\wbD_{12,n}(\bx)&=&\frac{1}{n}\sum_{i=1}^n \itK_{\bH_d}(\bX_i-\bx)\delta_i\left[\psi^\prime\left(\frac{Y_i-\breve{\bx}_{i,\alpha}\trasp\bH^{(\alpha)}\bbe(\bX_i)}{\sigma}\right)-\psi^\prime\left(\frac{Y_i-\breve{\bx}_{i,\alpha}\trasp\bH^{(\alpha)}\bbe(\bx)}{\sigma}\right)\right]\breve{\bx}_{i,\alpha}\breve{\bx}_{i,\alpha}\trasp\\
\wbD_{13,n}(\bx)&=&-\frac{1}{n}\sum_{i=1}^n \itK_{\bH_d}(\bX_i-\bx)\delta_i\psi^\prime\left(\frac{Y_i-\breve{\bx}_{i,\alpha}\trasp\bH^{(\alpha)}\bbe(\bX_i)}{\sigma}\right)\breve{\bx}_{i,\alpha}\breve{\bx}_{i,\alpha}\trasp\,.
\end{eqnarray*}
Using that $\psi^\prime$ is Lipschitz,  $\sup_{\bx\in\itS_Q}\|\bH^{(\alpha)}[\bxi_n(\bx)-\bbe(\bx)]\|=o_\prob(1)$, $\sup_{\bx\in\itC}\sum_{i=1}^n |\itK_{\bH_d}(\bX_i-\bx)|/n=O_\prob(1)$, $\sup_{|\itK_{\bH_d}(\bX_i-\bx)|\neq 0}|\breve{\bx}_{i,\alpha}|\leq 1$   we obtain that, for $1\leq j,m\leq q+1$, $\sup_{\bx\in\itS_Q}|\wD_{11,n,j,m}(\bx, \bxi_n)|=o_\prob(1)$, where $\wD_{11,n,j,m}(\bx, \bxi_n)$ is the $(j,m)-$th element of matrix $\wbD_{11,n}(\bx, \bxi_n)$. 

On the other hand, from the bound $\sup_{|\itK_{\bH_d}(\bX_i-\bx)|\neq 0}|\breve{\bx}_{i,\alpha}\trasp\bH^{(\alpha)}(\bbe(\bX_i)-\bbe(\bx))|\leq C(\wth+h_\alpha^{q+1})$, for $1\leq j,m\leq q+1$ we get that $\sup_{\bx\in\itS_Q}|\wD_{12,n,j,m}(\bx)|=o_\prob(1)$. 

Finally,  Lemma \ref{apend1}.2 entails that, for $1\leq j,m\leq q+1$,  $\sup_{\bx\in\itS_Q}|\wD_{13,n,j,m}(\bx)-\esp\wD_{13,n,j,m}(\bx)|=o_\prob(1)$, while standard arguments allow to show that  $\sup_{\bx\in\itS_Q}|\esp\wD_{13,n,j,m}(\bx)-D_{0,j,m}(\bx)|=o_\prob(1)$, concluding  the proof of (\ref{klx5alpha}) and so that of (\ref{klx4alpha}).

Observe that since the first element of the diagonal matrix $\bH^{(\alpha)}$ equals 1, we have that
$${\wg}_{\alpha, {\eme_{q,\alpha}}}(x_{\alpha})=\int\!\be_{1}\trasp\,\wbbe(x_{\alpha},\bu_\undalpha )q_\undalpha (\bu_\undalpha )\,d\bu_\undalpha=\int\!\be_{1}\trasp\,\bH^{(\alpha)}\wbbe(x_{\alpha},\bu_\undalpha )q_\undalpha (\bu_\undalpha )\,d\bu_\undalpha\,.$$
On the other hand, $\bbe(\bx)=(g(\bx),g_\alpha^{(1)}(x_\alpha),\dots,g_\alpha^{(q)}(x_\alpha))\trasp$, so using  (\ref{identif}) we get that 
$$\int\!\be_1\trasp\bH^{(\alpha)}\bbe(\bx) q_\undalpha(\bx_\undalpha)\,d\bx_\undalpha=\int\!g(\bx) q_\undalpha(\bx_\undalpha)\,d\bx_\undalpha= g_\alpha(x_\alpha)\,,$$
which implies that
\begin{eqnarray}
&&\left|\sqrt{nh_\alpha}(\wg_{\alpha,\eme_{q,\alpha}}(x_\alpha)-g_\alpha(x_\alpha))-\sqrt{nh_\alpha}\int\!\be_1\trasp\bH^{(\alpha)}[\wtbbe(\bx)-\bbe(\bx)]q_\undalpha(\bx_\undalpha)\,d\bx_\undalpha\right|\nonumber\\
&&=\left|\sqrt{nh_\alpha}\int\!\be_1\trasp\bH^{(\alpha)}[\wbbe(\bx)- \wtbbe(\bx) ]q_\undalpha(\bx_\undalpha)\,d\bx_\undalpha\right|\leq \sqrt{h_\alpha}\sqrt{n}\wD_n\convprob 0\,.\label{ecua3}
\end{eqnarray}
Let us denote as $\wg_{\alpha}(x_\alpha)=\int\!\be_1\trasp\bH^{(\alpha)}\wtbbe(\bx)q_\undalpha(\bx_\undalpha)\,d\bx_\undalpha$. Then, (\ref{ecua3}) implies that to obtain the asymptotic distribution of $\sqrt{nh_\alpha}(\wg_{\alpha,\eme_{q,\alpha}}(x_\alpha)-g_\alpha(x_\alpha))$ it is enough to derive that   of
$$\sqrt{nh_\alpha}\left[\wg_\alpha(x_\alpha)-g_\alpha(x_\alpha)\right]=\sqrt{nh_\alpha}\int\!\be_1\trasp\bH^{(\alpha)}[\wtbbe(\bx)-\bbe(\bx)]q_\undalpha(\bx_\undalpha)\,d\bx_\undalpha\,,$$ 
that is, we have reduced the problem to obtain the conclusion of Theorem \ref{ASMq}.1, when the scale is known.

\vskip0.1in
\noi \textbf{Step 2.} Using that $\bPsi_{n,\alpha}^{\star}(\wtbbe(\bx),\bx, \sigma)=\bcero_{q+1}$ and a first order Taylor's expansion of $\bPsi_{n,\alpha}^{\star}(\bb,\bx, \sigma) $ around $\bbe(\bx)$, it is easy to see that
\begin{equation}\label{413}
\bH^{(\alpha)}[\wtbbe(\bx)-\bbe(\bx)]=\sigma\,\wbA_{0,n}^{-1}(\bx)\,\wbA_{1,n}(\bx)
\end{equation} 
where $\wbA_{0,n}(\bx)=\wbA_{01,n}(\bx)+\wbA_{02,n}(\bx)$ with
\begin{eqnarray*}
\wbA_{01,n}(\bx)&=&\frac{1}{n}\sum_{i=1}^n \delta_i\itK_{\bH_d}(\bX_i-\bx)\psi^\prime\left(\frac{Y_i-\breve{\bx}_{i,\alpha}\trasp\bH^{(\alpha)} \bbe(\bx)}{\sigma}\right)\breve{\bx}_{i,\alpha}\breve{\bx}_{i,\alpha}\trasp\\
&=&\frac{1}{n}\sum_{i=1}^n \delta_i\itK_{\bH_d}(\bX_i-\bx)\psi^\prime\left(\eps_i+\frac{R(\bX_i,\bx)}{\sigma}\right)\breve{\bx}_{i,\alpha}\breve{\bx}_{i,\alpha}\trasp\\
\wbA_{02,n}(\bx)&=& \,-\,\frac{1}{2n}\sum_{i=1}^n \delta_i\itK_{\bH_d}(\bX_i-\bx)\psi^{\prime\prime}\left(\frac{Y_i-\breve{\bx}_{i,\alpha}\trasp \wbthe(\bx)}{\sigma}\right)\breve{\bx}_{i,\alpha}\breve{\bx}_{i,\alpha}\trasp\, \left(\breve{\bx}_{i,\alpha}\trasp \bH^{(\alpha)}\left[\wtbbe(\bx)-\bbe(\bx)\right]\right)\\
\wbA_{1,n}(\bx)&=&\frac{1}{n}\sum_{i=1}^n \delta_i\itK_{\bH_d}(\bX_i-\bx) \psi\left(\frac{Y_i-\breve{\bx}_{i,\alpha}\trasp\bH^{(\alpha)}\bbe(\bx)}{\sigma}\right)\breve{\bx}_{i,\alpha}
\end{eqnarray*}
where $\wbthe(\bx)$ is a midpoint between $\bH^{(\alpha)}\bbe(\bx)$ and $\bH^{(\alpha)}\wtbbe(\bx)$. Denote as $v(\bu)=p(\bu)f_\bX(\bu)$ and  $\bA_0(\bu)=v(\bu)A_0(\psi)\bS^{(\alpha)}$. Lemma \ref{apend1}.2 allow to show that $\sup_{\bx\in\itS_Q}|\wbA_{01,n}(\bx)- \bA_0(\bx)|=o_\prob(1)$. On the other hand, the fact that $\psi^{\prime\prime}$ is bounded, $\sup_{\bx\in\itS_Q}\left\|\bH^{(\alpha)}\left[\wtbbe(\bx)-\bbe(\bx)\right]\right\|=o_\prob(1)$ and that  that each component of $\breve{\bx}_{i,\alpha}$ is smaller or equal to 1 when $\itK_{\bH_d}(\bX_i-\bx)\ne 0$, imply that  $\sup_{\bx\in\itS_Q}|\wbA_{02,n}(\bx)|=o_\prob(1)$, so $\sup_{\bx\in\itS_Q}|\wbA_{0,n}(\bx)- \bA_0(\bx)|=o_\prob(1)$.

In \textbf{Step 2.1}, we study the asymptotic behaviour of
$$\wbB_n=\sigma\sqrt{nh_\alpha}\int\!\bA_0^{-1}(\bx)\wbA_{1,n}(\bx)q_\undalpha(\bx_\undalpha)\,d\bx_\undalpha$$
and we show that 
\begin{equation}
\wbB_n\convdist N_{q+1}\left(\bb_{q,\alpha}(x_\alpha), \bSi_{q,\alpha}(x_\alpha)\right)\,,
\label{asdistBn}
\end{equation}
\begin{eqnarray}
\bb_{q,\alpha}(x_\alpha)&=& \frac{\beta^{\frac{2q+3}2}}
{(q+1)!}\;g_\alpha^{(q+1)}(x_\alpha)(\bS^{(\alpha)})^{-1}\bese_q^{(\alpha)}
\label{sesgoBn}\\ 
\bSi_{q,\alpha}(x_\alpha)&=&\sigma^2 \frac{\esp \psi^2(\eps)}{A_0^2(\psi)}\int\!\frac{q_\undalpha^2(\bx_\undalpha)}{f_\bX(x_\alpha, \bx_\undalpha)p(x_\alpha, \bx_\undalpha)}\,d\bx_\undalpha (\bS^{(\alpha)})^{-1}\bV_\alpha (\bS^{(\alpha)})^{-1}\,.
\label{varianzaBn}
\end{eqnarray}
We will then show, in \textbf{Step 2.2}, that 
$$\sqrt{nh_\alpha}\left[\wg_\alpha(x_\alpha)-g_\alpha(x_\alpha)\right]- \be_1\trasp\wbB_n =\sigma \sqrt{nh_\alpha}  \int\!\be_1\trasp \,\left(\wbA_{0,n}^{-1}(\bx)-\bA_0^{-1}(\bx)\right)\,\wbA_{1,n}(\bx)q_\undalpha(\bx_\undalpha)\,d\bx_\undalpha  =o_\prob(1)\,,$$
which together with (\ref{asdistBn}) concludes the proof.

\vskip0.2in
\noi \textbf{Step 2.1.}
Recall that $Y_i-\breve{\bx}_{i,\alpha}\bH^{(\alpha)}\bbe(\bx)=\sigma \epsilon_i+R(\bX_i, \bx) $, so that 
\begin{equation}
\esp \left\{\psi\left(\frac{Y_i-\breve{\bx}_{i,\alpha}\bH^{(\alpha)}\bbe(\bx)}{\sigma}\right)|\bX_i\right\}= \lambda\left(\frac{R(\bX_i, \bx)}{\sigma}\right)\,.
\label{ecua4}
\end{equation}
Define $\wtbA_{1,n}(\bx)$ as in Lemma \ref{apend2}.1, i.e.,
$$\wtbA_{1,n}(\bx)=\frac{1}{n}\sum_{i=1}^n \itK_{\bH_d}(\bX_i- \bx)p(\bX_i)\lambda\left(\frac{R(\bX_i, \bx)}{\sigma}\right)\breve{\bx}_{i,\alpha}\;,$$ 
and note that  (\ref{ecua4}) entails that $\esp \wtbA_{1,n}(\bx)=\esp\wbA_{1,n}(\bx)$. Moreover, we have that $\wbB_n=\wbB_{n,1}+\wbB_{n,2}$ where
\begin{eqnarray*}
\wbB_{n,1}&=&\sigma\sqrt{nh_\alpha}\int\!\bA_0^{-1}(\bx)\wtbA_{1,n}(\bx)\,q_\undalpha(\bx_\undalpha)\,d\bx_{\undalpha}\\
\wbB_{n,2}&=&\sigma\sqrt{nh_\alpha}\int\!\bA_0^{-1}(\bx)\left[\wbA_{1,n}(\bx)-\wtbA_{1,n}(\bx)\right]q_\undalpha(\bx_\undalpha)\,d\bx_\undalpha\,.
\end{eqnarray*}
Then, to derive (\ref{asdistBn}), we have to show that
\begin{itemize}
\item[a)] $\wbB_{n,1}\convprob \bb_{q,\alpha}(x_\alpha)$ with $\bb_{q,\alpha}(x_\alpha)$ given in (\ref{sesgoBn}) .
\item[b)] $\wbB_{n,2}\convdist N_{q+1}(\bcero, \bSi_{q,\alpha}(x_\alpha))$ where $\bSi_{q,\alpha}(x_\alpha) $ is defined in (\ref{varianzaBn}). 
\end{itemize}

\noi  a) To show that $\wbB_{n,1}\convprob \bb_{q,\alpha}$, it is enough to see that $\esp \wbB_{n,1} \to \bb_{q,\alpha}$ and that for all $1\leq j\leq q+1$,  $\var (\wB_{n,1,j}) \to 0$. 

Lemma \ref{apend2}.1 together with the fact that $\bA_0(\bu)=v(\bu)A_0(\psi)\bS^{(\alpha)}$ and  $\sqrt{nh_\alpha}h_\alpha^{q+1}=\beta^{(2q+3)/2}$ entail that $\esp \wbB_{n,1}=\sigma\sqrt{nh_\alpha}\int\!\bA_0(\bx)^{-1}\esp\wtbA_{1,n}(\bx)q_\undalpha(\bx_\undalpha)\,d\bx_\undalpha=\bB_{11,n}+\bB_{12,n}$, where
\begin{eqnarray*}
\bB_{11,n}&=&\beta^{(2q+3)/2}\frac{1}{(q+1)!}g_\alpha^{(q+1)}(x_\alpha)(\bS^{(\alpha)})^{-1}\bese_q^{(\alpha)}\\
\bB_{12,n}&=&\sqrt{nh_\alpha}\int\!\bA_0^{-1}(\bx)\esp\left\{\bnu_n(\bx)\right\}q_\undalpha(\bx_\undalpha)\,d\bx_\undalpha\,.
\end{eqnarray*}
Hence, $\esp \wbB_{n,1}  \to \bb_{q,\alpha}$, since $\sup_{\bx\in\itS_Q}\|\bnu_n(\bx)\|=h_\alpha^{q+1}o(1)$ and $\sqrt{nh_\alpha}h_\alpha^{q+1}=\beta^{(2q+3)/2}$ entail that $\bB_{12,n}\to\bcero$.

We will now show that $\var (\wB_{n,1,j}) \to 0$, for $1\leq j\leq q+1$. denote as $\wtbB=A_0(\psi)\bS^{(\alpha)}\wbB_{n,1}/\sigma$ and $\wtB_j$ its $j-$th component. Then, it will be enough to show that the variance of $\wtB_j$ converges to  $0$. Note  that  $\bA_0(\bu)=v(\bu)A_0(\psi)\bS^{(\alpha)}$ implies that
\begin{eqnarray*}
\wtB_{j}&=&\sqrt{nh_\alpha}\int\!\frac{1}{v(\bx)}\wtA_{1,n,j}(\bx)q_\undalpha(\bx_\undalpha)\,d\bx_\undalpha\\
&=&\frac 1{\sqrt{n\,h_\alpha}} \sum_{i=1}^n  K_\alpha\left(\frac{X_{i,\alpha}-x_\alpha}{h_\alpha}\right)p(\bX_i)\zeta(\bH_{d-1},\bX_i,x_\alpha)\left(\frac{X_{i,\alpha}-x_\alpha}{h_\alpha}\right)^{j-1}
\end{eqnarray*}
with
\begin{eqnarray*}
\zeta(\bH_{d-1},\bX_i,x_\alpha)&=&\frac{1}{\wth^{d-1}}\int\!\prod_{s\neq\alpha}K_s\left(\frac{X_{i,s}-x_s}{\wth}\right)q_\undalpha(\bx_\undalpha)\lambda\left(\frac{R(\bX_i,\bx)}{\sigma}\right)\,d\bx_\undalpha\\
&=&\int\!\prod_{s\neq\alpha}K_s(u_s)q_\undalpha(\bX_{i,\undalpha}+\bH_{d-1}\bu_\undalpha)\lambda\left(\frac{R(\bX_i,(x_\alpha,\bX_{i,\undalpha}+\bH_{d-1}\bu_\undalpha))}{\sigma}\right)\,d\bu_\undalpha
\end{eqnarray*} 
where $\bH_{d-1}=\mbox{diag}(\wth,\dots,\wth)\in\real^{(d-1)\times (d-1)}$.
Thus, using that $\psi$ is bounded, $q_\undalpha$ is continuous and bounded, we get that $|\zeta(\bH_{d-1},\bX_i,x_\alpha)|\le C$, for all $i$.
Since  $p\le 1$  and  $|X_{1,\alpha}-x_\alpha|\le h_\alpha$ if $K_\alpha\left(({X_{1,\alpha}-x_\alpha})/{h_\alpha}\right)\ne 0$, we conclude  that
\begin{eqnarray*}
\var(\wtB_{j})&=&\frac{1}{h_\alpha}\var\left(K_\alpha\left(\frac{X_{1,\alpha}-x_\alpha}{h_\alpha}\right)p(\bX_i)\zeta(\bH_{d-1},\bX_i,x_\alpha)\left(\frac{X_{1,\alpha}-x_\alpha}{h_\alpha}\right)^{j-1}\right)\\
&\leq&\frac{1}{h_\alpha}\esp\left[K_\alpha^2\left(\frac{X_{1,\alpha}-x_\alpha}{h_\alpha}\right)p^2(\bX_i)\zeta^2(\bH_{d-1},\bX_i,x_\alpha)\left(\frac{X_{1,\alpha}-x_\alpha}{h_\alpha}\right)^{2(j-1)}\right]\,.\\
&\leq&\frac{1}{h_\alpha}\int\!K_\alpha^2\left(\frac{v_\alpha-x_\alpha}{h_\alpha}\right)\zeta^2(\bH_{d-1},\bv,x_\alpha)f_\bX(\bv)\,d\bv\\
&\leq&\int\!K_\alpha^2(u_\alpha)\zeta^2(\bH_{d-1},(x_\alpha+h_\alpha u_\alpha,\bv_\undalpha),x_\alpha)f_\bX(x_\alpha+h_\alpha u_\alpha,\bv_\undalpha)\,du_\alpha d\bv_\undalpha\,.
\end{eqnarray*}
Then, from the dominated convergence theorem  it follows that $\var(\wtB_j)\to 0$ since $\bH_d\to\bcero$ when $n\to \infty$ and  $\zeta^2(\bcero_{d-1},(x_\alpha,\bv_\undalpha),x_\alpha)=0$, since $R(\bx,\bx)=0$ and $\lambda(0)=0$, concluding the proof of a).

 \vskip0.2in
\noi b) Let $\bB_{n,2}=({A_0(\psi)}/{\sigma})\bS^{(\alpha)}\wbB_{n,2}$. To obtain b) it is enough to see that,   for any $\bc\in\real^{q+1}$, $\bc\neq\bcero$,  $\bc\trasp \bB_{n,2}=\sum_{i=1}^n \bc\trasp \bW_{i,n}\convdist N(0,\bc\trasp\bSi_{11}(x_\alpha)\bc)$,
where
$$\bSi_{11}(x_\alpha)=\esp \psi^2(\eps)\int\!\frac{q_\undalpha^2(\bx_\undalpha)}{f_\bX(x_\alpha,\bx_\undalpha)p(x_\alpha,\bx_\undalpha)}\,d\bx_\undalpha \bV_\alpha\;.$$
Denote as $\bH_{d-1}=\mbox{diag}(\wth,\dots,\wth)\in\real^{(d-1)\times (d-1)}$ and 
 \begin{eqnarray*}
 V(\epsilon_i, \bX_i,\bx)&=&\delta_i\psi\left(\eps_i+\frac{R(\bX_i,\bx)}{\sigma}\right)-p(\bX_i)\lambda\left(\frac{R(\bX_i,\bx)}{\sigma}\right)\,,\\ 
\gamma(\epsilon_i,\wth,\bX_i,x_\alpha)&=&\frac{1}{\wth^{d-1}}\int\! \frac{1}{v(\bx)}\prod_{j\neq\alpha}K_j\left(\frac{X_{i,j}-x_j}{\wth}\right)q_\undalpha(\bx_\undalpha)V(\epsilon_i,\bX_i,(x_\alpha,\bx_\undalpha))\,d\bx_\undalpha \\
&=&\int\!\frac{q_\undalpha(\bX_{i,\undalpha}+\bH_{d-1}\bu_\undalpha)}{v(x_\alpha,\bX_{i,\undalpha}+\bH_{d-1}\bu_\undalpha)}\prod_{j\neq\alpha}K_j(u_j) V(\epsilon_i,\bX_i,(x_\alpha,\bX_{i,\undalpha}+\bH_{d-1}\bu_\undalpha))\,d\bu_\undalpha\,,\\
\bW_{i,n}&=&\frac{1}{\sqrt{nh_\alpha}}K_\alpha\left(\frac{X_{i,\alpha}-x_\alpha}{h_\alpha}\right)\breve{\bx}_{i,\alpha} \gamma(\epsilon,\wth,\bX_i,x_\alpha)\,.
\end{eqnarray*}
Note that $ \gamma(\epsilon_i,0,\bX_i,x_\alpha)$ is well defined as
\begin{equation}
\gamma(\epsilon_i,0,\bX_i,x_\alpha)=\frac{q_\undalpha(\bX_{i,\undalpha})}{v(x_\alpha,\bX_{i,\undalpha})} V(\epsilon_i,\bX_i,(x_\alpha,\bX_{i,\undalpha})) \,.
\label{gammaen0}
\end{equation}
It is clear that
$$\wbA_{1,n}(\bx)-\wtbA_{1,n}(\bx)=\frac{1}{n}\sum_{i=1}^n \itK_{\bH_d}(\bX_i-\bx)\breve{\bx}_{i,\alpha}V(\epsilon_i,\bX_i,\bx)\,,$$
hence,
\begin{eqnarray*}
\bB_{n,2}&=&\sqrt{nh_\alpha}\int\! v^{-1}(\bx)\left[\wbA_{1,n}(\bx)-\wtbA_{1,n}(\bx)\right]q_\undalpha(\bx_\undalpha)\,d\bx_\undalpha\\
&=&\frac{1}{\sqrt{nh_\alpha}}\sum_{i=1}^n K_\alpha\left(\frac{X_{i,\alpha}-x_\alpha}{h_\alpha}\right)\breve{\bx}_{i,\alpha} \gamma(\epsilon_i,\wth,\bX_i,x_\alpha)=\sum_{i=1}^n \bW_{i,n}\;.
\end{eqnarray*}
 Let $\bc\in\real^{q+1}$, $\bc\neq\bcero$. Since $\esp (V(\epsilon_i,\bX_i,\bx)|\bX_i)=\bcero$, for all $\bx$, we have that $\esp \bW_{i,n}=\bcero$ so $\esp \bc\trasp\bW_{i,n}=0$. Besides, as $\psi$ and $p$ are continuous functions and $|\delta_i|\leq 1$ we have that $|V(\epsilon_{i},\bX_i,\bx)|\leq C$ for some   constant $C>0$ which entails that $\gamma(\epsilon_i,\wth,\bX_i,x_\alpha)$ is bounded since $\inf_{\bx \in \itS_Q} v(\bx)>0$ and $q_\undalpha$ is  bounded on its support. Therefore, using that   $|\breve{x}_{i,j,\alpha}|\le 1$ when $ \itK_{\bH_d}(\bX_i-\bx)\ne 0$, we obtain that,  for some general constant $C_1>0$,
\begin{eqnarray*}
\sum_{i=1}^n \esp|\bc\trasp\bW_{i,n}|^3&\leq& C_1 n\frac{1}{(nh_\alpha)^{3/2}}\esp\left|K_\alpha\left(\frac{X_{i,\alpha}-x_\alpha}{h_\alpha}\right)\right|^3=C_1\frac{1}{\sqrt{nh_\alpha}}\frac{1}{h_\alpha}\int\!\left|K_\alpha^3\left(\frac{u-x_\alpha}{h_\alpha}\right)\right|f_{X_\alpha}(u)\,du\\
&\leq& C_2\frac{1}{\sqrt{nh_\alpha}}\to 0\,.
\end{eqnarray*}
Hence,  applying the Lyapunov's central limit theorem to the triangular array of independent variables $\{\bc\trasp \bW_{i,n}\}_{i=1}^n$ the proof of b) follows if we show that $\lim_{n\to \infty} \var(\bc\trasp \sum_{i=1}^n \bW_{i,n})=\bc\trasp \bSi_{11}(x_\alpha)\bc$ or equivalently that $\lim_{n\to \infty} \var( \sum_{i=1}^n \bW_{i,n})=  \bSi_{11}(x_\alpha) $.
 
 Using that    $\bW_{1,n}, \dots, \bW_{n,n}$ are independent and that $\esp \bW_{1,n}=0$, we get that    $\var( \sum_{i=1}^n \bW_{i,n})=  n \var(   \bW_{1,n})=n\esp \left(\bW_{1,n}\bW_{1,n}\trasp\right)$. Given $1\leq s,m\leq q+1$,  denote as $E_{sm}=n\esp \left(W_{1,n,s}W_{1,n,m}\right)$ where $W_{1,n,m}$ is the $m-$th component of $\bW_{1,n}$. We have to show that $E_{sm}$ converges to the $(s,m)-$th element of $\bSi_{11}(x_\alpha) $. 
 
 Let $M(\wth, \bu, x_\alpha)=\esp\left[\gamma^2(\epsilon_1,\wth,\bu,x_\alpha)|\bX_1=\bu\right]$, then we have that
 \begin{eqnarray}
E_{sm}&=&\frac{1}{h_\alpha} \esp\left[K_\alpha^2\left(\frac{X_{1,\alpha}-x_\alpha}{h_\alpha}\right)\gamma^2(\epsilon_1,\wth,\bX_1,x_\alpha)\left(\frac{X_{1,\alpha}-x_\alpha}{h_\alpha}\right)^{s+m-2}\right]\nonumber\\
&=& \frac{1}{h_\alpha}\int\!K_\alpha^2\left(\frac{u_\alpha-x_\alpha}{h_\alpha}\right)M(\wth, \bu, x_\alpha) \left(\frac{u_\alpha-x_\alpha}{h_\alpha}\right)^{s+m-2}f_\bX(u_\alpha,\bu_\undalpha)\,d\bu\nonumber\\
&=&\int\!K_\alpha^2(u_\alpha)u_\alpha^{s+m-2} M(\wth, (u_\alpha h_\alpha+x_\alpha,\bu_\undalpha) , x_\alpha)   f_\bX(u_\alpha h_\alpha+x_\alpha,\bu_\undalpha)\,d\bu\;.
\label{cuenta2alpha}
\end{eqnarray}
Note that (\ref{gammaen0}) implies that $M(0, \bu, x_\alpha)$ is well defined and equals
 \begin{equation}
 M(0, \bu, x_\alpha)=\frac{q^2_\undalpha(\bu_{\undalpha})}{v^2(x_\alpha,\bu_{\undalpha})}\esp \left[V^2(\epsilon_1,\bu,(x_\alpha,\bu_{\undalpha}))|\bX_1=\bu\right]
 \label{Mnen0}
 \end{equation}
 Hence,   taking limit in (\ref{cuenta2alpha}) and using the dominated convergence theorem, we get that
$$
\lim_{n\to\infty} E_{sm} =\int\!K_\alpha^2(u_\alpha)u_\alpha^{s+m-2} M(0,(x_\alpha,\bu_\undalpha),x_\alpha)f_\bX(x_\alpha,\bu_\undalpha)\,d\bu\,.
$$
Using that 
$$\esp\left[\delta_1\psi\left(\eps_1+\frac{R(\bX_1,\bx)}{\sigma}\right)|\bX_1\right]=p(\bX_1)\lambda\left(\frac{R(\bX_1,\bx)}{\sigma}\right)\,,$$ 
we get that  
\begin{equation}
\esp \left[V^2(\epsilon_1,\bu,(x_\alpha,\bu_{\undalpha}))|\bX_1=\bu\right]= {p(\bu)} \wtlam_2\left(R(\bu,(x_\alpha,\bu_\undalpha))\right) -  p^2(\bu)\lambda^2\left(\frac{R(\bu,(x_\alpha,\bu_\undalpha))}{\sigma}\right) 
\label{Ven0}
\end{equation}
where  $\wtlam_2(a)=\esp\psi^2(\eps+a)$. Then, the fact that $R((x_\alpha,\bu_\undalpha),(x_\alpha,\bu_\undalpha))=0$, $\lambda(0)=0$ and $\wtlam_2(0)=\esp\psi^2(\eps)$, together with (\ref{Mnen0}) and (\ref{Ven0}) entail that 
$$M(0,(x_\alpha,\bu_\undalpha),x_\alpha)=\frac{q^2_\undalpha(\bu_{\undalpha})}{v^2(x_\alpha,\bu_{\undalpha})}\, p(x_\alpha,\bu_\undalpha) \esp\psi^2(\eps) $$
Hence, we have
\begin{eqnarray*}
\lim_{n\to\infty} E_{sm}&=&\int\!K_\alpha^2(u_\alpha)u_\alpha^{s+m-2}f_\bX(x_\alpha,\bu_\undalpha)\frac{q^2_\undalpha(\bu_\undalpha)}{v^2(x_\alpha,\bu_\undalpha)} \frac{r^2(x_\alpha,\bu_\undalpha)}{p(x_\alpha,\bu_\undalpha)}\esp\psi^2(\eps)\\
&=&\esp\psi^2(\eps)\int\!\frac{q_\undalpha^2(\bu_\undalpha)}{f_\bX(x_\alpha,\bu_\undalpha)p(x_\alpha,\bu_\undalpha)}\,d\bu_\undalpha v_{sm}^{(\alpha)}
\end{eqnarray*}
where $v_{sm}^{(\alpha)}$ is the $(s,m)$th element of the matrix $\bV_\alpha$, concluding the proof of b).

\vskip0.2in
\noi \textbf{Step 2.2} To conclude the proof, we have to show that 
\begin{equation}
\sqrt{nh_\alpha}\left[\wg_\alpha(x_\alpha)-g_\alpha(x_\alpha)\right]- \be_1\trasp\wbB_n =\sigma \sqrt{nh_\alpha}  \int\!\be_1\trasp\,\left(\wbA_{0,n}^{-1}(\bx)-\bA_0^{-1}(\bx)\right)\,\wbA_{1,n}(\bx)q_\undalpha(\bx_\undalpha)\,d\bx_\undalpha  =o_\prob(1)\,,
\label{aprobar}
\end{equation}
Note that $  (\wbA_{0,n}^{-1}(\bx)-\bA_0^{-1}(\bx))\wbA_{1,n}(\bx)=\wbD_1(\bx)+\wbD_2(\bx)$ with
\begin{eqnarray*}
\wbD_1(\bx)&=& (\wbA_{0,n}^{-1}(\bx)-\bA_0^{-1}(\bx))\esp\wbA_{1,n}(\bx)\\
\wbD_2(\bx)&=& (\wbA_{0,n}^{-1}(\bx)-\bA_0^{-1}(\bx))\left(\wbA_{1,n}(\bx)-\esp\wbA_{1,n}(\bx)\right)\,.
\end{eqnarray*}
We will show that, for all $1\leq j\leq q+1$
\begin{eqnarray}
\sqrt{nh_\alpha}\sup_{\bx\in\itS_Q}|\wD_{1,j}(\bx)|&\convprob&0\label{413alpha}\\
\sqrt{nh_\alpha}\sup_{\bx\in\itS_Q}|\wD_{2,j}(\bx)|&\convprob&0\;,\label{414alpha}
\end{eqnarray}
where $\wD_{\ell,j}(\bx)$ is the $j$th coordinate of $\wbD_{\ell}(\bx)$, $\ell=1,2$, which entails that (\ref{aprobar}) holds concluding the proof.

Fix $1\leq j\leq q+1$. In order to prove (\ref{413alpha}), observe that     Lemma \ref{apend2}.1, the fact that $\esp\wA_{1,n,j}(\bx)=\esp\wtA_{1,n,j}(\bx)$ and the Cauchy-Schwartz inequality entail that
$$\sup_{\bx\in\itS_Q}|\wD_{1,j}(\bx)|\leq  \sup_{\bx\in\itS_Q}\left\|\be_j\trasp(\wbA_{0,n}^{-1}(\bx)-\bA_0^{-1})\right\|o(h_\alpha^{q+1})\,,$$ where the term $o(h_\alpha^{q+1})$ does not depend on $\bx$ since $v$ and $g_\alpha^{(q+1)}$ are bounded. 
On the other hand, since $\bS^{(\alpha)}$ is non--singular, $\inf_{\bx\in\itS_Q}|v(\bx)|>0$, $A_0(\psi)\ne 0$ and $\sup_{\bx\in\itS_Q}\left\|\wbA_{0,n} (\bx)-\bA_0 (\bx)\right\|\convprob 0$ we get that $\sup_{\bx\in\itS_Q}\left\|\wbA_{0,n}^{-1}(\bx)-\bA_0^{-1}(\bx)\right\|\convprob 0$. Hence, since
$\sqrt{nh_\alpha}h_\alpha^{q+1}=\beta^{(2q+3)/2}$ we have that $\sqrt{nh_\alpha}\sup_{\bx\in\itS_Q}|\wD_{1,j}(\bx)|\convprob 0$ for all $1\leq j\leq q+1$, so the proof of (\ref{413alpha}) is concluded.

To prove (\ref{414alpha}), we will use Lemma \ref{apend1}.2 with $\theta_n=\sqrt{\log{n}/(nh_\alpha\wth^{d-1})}$ applied to each coordinate of vector $\wbA_{1,n}(\bx)=(\wA_{1,n,1}(\bx),\dots, \wA_{1,n,q+1}(\bx))\trasp$ obtaining that, for $1\leq j\leq q+1$,
$$\sup_{\bx\in\itS_Q}\left|\wA_{1,n,j}(\bx)-\esp\wA_{1,n,j}(\bx)\right|=O_\prob\left(\left(\frac{\log{n}}{nh_\alpha\wth^{d-1}}\right)^{1/2}\right)\,.$$
On the other hand, as above, from Lemma \ref{apend2}.1,  we get that $\sup_{\bx\in\itS_Q}\left|\esp\wA_{1,n,j}(\bx)\right|=o(h_\alpha^{q+1}) $. 
Then, using $\sup_{\bx\in\itS_Q}\left\|\wbA_{0,n}^{-1}(\bx)-\bA_0^{-1}(\bx)\right\|\convprob \bcero$ and $\inf_{\bx\in\itS_Q}\nu_1(\bA_0(\bx))>0$, we conclude that $\sup_{\bx\in\itS_Q}\nu_{q+1}(\wA_{0,n}^{-1}(\bx))=O_\prob(1)$. Hence, using (\ref{413}), we obtain that
\begin{equation}\label{415alpha}
\sup_{\bx\in\itS_Q}\left\|\bH^{(\alpha)}[\wtbeta(\bx)-\beta(\bx)]\right\|\leq o_\prob(h_\alpha^{q+1})+O_\prob\left(\left(\frac{\log{n}}{nh_\alpha\wth^{d-1}}\right)^{1/2}\right)\,.
\end{equation}
Let $\itK^{\star}(\bu)=|\itK(\bu)|/\int\!|\itK(\bu)|d\bu$, then,   Remark \ref{apend1}.1 implies that  $\wefe(\bx)=(1/n)\sum_{j=1}^n \itK_{\bH_d}^{\star}\left(\bx-\bX_j\right)$ converges uniformly and almost surely to $f_\bX$ (see (\ref{obs3lem3})). Hence, using  \textbf{A2}, we obtain that $\sup_{\bx\in \itC}\wefe(\bx)=O_{\as}(1)$ which together with the fact that $\psi^{\prime\prime}$ is bounded and that each component of $\breve{\bx}_{i,\alpha}$ is smaller or equal to 1 when $\itK_{\bH_d}(\bX_i-\bx)\ne 0$,  leads together with (\ref{415alpha})   to
$$
\sup_{\bx\in\itS_Q}\|\wbA_{02,n}(\bx)\|\le C \sup_{\bx\in\itS_Q}\left\|\bH^{(\alpha)}[\wtbeta(\bx)-\beta(\bx)]\right\| \,\sup_{\bx\in \itC}\wefe(\bx)\le o_\prob(h_\alpha^{q+1})+O_\prob\left(\left(\frac{\log{n}}{nh_\alpha\wth^{d-1}}\right)^{1/2}\right)\,,
$$
which together with the fact that $h_\alpha= \beta n^{-\frac{1}{2q+3}}$ and  $n^{\frac{q+1}{2q+3}}\wth^{d-1}/\log{n}\to\infty$ implies that
\begin{eqnarray}
\sup_{\bx\in\itS_Q}\| \wbA_{02,n}(\bx)  \|O_\prob\left(\left(\frac{\log{n}}{\wth^{d-1}}\right)^{1/2}\right) 
&\le&   n^{-\frac{q+1}{2q+3}}\left(\frac{\log{n}}{\wth^{d-1}}\right)^{1/2}o_\prob\left(1\right) + \left(\frac{\log{n}}{n^{\frac{q+1}{2q+3}}\wth^{d-1}}\right)O_\prob\left(1 \right) \nonumber\\
&\le & o_\prob(1)
\label{cotaA02}
\end{eqnarray}
Recall that $\sup_{\bx\in\itS_Q}\nu_{q+1}(\bA_0(\bx))<\infty$, since $\inf_{\bx\in\itS_Q}\nu_1(\bA_0(\bx))>0$. Therefore,  using the Cauchy-Schwartz inequality,   the fact that $\bA^{-1}_{0}(\bx)-\wbA_{0,n}^{-1}(\bx)= \wbA_{0,n}^{-1}(\bx)(\wbA_{0,n}(\bx)-\bA_{0}(\bx))\bA^{-1}_{0}(\bx)$,  $\wbA_{0,n}(\bx)=\wbA_{01,n}(\bx)+\wbA_{02,n}(\bx)$ and   $\sup_{\bx\in\itS_Q}\nu_{q+1}(\wA_{0,n}^{-1}(\bx))=O_\prob(1)$, we get that
\begin{eqnarray}
\sqrt{nh_\alpha}\sup_{\bx\in\itS_Q}|\wD_{2,j}(\bx)|&\leq&\sigma C_1 \sqrt{nh_\alpha}\sup_{\bx\in\itS_Q}\| \wbA_{0,n}(\bx)-\bA_{0}(\bx) \|\;\sup_{\bx\in\itS_Q}\left\|\wbA_{1,n}(\bx)-\esp\wbA_{1,n}(\bx)\right\|\nonumber\\
&\leq& C_2\sup_{\bx\in\itS_Q}\| \wbA_{01,n}(\bx)-\bA_{0}(\bx) \|O_\prob\left(\left(\frac{\log{n}}{\wth^{d-1}}\right)^{1/2}\right)\nonumber\\
&&+
C_2\sup_{\bx\in\itS_Q}\| \wbA_{02,n}(\bx)  \|O_\prob\left(\left(\frac{\log{n}}{\wth^{d-1}}\right)^{1/2}\right)\nonumber\\
&\leq& C_2\sup_{\bx\in\itS_Q}\| \wbA_{01,n}(\bx)-\bA_{0}(\bx) \|O_\prob\left(\left(\frac{\log{n}}{\wth^{d-1}}\right)^{1/2}\right)+ o_\prob(1)
\label{416alpha}
\end{eqnarray}
where the last inequality follows from (\ref{cotaA02}). 

Recall that $\lambda_1(t)=\esp\psi^{\prime}(\eps_1+t)$ and $\lambda_1(0)=A_0(\psi)$. Denote 
\begin{eqnarray*}
\bLam_{1,n}(\bx)&=&\esp\wbA_{01,n}(\bx)=\esp\left[\itK_{\bH_d}(\bX_1-\bx)r(\bX_1)\lambda_1\left(\frac{R(\bX_1,\bx)}{\sigma}\right)\breve{\bx}_{1,\alpha}\breve{\bx}_{1,\alpha}\trasp\right]\\
&=&\int\!\itK(\bu)v(\bx+\bH_d\bu)\lambda_1\left(\frac{R(\bx+\bH_d\bu,\bx)}{\sigma}\right)\breve{\bu}_\alpha\breve{\bu}_\alpha\trasp\,d\bu\,.
\end{eqnarray*}
Let $\wA_{01,j,m}(\bx)$ be the $(j,m)$th element of matrix $\wtbA_{01,n}$. Then,  analogous arguments to those considered in the proof of Lemma \ref{apend1}.2 allow to show that, for $1\leq j,m\leq q+1$
$$\sup_{\bx\in\itS_Q} \left|\wA_{01,j,m}(\bx)-\esp\wA_{01,j,m}(\bx)\right|=O_\prob\left(\left(\frac{\log{n}}{nh_\alpha\wth^{d-1}}\right)^{1/2}\right)\,.$$
Hence, 
\begin{equation}\label{417alpha}
\sup_{\bx\in\itS_Q}\| \wbA_{01,n}(\bx)-\bLam_{1,n}(\bx) \|O_\prob\left(\left(\frac{\log{n}}{\wth^{d-1}}\right)^{1/2}\right) \le n^{-\frac{q+1}{2q+3}}\left(\frac{\log{n}}{\wth^{d-1}}\right) O_\prob\left(1\right)  =o_\prob(1)
\end{equation}
Hence, (\ref{416alpha}) and (\ref{417alpha}) entail that to conclude the proof of  (\ref{414alpha}) we only have to show that 
\begin{equation}
\sup_{\bx\in\itS_Q}\| \bLam_{1,n}(\bx) - \bA_{0}(\bx)\|O_\prob\left(\left(\frac{\log{n}}{\wth^{d-1}}\right)^{1/2}\right)  =o_\prob(1)
\label{falta}
\end{equation}
Denote as  $A_{0,j,m}(\bx)$ and $\Lambda_{1,n, j,m}(\bx)$the $(j,m)$ element of $\bA_0(\bx)$ and $\bLam_{1,n}(\bx)$, respectively. Then, using that $v$ is $\ell$ times differentiable, $\lambda_1=\lambda^{\prime}$ is $\ell-1$ times differentiable, the kernel $L$ is of order $\ell$ and that $R(\bx+\bH_d\bu, \bx)=\sum_{j\ne \alpha}\wth^{\ell} u_j^\ell g_j^{(\ell)}(\xi_j)+ \sum_{r=1}^{\ell-1} \wth^r u_j^r g_j^{(r)}(x_j)+ h^{q+1} u_\alpha^{q+1} g_\alpha^{(q+1)}(\xi_\alpha)$, since $g_j$ and $g_\alpha^{(\ell)}$ are continuously differentiable functions, we obtain that, for $1\leq j,m\leq q+1$
\begin{eqnarray*}
\sup_{\bx\in\itS_Q}|\Lambda_{1,n, j,m}(\bx)-A_{0,j,m}(\bx)|&=&\sup_{\bx\in\itS_Q}\int\!\left| \itK(\bu)\left[v(\bx+\bH_d\bu)\lambda_1\left(\frac{R(\bx+\bH_d\bu,\bx)}{\sigma}\right)\right.\right.\nonumber\\
&&\left.\left.-v(\bx)\lambda_1\left(\frac{R(\bx,\bx)}{\sigma}\right)\right]\breve{u}_{j,\alpha}\breve{u}_{m,\alpha}\,d\bu\right|\nonumber\\
&\leq&C_2\left(\wth^{\ell}+h_\alpha^{q+1}\right)\leq C_3h_\alpha^{q+1}
\end{eqnarray*}
which allow to conclude that, for $1\leq j,m\leq q+1$,
$$\sup_{\bx\in\itS_Q}|\Lambda_{1,n, j,m}(\bx)-A_{0,j,m}(\bx)|O_\prob\left(\left(\frac{\log{n}}{\wth^{d-1}}\right)^{1/2}\right)\leq h_\alpha^{q+1}O_\prob\left(\left(\frac{\log{n}}{\wth^{d-1}}\right)^{1/2}\right)$$
 which combined with  the fact that $\log{n}/(n^{(q+1)/(2q+3)}\wth^{d-1})\to 0$ conclude the proof of (\ref{falta}) and also that of the Theorem \ref{ASMq}.1. \square

\noi \textbf{Proof of Theorem \ref{ASMq}.2}. The proof follows using similar arguments to those considered in the proof of Theorem \ref{ASMq}.1, noting that
\begin{eqnarray*}
\wg^{(\nu)}_{\alpha,\eme_{q,\alpha}}(x_\alpha)-g^{(\nu)}_\alpha(x_\alpha)&= &\nu!\int\!\be_{\nu+1}\trasp\,\left[\wbbe(x_{\alpha},\bu_\undalpha )-\bbe(x_{\alpha},\bu_\undalpha )\right]q_\undalpha (\bu_\undalpha )\,d\bu_\undalpha \\
&= &\nu!h_{\alpha}^{-\nu}\;\int\!\be_{\nu+1}\trasp\,\bH^{(\alpha)}\left[\wbbe(x_{\alpha},\bu_\undalpha )-\bbe(x_{\alpha},\bu_\undalpha )\right]q_\undalpha (\bu_\undalpha )\,d\bu_\undalpha \mbox{ . \square}
\end{eqnarray*}

\section*{References}
\small
 
\begin{description}
 \item
Alimadad, A. and Salibian-Barrera, M. (2012). An outlier-robust fit for generalized additive models with applications to disease outbreak detection. \textsl{Journal of American Statistical Association}, \textbf{106}, 719-731.

 \item
Baek, J. and Wehrly, T. (1993). Kernel estimation for additive models under dependence. \textsl{Stochastic Processes and their Applications}, \textbf{47}, 95-112.

  \item
Bianco A., and Boente, G. (1998). Robust Kernel Estimators for Additive Models with Dependent Observations. \textsl{The Canadian Journal of Statistics}, \textbf{6}, 239-255.

\item
Boente, G. and Fraiman, R. (1989). Robust nonparametric regression estimation. \textsl{Journal of Multivariate Analysis}, \textbf{29}, 180-198.

\item
Boente, G. and Mart\'{\i}nez, A. (2015). Estimating additive models with missing responses. \textsl{Communications in Statistics: Theory and Methods}. doi:10.1080/03610926.2013.815780

\item
  Boente, G., Mart\'{\i}nez, A. and Salibian--Barrera, M. (2015). Robust estimators for additive models using
backfitting. Technical report available at \url{http://www.stat.ubc.ca/~matias/RBF} 

\item
 Boente, G., Ruiz, M. and Zamar, R. (2010).   On a robust local estimator for the scale function in  heteroscedastic nonparametric regression. \textsl{Statistics and Probability Letters}, \textbf{80}, 1185-1195.

\item
Buja, A., Hastie, T. and Tibshirani, R. (1989). Linear smoothers and additive models (with discussion). \textsl{Annals of Statistics}, \textbf{17}. 453-555.

\item
 Chen, R., H\"ardle, W., Linton, O. B. and Serverance-Lossin, E.  (1996). Nonparametric estimation of additive separable regression models. \textsl{Statistical Theory and Computational Aspects of Smoothing, Proceedings of the COMPSTAT 94 Satellite Meeting}. Eds: Wolfgang H\"ardle, Michael G. Schimek. Springer,  pp 247-265.

\item
Croux, C., Gijbels, I. and Prosdocimi, I. (2011) Robust estimation of mean and dispersion functions in extended generalized additive models. \textsl{Biometrics}, \textbf{68}, 31-44.
\item
Ferraty, F. and Vieu, P. (2006). \textsl{Nonparametric Functional Data Analysis}. Springer Series in Statistics.

\item
Hastie, T.J. y Tibshirani, R.J. (1990). \textsl{Generalized Additive Models}.  Monographs on Statistics and Applied Probability No. 43. Chapman and Hall, London.

\item
Hengartner, N. y Sperlich, S. (2005). Rate optimal estimation with the integration method in the presence of many covariates. \textsl{Journal of Multivariate Analysis}, \textbf{95}, 246-272.

\item
 Kong, E.,  Linton, O. y Xia, Y. (2010). Uniform Bahadur representation for
local polynomial estimates of $M-$regression and its application to the additive model. \textsl{Econometric Theory}, \textbf{26}, 1529-1564.

\item
 Li, J., Zheng, Z. and Zheng, M. (2012) Robust estimation of additive models based on marginal integration. Available at \url{ http://www.math.pku.edu.cn:8000/var/preprint/7065.pdf}  

\item 
Linton, O. and Nielsen, J. (1995). A kernel method of estimating structured nonparametric regression based on marginal integration. \textsl{Biometrika}, \textbf{82}, 93-101.

\item 
Mart\'{\i}nez, A. (2014). \textsl{Inferencia en modelos aditivos}. PhD. dissertation, Universidad de Buenos Aires. Available at \url{http://cms.dm.uba.ar/academico/carreras/doctorado/TesisDoctorado_AlejandraMartinez.pdf}

\item 
Nielsen, J. P. and Linton, O. B. (1998). An optimization interpretation of integration and back-fitting estimators for separable nonparametric models. \textsl{Journal of the Royal Statistical Society}, \textbf{60}, 2017-222.

\item
Pollard, D. (1084) \textsl{Convergence of stochastinc processes}. Springer-Verlag, New York.

\item
 Severance-Lossin, E. and Sperlich, S. (1999). Estimation of derivatives for additive separable models, \textsl{Statistics}, \textbf{33}, 241-265.

\item
 Sperlich, S., Linton, O. and  H\"ardle, W. (1999). Integration and backfitting methods in additive models-finite sample properties and comparison. \textsl{TEST}, \textbf{8}, 419-458. 

\item
Stone, C.J. (1980). Optimal rates of convergence for nonparametric estimators. \textsl{Annals of Statistics}, \textbf{8}, 1348-1360.

\item
Stone, C.J. (1980). Optimal global rates of convergence for nonparametric regression. \textsl{Annals of Statistics}, \textbf{10}, 1040-1053.

\item
Stone, C.J. (1985). Additive regression and other nonparametric models. \textsl{Annals of Statistics}, \textbf{13}, 689-705.

\item
Tj{\o}stheim, D.  and Auestad, B. (1994) Nonparametric identification of nonlinear time series: Selecting significant lags. \textsl{Journal of the American Statistical Association}, \textbf{89}, 1410-1430.

\item
Wong, R. K. W., Yao, F. and L. T. C. M. (2014). Robust estimation for generalized additive models. \textsl{Journal of Computational and Graphical Statistics}, \textbf{23}, 270-289.

\end{description}

\end{document}

%% file: definiciones.tex
\newcommand\bb {\mathbf b}
\newcommand\bc {\mathbf c}

\newcommand\be {\mathbf e}

\newcommand\bh {\mathbf h}

\newcommand\beme {\mathbf m}

\newcommand\br {\mathbf r}
\newcommand\bese {\mathbf s}
\newcommand\bt {\mathbf t}
\newcommand\bu {\mathbf u}
\newcommand\bv {\mathbf v}

\newcommand\bx {\mathbf x}
\newcommand\by {\mathbf y}

\newcommand\bA {\mathbf A}
\newcommand\bB {\mathbf B}

\newcommand\bD {\mathbf D}

\newcommand\bJ {\mathbf J}

\newcommand\bH {\mathbf H}
\newcommand\bM {\mathbf M}

\newcommand\bS {\mathbf S}

\newcommand\bV {\mathbf V}
\newcommand\bW {\mathbf W}
\newcommand\bX {\mathbf X}

\newcommand\indica {\mathbb{I}}

\newcommand\wa {\widehat{{a}}}

\newcommand\wefe {\widehat{f}}

\newcommand\wg {\widehat{g}}

\newcommand\wese {\widehat{s}}

\newcommand\wA {\widehat{A}}
\newcommand\wbA {\widehat{\bA}}
\newcommand\wbB {\widehat{\bB}}
\newcommand\wB {\widehat{B}}

\newcommand\wD {\widehat{D}}
\newcommand\wbD {\widehat{\bD}}

\newcommand\wM {\widehat{M}}
\newcommand\wbM {\widehat{\bM}}

\newcommand\wV {\widehat{V}}

\newcommand\wZ {\widehat{Z}}

\newcommand\wtg {\widetilde{g}}

\newcommand\wth {\widetilde{h}}

\newcommand\wtm {\widetilde{m}}

\newcommand\wtbr {\widetilde{\br}}

\newcommand\wtbA {\widetilde{\bA}}
\newcommand\wtbB {\widetilde{\bB}}
\newcommand\wtA {\widetilde{A}}
\newcommand\wtB {\widetilde{B}}

\newcommand\wtS {\widetilde{S}}

\newcommand\wttS {\widetilde{\wtS}}

\newcommand\itB {{\mathcal{B}}}
\newcommand\itC {{\mathcal{C}}}
\newcommand\itD {{\mathcal{D}}}

\newcommand\itG {{\mathcal{G}}}
\newcommand\itH {{\mathcal{H}}}
\newcommand\itI {{\mathcal{I}}}

\newcommand\itK {{\mathcal{K}}}

\newcommand\itN {{\mathcal{N}}}

\newcommand\itR {{\mathcal{R}}}
\newcommand\itS {{\mathcal{S}}}

\newcommand\itV {{\mathcal{V}}}

\newcommand{\eps}{\varepsilon}

\newcommand\bbe {\mbox{\boldmath $\beta$}}

\newcommand\bnu {\mbox{\boldmath $\nu$}}

\newcommand\bthe {\mbox{\boldmath $\theta$}}

\newcommand\bxi {\mbox{\boldmath $\xi$}}

\newcommand\bLam {\mbox{\boldmath $\Lambda$}}

\newcommand\bPsi {\mbox{\boldmath $\Psi$}}
\newcommand\bSi {\mbox{\boldmath $\Sigma$}}

\newcommand\wbeta {\widehat{\beta}}
\newcommand\wbbe {\widehat{\bbe}}

\newcommand\wmu {\widehat{\mu}}

\newcommand\wtheta {\widehat{\theta}}
\newcommand\wbthe {\widehat{\bthe}}

\newcommand\wzeta {\widehat{\zeta}}

\newcommand\wDelta {\widehat{\Delta}}

\newcommand\wUps  {\widehat{\Upsilon}}

\newcommand\wtbbe {\widetilde{\bbe}}

\newcommand\wtbeta {\widetilde{\beta}}

\newcommand\wtlam {\widetilde{\lambda}}

\newcommand\undalpha {{\underline{\alpha}}}

\newcommand{\hub}{\mbox{\scriptsize \sc h}}
\newcommand{\tuk}{\mbox{\scriptsize \sc t}}

\newcommand{\eme}{\mbox{\scriptsize \sc m}}

\newcommand{\rob}{\mbox{\footnotesize \sc r}}
\newcommand{\cla}{\mbox{\footnotesize \sc c}}

\def\real{\mathbb{R}}
\def\natu{\mathbb{N}}

\newcommand{\esp}{\mathbb{E}}
\newcommand{\prob}{\mathbb{P}}

\newcommand{\var}{\mbox{\sc Var}}

\newcommand{\aco}{\mbox{\footnotesize a.co.}}
\newcommand{\as}{\mbox{\footnotesize a.s.}}

\newcommand{\convpp}{ \buildrel{a.s.}\over\longrightarrow}

\newcommand{\convprob  }{ \buildrel{p}\over\longrightarrow}

\newcommand{\convdist}{ \buildrel{D}\over\longrightarrow}

\newcommand{\trasp}{^{\mbox{\footnotesize \sc t}}}

\newcommand\bcero {{\bf{0}}}

\def\mad{\mathop{\mbox{\sc mad}}}

\def\argmin{\mathop{\mbox{argmin}}}

\newcommand\noi{\noindent}

\def\square{\ifmmode\sqr\else{$\sqr$}\fi}
\def\sqr{\vcenter{
         \hrule height.1mm
         \hbox{\vrule width.1mm height2.2mm\kern2.18mm
\vrule width.1mm}
         \hrule height.1mm}}

\newcommand\ls {\mbox{\scriptsize\sc ls}}